\documentclass[11pt]{article}%
\usepackage{amsmath}
\usepackage{amsfonts}
\usepackage{amssymb}
\usepackage{graphicx}
\usepackage{longtable}
\usepackage{pdflscape}
\usepackage{multirow}
\usepackage[normalem]{ulem}
\usepackage{natbib}
\usepackage[letterpaper,tmargin=1in,bmargin=1in,lmargin=1in,rmargin=1in,dvips]%
{geometry}
\usepackage[onehalfspacing]{setspace}%
\numberwithin{equation}{section}
\providecommand{\U}[1]{\protect\rule{.1in}{.1in}}
\newtheorem{theorem}{Theorem}

\newtheorem{corollary}[theorem]{Corollary}

\newtheorem{lemma}{Lemma}

\newtheorem{proposition}[theorem]{Proposition}

\ifx\pdfoutput\relax\let\pdfoutput=\undefined\fi
\newcount\msipdfoutput
\ifx\pdfoutput\undefined\else
\ifcase\pdfoutput\else
\ifx\paperwidth\undefined\else
\ifdim\paperheight=0pt\relax\else\pdfpageheight\paperheight\fi
\ifdim\paperwidth=0pt\relax\else\pdfpagewidth\paperwidth\fi
\fi\fi\fi
\begin{document}

\title{\textbf{A Powerful Subvector Anderson-Rubin Test in Linear Instrumental
Variables Regression with Conditional Heteroskedasticity}\thanks{We would like
to thank the Editor, Peter Phillips, the Co-Editor, Michael Jansson, and two referees for very helpful comments.
Guggenberger thanks the European University Institute for its hospitality
while parts of this paper were drafted. Mavroeidis gratefully acknowledges the
research support of the European Research Council via Consolidator grant
number 647152. We would like to thank Donald Andrews for detailed comments and for providing his Gauss
code of, and explanations about, \citet{Andrews2017} and Lixiong Li for
outstanding research assistance for the Monte Carlo study. We thank seminar
participants in Amsterdam, Bologna, Bristol, Florence (EUI), Indiana,
Konstanz, Manchester, Mannheim, Paris (PSE), Pompeu Fabra, Regensburg,
Rotterdam, Singapore (NUS and SMU), Tilburg, Toulouse, T\"{u}bingen, and
Zurich, and conference participants at the Institute for Fiscal Studies
(London) for helpful comments.}}
\author{$%
\begin{array}
[c]{c}%
\text{Patrik Guggenberger}\\
\text{Department of Economics}\\
\text{Pennsylvania State University}%
\end{array}
$
\and $%
\begin{array}
[c]{c}%
\text{Frank Kleibergen}\\
\text{Department of Quantitative Economics}\\
\text{University of Amsterdam}%
\end{array}
$
\and \vspace{0.25in}$%
\begin{array}
[c]{c}%
\text{Sophocles Mavroeidis}\\
\text{Department of Economics}\\
\text{University of Oxford}%
\end{array}
$}
\date{\today\vspace{0.25in}}
\maketitle

\begin{abstract}
We introduce a new test for a two-sided hypothesis involving a subset of the
structural parameter vector in the linear instrumental variables (IVs) model.
\citet{GuggenbergerKleibergenMavroeidis2019}, GKM19 from now on, introduce a
subvector Anderson-Rubin (AR) test with data-dependent critical values that
has asymptotic size equal to nominal size for a parameter space that allows
for arbitrary strength or weakness of the IVs and has uniformly nonsmaller
power than the projected AR test studied in
\citet{GuggenbergerKleibergenMavroeidisChen2012}. However, GKM19 imposes the
restrictive assumption of conditional homoskedasticity. The main contribution
here is to robustify the procedure in GKM19 to arbitrary forms of conditional
heteroskedasticity. We first adapt the method in GKM19 to a setup where a
certain covariance matrix has an approximate Kronecker product (AKP) structure
which nests conditional homoskedasticity. The new test equals this adaption
when the data is consistent with AKP structure as decided by a model selection
procedure. Otherwise the test equals the AR/AR test in \citet{Andrews2017}
that is fully robust to conditional heteroskedasticity but less powerful than
the adapted method. We show theoretically that the new test has asymptotic
size bounded by the nominal size and document improved power relative to the
AR/AR test in a wide array of Monte Carlo simulations when the covariance
matrix is not too far from AKP.\medskip



\end{abstract}

\section{Introduction}

Robust and powerful subvector inference constitutes an important problem in
Econometrics. For instance, it is standard practice to report confidence
intervals on each of the coefficients in a linear regression model. By robust
we mean a testing procedure for a hypothesis of (or a confidence region for) a
subset of the structural parameter vector such that the asymptotic size is
bounded by the nominal size for a parameter space that allows for weak or
partial identification. Recent contributions to robust subvector inference
have been made in the context of the linear instrumental variables (IVs from
now on) model (see, for example, \citet{DufourTaamouti2005},
\citet{GuggenbergerKleibergenMavroeidisChen2012} (GKMC from now on),
\citet{GuggenbergerKleibergenMavroeidis2019} (GKM19 from now on), and
\citet{Kleibergen2021}), GMM models (see, for example,
\citet{ChaudhuriZivot2011}, \citet{AndrewsCheng2014},
\citet{AndrewsMikusheva2016}, \citet{Andrews2017}, and
\citet{HanMcCloskey2019}), and also models defined by moment (in)equalities
(see, for example, \citet{BugniCanayShi2017}, \citet{Gafarov2017}, and
\citet{KaidoMolinariStoye2019}). GKM19 introduce a new subvector test that
compares the AR subvector statistic to conditional critical values that adapt
to the strength or weakness of identification and verify that the resulting
test has correct asymptotic size for a parameter space that imposes
conditional homoskedasticity (CHOM from now on) and uniformly improves on the
power of the projected AR test studied in \citet{DufourTaamouti2005}.

The contribution of the current paper is to provide a robust subvector test
that improves the power of another robust subvector test by combining it with
a more powerful test that is robust for only a smaller parameter space. More
specifically, in the context of the linear IV model, we first provide a
modification of the subvector AR test of GKM19, called the AR$_{AKP,\alpha}$
test, where $\alpha$ denotes the nominal size. We verify that it has correct
asymptotic size for a parameter space that nests the setup with CHOM and also
allows for particular cases of conditional heteroskedasticity (CHET from now
on), namely setups where a particular covariance matrix has a Kronecker
product (KP from now on) structure. For example, the data generating process
(DGP from now on) has a KP structure if the vector of structural and
reduced-form errors equals a random vector independent of the IVs times a
scalar function of the IVs. In particular then, the variances of all the
errors depend on the IVs by the same multiplicative constant given as a scalar
function of the IVs. In the companion paper
\citet{GuggenbergerKleibergenMavroeidis2022} (GKM22 from now on) we find that
KP structure is not rejected at the 5\% nominal size in more than 63\% of
empirical data sets we studied of several recently published empirical papers
(namely, 38 of 60 specifications are not rejected; and, including cases with
clustering, 56 out of 118 are not rejected). For comparison, CHOM is rejected
for 57 of the 60 specifications considered at the 5\% nominal size, using the
test in Kelejian (1982). Of course, these findings do not prove that empirical
data sets do have KP\ structure as the low number of rejections of KP
structure may be due to type II errors of the test. However, coupled with the
quite favorable finite sample power results of the test of KP structure
reported in GKM22 (for sample sizes of $n=200$) we believe that KP\ structure
might be compatible with a sizable subset of empirical data sets.

Second, depending on a model selection mechanism that determines whether the
data are compatible with KP, the recommended test then equals the
AR$_{AKP,\alpha}$ test or the AR/AR test in \citet{Andrews2017} that is robust
to arbitrary forms of CHET. We show that the recommended test has correct asymptotic size. An important ingredient in establishing that is showing that the AR$_{AKP,\alpha}$ test does not
reject less often under the null hypothesis than the AR/AR test when the data
are close to KP structure.

We propose two different model selection methods. One is based on the KPST
test statistic introduced in GKM22 for testing the null hypothesis that a
covariance matrix has KP structure. The other one is based on the standardized
norm of the distance between the covariance matrix estimator and its closest
KP approximation. As in the model selection method proposed in
\citet{AndrewsSoares2010}, we compare the test statistic to a user chosen
threshold that, in the asymptotics, is let go to infinity. The thresholds can
be chosen differently depending on the number of IVs $k$ and parameters not
under test. Based on comprehensive finite sample simulations we provide
choices for the thresholds for several values of $k$ that lead to good control
of the finite sample size.

As the main contribution of the paper, we verify that the resulting test,
called $\varphi_{MS-AKP,\alpha}$ test, has asymptotic size bounded by the
nominal size $\alpha$ under certain conditions on the selection mechanism and
implementation of the AR/AR test at nominal size $\alpha-\delta$ for some
arbitrarily small $\delta>0$.

In a Monte Carlo study, we compare the suggested new test $\varphi
_{MS-AKP,\alpha}$ with several alternatives given in \citet{Andrews2017}, in
particular, the AR/AR and the AR/QLR1 tests. \citet{Andrews2017} fills a very
important gap in the literature on subvector inference by providing two-step
Bonferroni-like methods\footnote{See \citet{McCloskey2017} for a general reference
on Bonferroni methods in nonstandard testing setups.} for a rich class of
models that nests GMM, that i) control the asymptotic size under relatively
mild high-level conditions that allow for CHET, ii) are asymptotically
non-conservative (in contrast to standard Bonferroni methods) and iii) for the
case of AR/QLR1 is asymptotically efficient under strong identification (while
the AR/AR test is not asymptotically efficient under strong identification in
overidentified situations). In contrast, the test considered here,
$\varphi_{MS-AKP,\alpha}$, can only be used in the linear IV model and is not
asymptotically efficient under strong identification. The Monte Carlo study
finds that $\varphi_{MS-AKP,\alpha}$ has uniformly higher rejection
probabilities than the AR/AR test for all the DGPs considered. That includes
the null rejection probabilities (NRPs from now on) with the $\varphi
_{MS-AKP,\alpha}$ test having finite sample size of 6\% versus the 5.4\% of
the AR/AR test at nominal size 5\%. Based on the Monte Carlo study we conclude
that relative to the AR/QLR1 test, $\varphi_{MS-AKP,\alpha}$ can be a useful
alternative in terms of power in situations of weak or mixed identification
strengths when the degree of overidentification is small and the covariance
matrix of the data is not too far from KP structure. Whenever the data are
compatible with KP structure, it also offers an important computational
advantage because the AR$_{AKP,\alpha}$ test is given in closed form. In
contrast, implementation of the two-step Bonferroni-like methods require
minimization of a statistic over a set that has dimension equal to the number
of parameters not under test. The computation time should grow exponentially
in the dimension of that set which constitutes a computational challenge
especially when an applied researcher uses the proposed methods for the
construction of a confidence region by test inversion. This being said, an
applied researcher who uses the $\varphi_{MS-AKP,\alpha}$ test has to be ready
to implement the AR/AR test in case it is determined that KP structure is not
compatible with the data. Given the construction of the AR$_{AKP,\alpha}$ test
it is not surprising to find the relative best performance of the
$\varphi_{MS-AKP,\alpha}$ test to occur under weak identification. Namely, the
critical values of the former test adapt to the strength of identification and
can be substantially lower than the corresponding chi-square critical values
when identification is deemed to be weak.

The rest of the paper is organized as follows. In Section \ref{AKP} we
introduce a version of a subvector \citet{AndersonRubin1949} test that has
correct asymptotic size for a parameter space that imposes an approximate
Kronecker product (AKP) structure for the covariance matrix. In Section
\ref{Het} we introduce the new test that has correct asymptotic size for a
parameter space that does not impose any structure on the covariance matrix
and therefore, in particular, allows for arbitrary forms of conditional
heteroskedasticity. Finally, in Section \ref{MC section} we study the finite
sample properties of the test. Proofs are given in the Appendix at the end.

Notation: Throughout the paper, we denote by \textquotedblleft$\otimes
$\textquotedblright\ the KP of two matrices, by $vec(\cdot)$ the column
vectorization of a matrix, and by $||\cdot||$ the Frobenius
norm.\footnote{Recall the Frobenius norm for a matrix $A=(a_{ij})\in
\Re^{m\times n}$ is defined as $||A||^{2}:=%
{\textstyle\sum\nolimits_{i=1}^{m}}
{\textstyle\sum\nolimits_{j=1}^{n}}
a_{ij}^{2}$. When $A$ is a vector the Frobenius and the Euclidean norm are
numerically equivalent.} We use the notation $M_{A}:=I_{n}-P_{A}$ and
$P_{A}:=A(A^{\prime}A)^{-1}A^{\prime}$ for any full rank matrix $A\in
\Re^{n\times k}.$

\section{Subvector AR Test under Approximate Kronecker Product
Structure\label{AKP}}

Assume the linear IV model is given by the equations%
\begin{align}
y\hspace{-0.03in}  &  =\hspace{-0.03in}Y\beta+W\gamma+\varepsilon\nonumber\\
Y\hspace{-0.03in}  &  =\hspace{-0.03in}\overline{Z}\Pi_{Y}+V_{Y}\nonumber\\
W\hspace{-0.03in}  &  =\hspace{-0.03in}\overline{Z}\Pi_{W}+V_{W},
\label{struc}%
\end{align}
where $y\in\Re^{n},$ $Y\in\Re^{n\times m_{Y}},$ $W\in\Re^{n\times m_{W}},$ and
$\overline{Z}\in\Re^{n\times k}.$ Here, $W$ contains endogenous regressors,
while the regressors $Y$ may be endogenous or exogenous. We assume that
$k-m_{W}\geq1$ and $m_{W}\geq1.$ The reduced form can be written as%
\begin{equation}
\left(
\begin{array}
[c]{ccc}%
y & Y & W
\end{array}
\right)  =\overline{Z}\left(
\begin{array}
[c]{cc}%
\Pi_{Y} & \Pi_{W}%
\end{array}
\right)  \left(
\begin{array}
[c]{ccc}%
\beta & I_{m_{Y}} & 0^{m_{Y}\times m_{W}}\\
\gamma & 0^{m_{W}\times m_{Y}} & I_{m_{W}}%
\end{array}
\right)  +\underbrace{\left(
\begin{array}
[c]{ccc}%
v_{y} & V_{Y} & V_{W}%
\end{array}
\right)  }_{V}, \label{eq: RF}%
\end{equation}
where $v_{y}:=V_{Y}\beta+$ $V_{W}\gamma+\varepsilon$ (which depends on the
true $\beta$ and $\gamma),$ $V_{W}^{\prime}=(V_{W,1},\ldots,V_{W,n}),$
$V_{Y}^{\prime}=(V_{Y,1},\ldots,V_{Y,n}),$ $\overline{Z}^{\prime}%
=(\overline{Z}_{1},\ldots,\overline{Z}_{n}).$ By $V_{i},$ for $i=1,...,n,$ we
denote the $i$-th row of $V$ written as a column vector and similarly for
other matrices.

The objective is to test the subvector hypothesis%
\begin{equation}
H_{0}:\beta=\beta_{0}\text{ against\ }H_{1}:\beta\neq\beta_{0},
\label{eq: hypotheses}%
\end{equation}
using tests whose size, i.e. the highest NRP over a large class of
distributions for $(\varepsilon_{i},\overline{Z}_{i}^{\prime},V_{Y,i}^{\prime
},V_{W,i}^{\prime})$ and the unrestricted nuisance parameters $\Pi_{Y},$
$\Pi_{W},$ and $\gamma$, equals the nominal size $\alpha$, at least
asymptotically. In particular, weak identification and non-identification of
$\beta$ and $\gamma$ are allowed for. The setup allows testing the
coefficients of exogenous or endogenous regressors $Y$ in the presence of
endogenous regressors $W$. We impose the following assumption as in GKM19
(from where the name of the assumption is inherited).\medskip

\noindent\textbf{Assumption B:} The random vectors $(\varepsilon_{i}%
,\overline{Z}_{i}^{\prime},V_{Y,i}^{\prime},V_{W,i}^{\prime})$ for $i=1,...,n$
in (\ref{struc}) are i.i.d. with distribution $F.$\smallskip

For a given sequence $a_{n}=o(1)$ in $\Re_{\geq0},$ we define a sequence of
parameter spaces $\mathcal{F}_{AKP,a_{n}}$ for $(\gamma,\Pi_{W},\Pi_{Y},F)$
under the null hypothesis $H_{0}:\beta=\beta_{0}$ that is larger than the
corresponding ones in GKMC and GKM19 in that general forms of AKP structures
for the variance matrix
\begin{equation}
\overline{R}_{F}:=E_{F}(vec(\overline{Z}_{i}U_{i}^{\prime})(vec(\overline
{Z}_{i}U_{i}^{\prime}))^{\prime})\in\Re^{kp\times kp} \label{def R_F}%
\end{equation}
are allowed for.\footnote{Regarding the notation $(\gamma,\Pi_{W},\Pi_{Y},F)$
and elsewhere, note that we allow as components of a vector column vectors,
matrices (of different dimensions), and distributions.} Namely, for
$U_{i}:=(\varepsilon_{i}+V_{W,i}^{\prime}\gamma,V_{W,i}^{\prime})^{\prime}$
(which equals $(v_{yi}-V_{Y,i}^{\prime}\beta,V_{W,i}^{\prime})^{\prime}$),
$p:=1+m_{W},$ and $m:=m_{Y}+m_{W}$ let%
\begin{align}
\mathcal{F}_{AKP,a_{n}}  &  =\{(\gamma,\Pi_{W},\Pi_{Y},F):\gamma
\overset{}{\in}\Re^{m_{W}},\Pi_{W}\overset{}{\in}\Re^{k\times m_{W}},\Pi
_{Y}\overset{}{\in}\Re^{k\times m_{Y}},\nonumber\\
&  \text{ }E_{F}(||T_{i}||^{2+\delta_{1}})\overset{}{\leq}B,\text{ for }%
T_{i}\overset{}{\in}\{vec(\overline{Z}_{i}U_{i}^{\prime}),||\overline{Z}%
_{i}||^{2}\},\nonumber\\
&  E_{F}(\overline{Z}_{i}V_{i}^{\prime})\overset{}{=}0^{k\times(m+1)},\text{
}\overline{R}_{F}\overset{}{=}G_{F}\otimes\overline{H}_{F}+\Upsilon_{n},\text{
}\nonumber\\
&  \kappa_{\min}(A)\overset{}{\geq}\delta_{2}\text{ for }A\overset{}{\in
}\{E_{F}(\overline{Z}_{i}^{\prime}\overline{Z}_{i}),G_{F},\overline{H}_{F}\}\}
\label{def par spa}%
\end{align}
for symmetric matrices $\Upsilon_{n}\in\Re^{kp\times kp}$ such that
\begin{equation}
||\Upsilon_{n}||\leq a_{n}, \label{approximate}%
\end{equation}
positive definite (pd from now on) symmetric matrices $G_{F}\in\Re^{p\times
p}$ (whose upper left element is normalized to 1) and $\overline{H}_{F}\in
\Re^{k\times k},$ $\delta_{1},\delta_{2}>0,$ $B<\infty$. Note that the factors
in the KP $G_{F}\otimes\overline{H}_{F}$ are not uniquely defined due to the
summand $\Upsilon_{n}$. Note that no restriction is imposed on the variance
matrix of $vec(\overline{Z}_{i}V_{Y,i}^{\prime})$ and, in particular,
$E_{F}(vec(\overline{Z}_{i}V_{Y,i}^{\prime})(vec(\overline{Z}_{i}%
V_{Y,i}^{\prime}))^{\prime})$ does not need to factor into a KP.

The factorization of the covariance matrix into an AKP in line three of
(\ref{def par spa}) is a weaker assumption than CHOM. Under CHOM, we have
$G_{F}=E_{F}\left(  U_{i}U_{i}^{\prime}\right)  $ and $\overline{H}_{F}%
=E_{F}(\overline{Z}_{i}^{\prime}\overline{Z}_{i})$ (prior to the normalization
of the upper left element of $G_{F}$) and $\Upsilon_{n}=0^{kp\times kp}.$ The
AKP structure allowed for here (but not in GKMC and GKM19) also covers some
important cases of CHET involving $vec(\overline{Z}_{i}U_{i}^{\prime})$.

\textbf{Examples. i)\ }Consider the case in (\ref{struc}) where
$(\widetilde{\varepsilon}_{i},\widetilde{V}_{W,i}^{\prime})^{\prime}\in\Re
^{p}$ are i.i.d. zero mean with a pd variance matrix, independent of
$\overline{Z}_{i},$ and $(\varepsilon_{i},V_{W,i}^{\prime})^{\prime
}:=f(\overline{Z}_{i})(\widetilde{\varepsilon}_{i},\widetilde{V}_{W,i}%
^{\prime})^{\prime}$ for some scalar valued function $f$ of $\overline{Z}%
_{i}.$\footnote{For example, \citet{Andrews2017} considers $f(Z_{i}%
)=||Z_{i}||/k^{1/2}.$} In that case, the covariance matrix $\overline{R}_{F}$
can be written%
\begin{align}
&  E_{F}(vec(\overline{Z}_{i}U_{i}^{\prime})(vec(\overline{Z}_{i}U_{i}%
^{\prime}))^{\prime})\nonumber\\
\hspace{-0.03in}  &  =\hspace{-0.03in}E_{F}\left(  U_{i}U_{i}^{\prime}%
\otimes\overline{Z}_{i}\overline{Z}_{i}^{\prime}\right) \nonumber\\
\hspace{-0.03in}  &  =\hspace{-0.03in}E_{F}\left(  (\varepsilon_{i}%
+V_{W,i}^{\prime}\gamma,V_{W,i}^{\prime})^{\prime}(\varepsilon_{i}%
+V_{W,i}^{\prime}\gamma,V_{W,i}^{\prime})\otimes\overline{Z}_{i}\overline
{Z}_{i}^{\prime}\right) \nonumber\\
\hspace{-0.03in}  &  =\hspace{-0.03in}E_{F}\left(  (\widetilde{\varepsilon
}_{i}+\widetilde{V}_{W,i}^{\prime}\gamma,\widetilde{V}_{W,i}^{\prime}%
)^{\prime}(\widetilde{\varepsilon}_{i}+\widetilde{V}_{W,i}^{\prime}%
\gamma,\widetilde{V}_{W,i}^{\prime})\right)  \otimes E_{F}\left(
f(\overline{Z}_{i})^{2}\overline{Z}_{i}\overline{Z}_{i}^{\prime}\right)
\label{example KPstructure}%
\end{align}
and thus has KP structure even though, obviously, CHOM is not satisfied
because
\begin{equation}
E_{F}(U_{i}U_{i}^{\prime}|\overline{Z}_{i})=f(\overline{Z}_{i})^{2}%
E_{F}(\widetilde{\varepsilon}_{i}+\widetilde{V}_{W,i}^{\prime}\gamma
,\widetilde{V}_{W,i}^{\prime})^{\prime}(\widetilde{\varepsilon}_{i}%
+\widetilde{V}_{W,i}^{\prime}\gamma,\widetilde{V}_{W,i}^{\prime})
\label{cond homo fails}%
\end{equation}
depends on $\overline{Z}_{i}.$

We can construct illustrative examples where the proportionality $(\varepsilon_{i},V_{W,i}^{\prime})^{\prime
}:=f(\overline{Z}_{i})(\widetilde{\varepsilon}_{i},\widetilde{V}_{W,i}%
^{\prime})^{\prime}$ (that would imply KP\ structure) holds. Consider e.g. the model%
\begin{align*}
y_{i}& =Y_{i}\beta _{i}+W_{i}\gamma \\
W_{i}& =Y_{i}\phi _{i}+X_{i}\Phi _{X}
\end{align*}%
where the covariates $Y_{i}$ and $X_{i}$ are exogenous, the variables $%
y_{i},W_{i}$ are endogenous, and $Y_{i}$ has heterogeneous causal effects on 
$y_{i},W_{i},$ denoted by $\beta _{i},\phi _{i},$ respectively. Let $\beta
:=E\left( \beta _{i}\right) ,$ $\phi :=E\left( \phi _{i}\right) ,$ define $%
\widetilde{\varepsilon }_{i}:=\beta _{i}-\beta $ and $\widetilde{V}%
_{Wi}:=\phi _{i}-\phi ,$ and assume that $\widetilde{\varepsilon }_{i},%
\widetilde{V}_{Wi}$ are orthogonal to $Z_{i}:=\left( Y_{i},X_{i}\right) .$
Then, this fits exactly into the setup above with $f\left(
Z_{i}\right) =Y_{i}.$ In other words, KP structure can result as a special case of
heterogeneous causal effects.

\textbf{ii) }In a wage regression to assess the effect of "years of
education", the assumption of CHOM would require that e.g. the variance of
"wage" does not depend on the included regressor "race". This assumption is
incompatible with recent US data where the wage dispersion is largest for
Asians. Instead, the construction $(\varepsilon_{i},V_{W,i}^{\prime})^{\prime
}:=f(\overline{Z}_{i})(\widetilde{\varepsilon}_{i},\widetilde{V}_{W,i}%
^{\prime})^{\prime}$ in i) allows for dependence of the variances of the
regressand and all endogenous regressors on a scalar function of $\overline
{Z}_{i}.$ The maintained restriction is that all these variances are affected
approximately by the \emph{same} scalar function of $\overline{Z}_{i}.$ In the
related paper, GKM22, we test the null hypothesis of KP structure for 118
specifications in about a dozen highly cited papers and find that at the 5\%
nominal size in 47.5\% of the cases the null is not rejected.\bigskip

In this section we will introduce a new conditional subvector AR$_{AKP}$ test
and show it has asymptotic size with respect to the parameter space
$\mathcal{F}_{AKP,a_{n}}$ equal to the nominal size. We next define the new
test statistic and the critical value for the case considered here of
AKP\ structure. \medskip

\textbf{Estimation of the two factors in the AKP structure:} Define
\begin{equation}
Z_{i}:=(n^{-1}\overline{Z}^{\prime}\overline{Z})^{-1/2}\overline{Z}_{i}\in
\Re^{k} \label{transformed IVs}%
\end{equation}
and $Z\in\Re^{n\times k}$ with rows given by $Z_{i}^{\prime}$ for
$i=1,...,n.\footnote{For simplicity, we do not use the more precise notation
$Z_{in}$ for $Z_{i}.$ It is explained in detail in Comment 3 below Theorem
\ref{correct asy size} why we introduce $Z_{i},$ namely to obtain invariance
of the testing procedure with respect to nonsingular transformations of the
IVs.}$ Define an estimator of the matrix
\begin{equation}
R_{F}=(I_{p}\otimes(E_{F}\overline{Z}_{i}\overline{Z}_{i}^{\prime}%
)^{-1/2})\overline{R}_{F}(I_{p}\otimes(E_{F}\overline{Z}_{i}\overline{Z}%
_{i}^{\prime})^{-1/2})\in\Re^{kp\times kp} \label{def RF}%
\end{equation}
by
\begin{align}
\widehat{R}_{n}  &  :=n^{-1}%
{\textstyle\sum\nolimits_{i=1}^{n}}
f_{i}f_{i}^{\prime}\in\Re^{kp\times kp},\text{ where}\nonumber\\
f_{i}  &  :=((M_{Z}\overline{Y}_{0})_{i},(M_{Z}W)_{i}^{\prime})^{\prime
}\otimes Z_{i}\in\Re^{kp},\text{ and }\overline{Y}_{0}:=y-Y\beta_{0}.
\label{fi}%
\end{align}
Note that $\widehat{R}_n$ is automatically a centered estimator because, as
straightforward calculations show, $n^{-1}%
{\textstyle\sum\nolimits_i}
f_i=0.$ From $\overline{R}_F=G_F\otimes\overline{H}_F+\Upsilon_n,$ it follows
that $R_F=G_F\otimes H_F+o(1)$ for
\begin{equation}
H_{F}:=(E_{F}\overline{Z}_{i}\overline{Z}_{i}^{\prime})^{-1/2}\overline{H}%
_{F}(E_{F}\overline{Z}_{i}\overline{Z}_{i}^{\prime})^{-1/2}. \label{defH}%
\end{equation}
Let%
\begin{equation}
(\widehat{G}_{n},\widehat{H}_{n})=\arg\min||G\otimes H-\widehat{R}_{n}||,
\label{minprob}%
\end{equation}
where the minimum is taken over $(G,H)$ for $G\in\Re^p\times p,$ $H\in
\Re^k\times k$ being pd, symmetric matrices, and normalized such that the
upper left element of $G$ equals 1.

It can be shown that $(\widehat{G}_{n},\widehat{H}_{n})$ are given in closed
form by the following construction.\footnote{This follows from a combination
of Lemma \ref{frobenius norm min} below and Theorem 5.8 in
\citet[Corollary 2.2]{VanLoanPitsianis1993}.} First, for a pd matrix $A\in
\Re^{kp\times kp}$ define the rearrangement of $A$ as%
\begin{align}
\mathcal{R}(A)  &  :=\left(
\begin{array}
[c]{c}%
A_{1}\\
...\\
A_{p}%
\end{array}
\right)  \in\Re^{pp\times kk},\text{ where }\nonumber\\
A_{j}  &  :=\left(
\begin{array}
[c]{c}%
(vec(A_{1j}))^{\prime}\\
...\\
(vec(A_{pj}))^{\prime}%
\end{array}
\right)  \in\Re^{p\times kk}\text{ for }j=1,...,p, \label{rearrangement}%
\end{align}
where $A_{lj}\in\Re^{k\times k}$ denotes the $(l,j)$ submatrix of dimensions
$k\times k,$ where $l,j=1,...,p.$ Second, denote by
\begin{equation}
\widehat{L}^{\prime}\mathcal{R}(A)\widehat{N}=diag(\widehat{\sigma}_{l})\in
\Re^{pp\times kk} \label{SVD vLP}%
\end{equation}
a singular value decomposition of $\mathcal{R}(A)$,\footnote{In
\citet[Corollary 2]{VanLoanPitsianis1993}, the orthogonal matrices
$\widehat{L}\in\Re^{pp\times pp}$ and $\widehat{N}\in\Re^{kk\times kk}$ are
called $U$ and $V,$ respectively, notation that we have already used for other
objects.} where the singular values $\widehat{\sigma}_{l}$ for $l=1,...,p^{2}$
are ordered non-increasingly. Finally, denote by $\widehat{L}(:,1)$ and
$\widehat{N}(:,1)$ singular vectors corresponding to the largest singular
value $\widehat{\sigma}_{1}$ and let $\widehat{L}(1,1)$ denote the first
component of $\widehat{L}(:,1).$ Then, letting the role of $A$ be played by
$\widehat{R}_{n}$ in (\ref{SVD vLP}), minimizers $(\widehat{G}_{n}%
,\widehat{H}_{n})$ to (\ref{minprob}) are defined by
\begin{equation}
vec(\widehat{G}_{n})=\widehat{L}(:,1)/\widehat{L}(1,1)\text{ and
}vec(\widehat{H}_{n})=\widehat{\sigma}_{1}\widehat{L}(1,1)\widehat{N}(:,1),
\label{def of KP factors}%
\end{equation}
where $\widehat{L}(1,1)>0$ whenever $\widehat{R}_{n}$ is pd. By Lemma
\ref{CONSISTENCY GHATS} below, the definition given in
(\ref{def of KP factors}) is unique for all large enough $n$ wp1\footnote{Note
that it would not be unique if the eigenspace associated with the largest
singular value had dimension larger than 1.} and
\begin{equation}
\widehat{G}_{n}-G_{F_{n}}\rightarrow0^{p\times p}\text{ and }\widehat{H}%
_{n}-H_{F_{n}}\rightarrow0^{k\times k}\text{ a.s.} \label{consistency}%
\end{equation}
under certain sequences $F_{n}$ as defined in $\mathcal{F}_{AKP,a_{n}}$ for
which $R_{F_{n}}=G_{F_{n}}\otimes H_{F_{n}}+o(1)$ (where $R_{F_{n}}$ is
defined in (\ref{def RF}) with $F$ replaced by $F_{n}$), $H_{F_{n}}%
:=(E_{F_{n}}\overline{Z}_{i}\overline{Z}_{i}^{\prime})^{-1/2}\overline
{H}_{F_{n}}(E_{F_{n}}\overline{Z}_{i}\overline{Z}_{i}^{\prime})^{-1/2}$ (as
defined in (\ref{defH})), and the upper left element of $G_{F_{n}}$ is
normalized to 1.\bigskip

\textbf{Definition of the conditional subvector test:}\ We denote the
subvector AR statistic when the variance matrix has AKP structure by
$AR_{AKP,n}(\beta_{0})$ and define it as the smallest root $\hat{\kappa}_{pn}$
of the roots $\hat{\kappa}_{in},$ $i=1,...,p$ (ordered nonincreasingly) of the
characteristic polynomial
\begin{equation}
\left\vert \hat{\kappa}I_{p}-n^{-1}\widehat{G}_{n}^{-1/2}\left(  \overline
{Y}_{0},W\right)  ^{\prime}Z\widehat{H}_{n}^{-1}Z^{\prime}\left(  \overline
{Y}_{0},W\right)  \widehat{G}_{n}^{-1/2}\right\vert =0.
\label{eq: eigenprob ARhat}%
\end{equation}
The conditional subvector test AR$_{AKP,\alpha}$ rejects $H_{0}$ at nominal
size $\alpha$ if
\begin{equation}
AR_{AKP,n}(\beta_{0})>c_{1-\alpha}(\hat{\kappa}_{1n},k-m_{W}),
\label{rej condition in asy model}%
\end{equation}
where $c_{1-\alpha}\left(  \cdot,\cdot\right)  $ is defined as follows.
\citet{Muirhead1978}, in the case where $m_{W}=1$ and assuming normality,
provides an approximate, nuisance parameter free, conditional density of the
smaller eigenvalue $\hat{\kappa}_{2n}$ given the larger one $\hat{\kappa}%
_{1n}$ for any degree of overidentification $k-m_{W},$ see (2.12) in GKM19 for
the conditional pdf$.$ For given $\hat{\kappa}_{1n}$ and arbitrary $m_{W},$
$c_{1-\alpha}(\hat{\kappa}_{1n},k-m_{W})$ denotes the $1-\alpha$-quantile of
that approximation. GKM19 (Table 1 and Supplement C) provide $c_{1-\alpha
}(\hat{\kappa}_{1n},k-m_{W})$ for $\alpha=1,5,10\%,$ $k-m_{W}=1,...,20$ and a
fine grid of values for $\hat{\kappa}_{1n},$ say $\hat{\kappa}_{1,1}%
\leq...\leq\hat{\kappa}_{1,j}\leq...\leq\hat{\kappa}_{1,J}$ for some large
$J.$ We reproduce Table 1 (that covers the case $\alpha=5\%$ and $k-m_{W}=4)$
from GKM19 below. Conditional critical values for values of $\hat{\kappa}%
_{1n}$ not reported in the tables are obtained by linear interpolation.
Specifically, let $q_{1-\alpha,j}(k-1)$ denote the $1-\alpha$ quantile of the
distribution whose density is given by (2.12) in GKM19 with $\hat{\kappa}%
_{1n}$ replaced by $\hat{\kappa}_{1,j}$. The end point of the grid
$\hat{\kappa}_{1,J}$ should be chosen high enough so that $q_{1-\alpha
,J}(k-m_{W})\approx\chi_{k-m_{W},1-\alpha}^{2}$. For any realization of
$\hat{\kappa}_{1n}\leq\hat{\kappa}_{1,J}$, find $j$ such that $\hat{\kappa
}_{1n}\in\left[  \hat{\kappa}_{1,j-1},\hat{\kappa}_{1,j}\right]  $ with
$\hat{\kappa}_{1,0}=0$ and $q_{1-\alpha,0}\left(  k-m_{W}\right)  =0$, and
let
\begin{equation}
c_{1-\alpha}\left(  \hat{\kappa}_{1n},k-m_{W}\right)  :=\frac{\hat{\kappa
}_{1,j}-\hat{\kappa}_{1n}}{\hat{\kappa}_{1,j}-\hat{\kappa}_{1,j-1}}%
q_{1-\alpha,j-1}\left(  k-m_{W}\right)  +\frac{\hat{\kappa}_{1n}-\hat{\kappa
}_{1,j-1}}{\hat{\kappa}_{1,j}-\hat{\kappa}_{1,j-1}}q_{1-\alpha,j}\left(
k-m_{W}\right)  . \label{def critical value}%
\end{equation}
\bigskip Table 1: $cv=c_{1-\alpha}(\hat{\kappa}_{1},k-m_{W})$ for
$\alpha=5\%,$ $k-m_{W}=4$ for various values of $\hat{\kappa}_{1}$

{\footnotesize
\begin{tabular}
[c]{rr|rr|rr|rr|rr|rr|rr|rr|rr}\hline\hline
\multicolumn{18}{c}{}\\\hline
$\hat{\kappa}_{1}$ & cv & $\hat{\kappa}_{1}$ & cv & $\hat{\kappa}_{1}$ & cv &
$\hat{\kappa}_{1}$ & cv & $\hat{\kappa}_{1}$ & cv & $\hat{\kappa}_{1}$ & cv &
$\hat{\kappa}_{1}$ & cv & $\hat{\kappa}_{1}$ & cv & $\hat{\kappa}_{1}$ &
cv\\\hline
1.2 & 1.1 & 2.1 & 1.9 & 3.2 & 2.9 & 4.5 & 3.9 & 5.9 & 4.9 & 7.4 & 5.9 & 9.4 &
6.9 & 12.5 & 7.9 & 20.9 & 8.9\\
1.3 & 1.2 & 2.3 & 2.1 & 3.5 & 3.1 & 4.7 & 4.1 & 6.2 & 5.1 & 7.8 & 6.1 & 9.9 &
7.1 & 13.4 & 8.1 & 26.5 & 9.1\\
1.4 & 1.3 & 2.5 & 2.3 & 3.7 & 3.3 & 5.0 & 4.3 & 6.5 & 5.3 & 8.2 & 6.3 & 10.5 &
7.3 & 14.5 & 8.3 & 39.9 & 9.3\\
1.6 & 1.5 & 2.7 & 2.5 & 4.0 & 3.5 & 5.3 & 4.5 & 6.8 & 5.5 & 8.6 & 6.5 & 11.1 &
7.5 & 15.9 & 8.5 & 57.4 & 9.4\\
1.8 & 1.7 & 3.0 & 2.7 & 4.2 & 3.7 & 5.6 & 4.7 & 7.1 & 5.7 & 9.0 & 6.7 & 11.7 &
7.7 & 17.9 & 8.7 & 1000 & 9.48\\\hline\hline
\end{tabular}
}\bigskip

Denote by $P_{(\gamma,\Pi_{W},\Pi_{Y},F)}(\cdot)$ the probability of an event
under the null hypothesis when the true values of the structural and reduced
form parameters and the distribution of the random variables are given by
$(\gamma,\Pi_{W},\Pi_{Y},F).$ Recall the definition of the parameter space
$\mathcal{F}_{AKP,a_{n}}$ in (\ref{def par spa}). We can now formulate the
main result of this section.

\begin{theorem}
\label{correct asy size}Under Assumption B, the conditional subvector test
AR$_{AKP,\alpha}$ defined in $($\ref{rej condition in asy model}$)$
implemented at nominal size $\alpha$ has asymptotic size, i.e.%
\[
\lim\sup_{n\rightarrow\infty}\sup_{(\gamma,\Pi_{W},\Pi_{Y},F)\in
\mathcal{F}_{AKP,a_{n}}}P_{(\gamma,\Pi_{W},\Pi_{Y},F)}(AR_{AKP,n}(\beta
_{0})>c_{1-\alpha}(\hat{\kappa}_{1n},k-m_{W}))
\]
equal to $\alpha$ for $\alpha\in\left\{  1\%,5\%,10\%\right\}  $ and
$k-m_{W}\in\left\{  1,...,20\right\}  .$
\end{theorem}

\textbf{Comment.} 1. The conditional subvector test AR$_{AKP,\alpha}$ adapts
the test in GKM19 from a setup under CHOM to AKP structure. The modification
involves replacing the matrices $\left(  \overline{Y}_{0},W\right)  ^{\prime
}M_{\overline{Z}}\allowbreak \left(  \overline{Y}_{0},W\right)  /(n-k)$ and $n^{-1}%
\overline{Z}^{\prime}\overline{Z}$ in GKM19 by the matrices $\widehat{G}_{n}$
and $\widehat{H}_{n},$ respectively, in (\ref{eq: eigenprob ARhat}) to account
for the more general structure of the covariance matrix. Some portions of the
proof follow similar steps as the proof of Theorem 5 in GKM19. In particular,
one portion of the proof relies on a one-dimensional simulation exercise to
prove that the NRPs are bounded by the nominal size. This exercise could be
extended to choices of $\alpha$ and $k-m_{W}$ beyond those in the theorem and
likely the theorem would extend to many more choices.

2. Trivially, under the same assumptions as in Theorem \ref{correct asy size},
we obtain that
\[
\lim\sup_{n\rightarrow\infty}\sup_{(\gamma,\Pi_{W},\Pi_{Y},F)\in
\mathcal{F}_{AKP,a_{n}}}P_{(\gamma,\Pi_{W},\Pi_{Y},F)}(AR_{AKP,n}(\beta
_{0})>\chi_{k-m_{W},1-\alpha}^{2})=\alpha.
\]
That is, the generalization of the subvector test in GKMC to AKP structure has
correct asymptotic size. This result is obtained fully analytically; its proof
does not require any simulations.

3. \textbf{Invariance with respect to nonsingular transformations of the IV
matrix.} The identifying power of the model comes from the moment condition
$E_{F}\varepsilon_{i}\overline{Z}_{i}=E_{F}(y_{i}-Y_{i}^{\prime}\beta
-W_{i}^{\prime}\gamma)\overline{Z}_{i}=0.$ This moment condition obviously
still holds when the instrument vector is premultiplied by a nonrandom
nonsingular matrix $A\in\Re^{k\times k},$ i.e. $E_{F}\varepsilon_{i}%
A\overline{Z}_{i}=0.$ It then seems reasonable to look for testing procedures
whose outcome is invariant to such nonsingular transformations. In the weak IV
literature, e.g. \citet{AndrewsMoreiraStock06} and \citet{AndrewsMarmerYu2019}
and references therein, the class of (similar) invariant tests to orthogonal
transformations $A$, that is, changes of the coordinate system, has been
studied. The transformation of the IVs in (\ref{transformed IVs}) is performed
in order for the test to be invariant to nonsingular transformations of the IVs.

If the conditional subvector AR$_{AKP}$ test defined in
(\ref{rej condition in asy model}) (and $\widehat{R}_{n}$ in (\ref{fi})) was
defined with $\overline{Z}_{i}$ in place of $Z_{i}$ it would be invariant to
orthogonal transformations but not necessarily to nonsingular ones. To see the
former, denote by $\widehat{R}_{nA}$ the matrix $\widehat{R}_{n}$ when the
instrument vector has been transformed to $A\overline{Z}_{i}$ (and
consequently $\overline{Z}$ is changed to $\overline{Z}A^{\prime}).$ Then the
claim follows from $\mathcal{R}(\widehat{R}_{nA})=\mathcal{R}(\widehat{R}%
_{n})(A^{\prime}\otimes A^{\prime})$ (which holds for any nonsingular matrix
$A$ by straightforward calculations using $vec(ABC)=(C^{\prime}\otimes
A)vec(B)$ for any conformable matrices $A,B,$ and $C$ and $M_{\overline{Z}%
}=M_{\overline{Z}A^{\prime}})$ which implies $\widehat{G}_{nA}=\widehat{G}%
_{n}$ and $\widehat{H}_{nA}=A\widehat{H}_{n}A^{\prime}$ when $A$ is
orthogonal, where again $\widehat{G}_{nA}$ and $\widehat{H}_{nA}$ denote the
matrices $\widehat{G}_{n}$ and $\widehat{H}_{n}$ when the instrument vector
$\overline{Z}_{i}$ has been transformed to $A\overline{Z}_{i}$. It then
follows that the matrix $n^{-1}\widehat{G}_{n}^{-1/2}\left(  \overline{Y}%
_{0},W\right)  ^{\prime}\overline{Z}\widehat{H}_{n}^{-1}\overline{Z}^{\prime
}\left(  \overline{Y}_{0},W\right)  \widehat{G}_{n}^{-1/2}$ in
(\ref{eq: eigenprob ARhat}) (and thus its eigenvalues) remain invariant under
orthogonal transformations $\overline{Z}_{i}\rightarrow A\overline{Z}_{i}$ of
the instrument matrix. This test however is not invariant in general to
arbitrary nonsingular transformations.

But with the replacement of $\overline{Z}_{i}$ by $Z_{i}$ as done in
(\ref{fi}) and, correspondingly, $\overline{Z}$ by $\overline{Z}%
(n^{-1}\overline{Z}^{\prime}\overline{Z})^{-1/2}$ in
(\ref{eq: eigenprob ARhat}), the test is invariant against nonsingular
transformations $A$. The invariance of this test to arbitrary nonsingular
transformations $\overline{Z}_{i}\rightarrow A\overline{Z}_{i}$ of the
instrument matrix (which leads to a transformation of $Z_{i}$ to
$(A\overline{Z}^{\prime}\overline{Z}A^{\prime})^{-1/2}A\overline{Z}_{i}$)
follows from straightforward calculations and the fact that the matrix
\begin{equation}
T_{A}:=(\overline{Z}^{\prime}\overline{Z})^{1/2}A^{\prime}(A\overline
{Z}^{\prime}\overline{Z}A^{\prime})^{-1/2}\in\Re^{k\times k} \label{TA}%
\end{equation}
is orthogonal. In particular, one can easily show that the matrices
$\mathcal{R}(\widehat{R}_{n}),$ $\widehat{G}_{n},$ and $\widehat{H}_{n}$ that
appear as ingredients in the conditional subvector test AR$_{AKP,\alpha}$ with
$A=I_{k}$ are related to the corresponding matrices $\mathcal{R}%
(\widehat{R}_{nA}),$ $\widehat{G}_{nA},$ and $\widehat{H}_{nA}$, when $A$ is
an arbitrary nonsingular matrix, via
\begin{equation}
\mathcal{R}(\widehat{R}_{nA})=\mathcal{R}(\widehat{R}_{n})\left(  T_{A}\otimes
T_{A}\right)  ,\text{ }\widehat{G}_{nA}=\widehat{G}_{n},\text{ and
}\widehat{H}_{nA}=T_{A}^{\prime}\widehat{H}_{n}T_{A}
\label{nonsingular transforms}%
\end{equation}
which immediately implies the desired invariance result.

4. The conditional subvector test can be generalized to a stationary time
series setting, see the Appendix, Section \ref{time series section}, for
details. In the context of a time series setting we offer another example of
AKP structure. Namely, consider a structural vector autoregression
$AX_{t}=BX_{t-1}+\eta_{t},$ where $\dim X_{t}=\dim\eta_{t}=n,$ $E\left(
\eta_{t}|X_{t-1}\right)  =0$ and suppose that $var\left(  \eta_{t}%
|X_{t-1}\right)  =\allowbreak var\left(  \eta_{t}\right)  =\allowbreak
\Sigma_{t}=\allowbreak diag\left(  \sigma_{1t}^{2},...,\sigma_{nt}^{2}\right)
$. If $\sigma_{it}^{2}=a_{t}\sigma_{i}^{2}$ for some scalar function of time
$a_{t},$ i.e., the volatilities of all the shocks change over time in a
proportional manner, then the variance of $X_{t-1}\eta_{t}$ has KP structure.
In this model, identification can be achieved by exclusion restrictions
\citep{Sims80} that render some of $X_{t-1}$ valid instruments. It can also be
achieved with external instruments if available \citep{StockWatson2018}.
Time-variation in volatilities has been reported in many contexts. For
instance, the `great moderation' is a well-documented phenomenon of a fall in
macroeconomic volatility in the US in the early 1980s (cf.
\citet{Bernanke2004}, ch. 4). AKP would result if the fall in the volatilities
were similar across variables.

5. Note that under the null hypothesis the test does not depend on the value
of the reduced form matrix $\Pi_{Y}$ because the test statistic and the
critical value are affected by $Y$ only through $\overline{Y}_{0}=y-Y\beta
_{0}.$

6. GKM19 establish that the conditional subvector AR test introduced there
enjoys near optimality properties in the linear IV\ model with conditional
homoskedasticity in a certain class of tests that depend on the data only
through the roots $\hat{\kappa}_{in},$ $i=1,...,p$ when $k-m_{W}=1.$ On the
other hand, when $k-m_{W}$ gets bigger the test may be quite conservative. The
power gains over the projected AR subvector test discussed in
\citet{DufourTaamouti2005} arise in weakly identified scenarios while under
strong identification these two tests become identical. Similarly, we expect
the power properties of the new conditional subvector test AR$_{AKP,\alpha}$
to be most competitive for small $k-m_{W},$ in particular, when $k-m_{W}=1,$ in
weakly identified situations.

Intuition behind the result derived in GKM19 that conditioning on the largest
eigenvalue when $m_{W}>1$ leads to a test with correct size is based on i) the
corresponding result for $m_{W}=1$ and ii) the so-called "inclusion principle"
that provides a ranking of the corresponding eigenvalues of a Hermitian matrix
and its principal submatrices (see GKM19 bottom p.499-500, in particular eq (2.23)).

\section{Subvector Testing under Arbitrary Forms of Conditional
Heteroskedasticity\label{Het}}

We now allow for arbitrary forms of CHET, that is, the parameter space does
not impose an AKP structure for $\overline{R}_{F}$. We describe a testing
procedure under high level assumptions that we then verify in the next
subsections for particular implementations of the test. In particular, Lemma
\ref{ARAR test} below verifies Assumptions RT and RP below for a particular
implementation of the AR/AR test.

In what follows, $\mathcal{F}_{Het}$ is a generic parameter space for
$(\gamma,\Pi_{W},\Pi_{Y},F)$ that does not impose an AKP structure, but if the
restriction $\overline{R}_{F}=G_{F}\otimes\overline{H}_{F}+\Upsilon_{n}$ as in
$\mathcal{F}_{AKP,a_{n}}$ in (\ref{def par spa}) was added to the conditions
in $\mathcal{F}_{Het}$ then $\mathcal{F}_{Het}\subset\mathcal{F}_{AKP,a_{n}}.$
For example, the null parameter space $\mathcal{F}_{Het}$ may impose stronger
moment conditions than $\mathcal{F}_{AKP,a_{n}}$ so that certain Lyapunov CLTs
apply. See the definitions of $\mathcal{F}_{Het}$ in the next subsections. We
summarize the restrictions on the parameter space (PS) in the following
assumption.\smallskip

\textbf{Assumption PS:} $\mathcal{F}_{Het}\subset\widetilde{\mathcal{F}%
}_{AKP,a_{n}},$ where $\widetilde{\mathcal{F}}_{AKP,a_{n}}$ is equal to
$\mathcal{F}_{AKP,a_{n}}$ without the condition $\overline{R}_{F}=G_{F}%
\otimes\overline{H}_{F}+\Upsilon_{n}$ (AKP structure) and without the
assumptions $\kappa_{\min}(A)\geq\delta_{2}$ for $A\in\{G_{F},\overline{H}%
_{F}\}.$\smallskip

We assume there exists a robust test (RT) $\varphi_{Rob,\alpha}$ that has
asymptotic size for the parameter space $\mathcal{F}_{Het}$ bounded by the
nominal size $\alpha$. For example, in the next subsection we consider a
particular implementation of the AR/AR test in \citet{Andrews2017}. In
general, we think of $\varphi_{Rob,\alpha}$ as a test whose power can be
substantially improved on by the test $\varphi_{AKP,\alpha}$ when
$\overline{R}_{F}$ has AKP structure. \smallskip

\noindent\textbf{Assumption RT:} The test $\varphi_{Rob,\alpha}$ of
(\ref{eq: hypotheses}) has asymptotic size bounded by the nominal size
$\alpha$ for the parameter space $\mathcal{F}_{Het}$.\smallskip

We now define a new test that, roughly speaking, coincides with $\varphi
_{AKP,\alpha}$ or $\varphi_{Rob,\alpha}$ depending on whether the data seems
consistent or not with AKP structures. We now provide the details.

Consider a given sequence of constants $c_{n}$ such that
\begin{equation}
c_{n}\rightarrow\infty\text{ and }c_{n}/n^{1/2}\rightarrow0
\label{sequence cn}%
\end{equation}
e.g. $c_{n}=cn^{1/2}/\ln(n)$ or $c_{n}=cn^{1/2}/\ln\ln(n)$ for some constant
$c>0$ and define
\begin{equation}
\lambda_{9n}:=\min||R_{F_{n}}^{-1/2}(G\otimes H-R_{F_{n}})R_{F_{n}}%
^{-1/2}||/c_{n}, \label{lambda9}%
\end{equation}
where the minimum (here and in analogous expressions below) is taken over
$(G,H)$ for $G\in\Re^{p\times p},$ $H\in\Re^{k\times k}$ being pd, symmetric
matrices, normalized such that the upper left element of $G$ equals
1.\footnote{The expression $G\otimes H-R_{F_{n}}$ is pre- and postmultiplied
by $R_{F_{n}}^{-1/2}$ for invariance reasons.} The quantity $\lambda_{9n}$
measures how far from KP structure the covariance matrix $R_{F_{n}}$ in
(\ref{def RF}) when $F=F_{n}$ is. To show that the new test $\varphi
_{MS-AKP,\alpha}$ defined below has asymptotic significance level $\alpha$, it
is sufficient (as proven in the Appendix) to consider two types of drifting
sequences of DGPs in $\mathcal{F}_{Het}$ and to establish that the test has
limiting NRP bounded by the nominal size $\alpha$ in each case. The first type
of sequences are those for which
\begin{equation}
n^{1/2}\lambda_{9n}\rightarrow h_{9}=\infty, \label{hetero case}%
\end{equation}
that is sequences where the covariance matrix $R_{F_{n}}$ is "far away" from
KP structure. We assume that there is a model selection (MS) method
$\varphi_{MS,c_{n}}\in\{0,1\}$ such that when $R_{F_{n}}$ is "far from" KP
structure\ it will choose the robust test wpa1. The next assumption makes that
statement more precise. To properly formulate the assumption we require
terminology that is provided in the Appendix because it requires a lot of
space. In particular, we need to consider particular sequences of drifting
parameters $\lambda_{w_{n},h}$ (defined in (\ref{Defn lambda n,h}) in the
Appendix) where $w_{n}$ denotes a subsequence of $n$. \smallskip

\noindent\textbf{Assumption MS:} The model selection method $\varphi
_{MS,c_{n}}\in\{0,1\}$ satisfies $\varphi_{MS,c_{n}}=1$ wpa1 under parameter
sequences $\lambda_{w_{n},h}$ (with underlying parameter space $\mathcal{F}%
_{Het})$ with $h_{9}=\infty$.\smallskip

By definition, along $\lambda_{w_{n},h},$ $w_{n}^{1/2}\lambda_{9w_{n}%
}\rightarrow h_{9}$ and thus when $h_{9}=\infty$ the sequence is not local to
KP structure.\smallskip

\textbf{Definition of the fully robust test:} Let $\delta\geq0.$ The new
suggested test $\varphi_{MS-AKP,\delta,c_{n},\alpha}$ of nominal size $\alpha$
of the null hypothesis (\ref{eq: hypotheses}) is defined as
\begin{equation}
\varphi_{MS,c_{n}}\varphi_{Rob,\alpha-\delta}+(1-\varphi_{MS,c_{n}}%
)\varphi_{AKP,\alpha}. \label{new test}%
\end{equation}
We typically write $\varphi_{MS-AKP,\alpha}$ rather than $\varphi
_{MS-AKP,\delta,c_{n},\alpha}$ to simplify notation. Ideally, $\delta=0$ can
be chosen in this construction. To verify Assumption RP below using the AR/AR
test as $\varphi_{Rob,\alpha-\delta}$ we need to have $\delta>0.$
(Potentially, Assumption RP may hold with $\delta=0$ but our current proof
technique does not allow verifying it).\smallskip

By Assumption MS, $\varphi_{MS-AKP,\alpha}=\varphi_{Rob,\alpha-\delta}$ wpa1
in case (\ref{hetero case}). Thus, by Assumption RT, the new test
$\varphi_{MS-AKP,\alpha}$ has limiting NRP bounded by $\alpha-\delta$ of the
test in that case.

For the model selection methods introduced below, the sequence of constants
$c_{n}$ reflects a trade-off between size and power. Large values of $c_{n}$
will imply frequent use of $\varphi_{AKP,\alpha}$ which should translate into
good power properties. On the other hand, use of $\varphi_{AKP,\alpha}$ could
distort the NRPs in finite samples if the test is used in a scenario where the
covariance matrix does not have AKP structure. Below we make a recommendation
regarding the choice of $c_{n}$ based on comprehensive Monte Carlo studies.
Note that $c_{n}$ can also depend on observed nonrandom quantities such as
e.g. $k$ and $m_{W}$ but for the sake of notational simplicity we do not make
that explicit.

To guarantee correct asymptotic significance level $\alpha$ of the test
$\varphi_{MS-AKP,\alpha}$ and to rule out any potential pretesting issue, we
have to implement the test $\varphi_{Rob,\alpha}$ at a nominal size
infinitesimally smaller than $\alpha.$ For example, we can pick $\delta
=10^{-6},$ which should not make any practical difference in terms of power
relative to using the test with $\delta=0.$

In addition, we have to impose one additional assumption regarding the
relative NRPs (Assumption RP below) of the robust test $\varphi_{Rob,\alpha
-\delta}$ and $\varphi_{AKP,\alpha}$ under sequences with AKP structure in
order to make sure that $\varphi_{MS-AKP,\alpha}$ has limiting NRP bounded by
$\alpha.$ More precisely, consider a sequence of DGPs in $\mathcal{F}_{Het}$
such that
\begin{equation}
n^{1/2}\lambda_{9n}\rightarrow h_{9}\in\lbrack0,\infty).\label{KPS case}%
\end{equation}
Using $n^{1/2}/c_{n}\rightarrow\infty,$ one can then show that $\min||G\otimes
H-R_{F_{n}}||\rightarrow0$ and the sequences are of AKP structure. Therefore,
under such sequences the test $\varphi_{AKP,\alpha}$ has limiting NRP bounded
by $\alpha$. The notation $P_{\lambda_{w_{n},h}}(A)$ denotes probability of an
event $A$ when the true DGP is characterized by $\lambda_{w_{n},h}$. By
definition, along $\lambda_{w_{n},h},$ $w_{n}^{1/2}\lambda_{9w_{n}}\rightarrow
h_{9}$ and thus when $h_{9}<\infty$ the sequence is local to KP
structure.\smallskip

\noindent\textbf{Assumption RP:} Under sequences of DGPs $(\gamma_{w_{n}}%
,\Pi_{Ww_{n}},\Pi_{Yw_{n}},F_{w_{n}})$ in $\mathcal{F}_{Het}$ for subsequences
$w_{n}$ for which $\lambda_{w_{n},h}$ satisfies $h_{9}\in\lbrack0,\infty),$
$P_{\lambda_{w_{n},h}}(\varphi_{Rob,\alpha-\delta}\leq\varphi_{AKP,\alpha
})\rightarrow1.$ \smallskip

Assumption RP says that under null sequences local to KP\ structure the robust
test $\varphi_{Rob,\alpha-\delta}$ has critical region that is contained in
the critical region of $\varphi_{AKP,\alpha}$ with probability going to one.
Even when $\delta=0$ this does not need to imply that the two tests are
asymptotically identical because the robust test may have limiting NRP
strictly smaller than $\alpha$ and may be more conservative than
$\varphi_{AKP,\alpha}.$ Under Assumption RP one can show that in case
(\ref{KPS case}) (i.e. under drifting sequences of DGPs $\lambda_{w_{n},h}$
with finite $h_{9})$ $\varphi_{MS-AKP,\alpha}$ has limiting NRP bounded by the
nominal size of the test (because from the proof of Theorem
\ref{correct asy size} the test $\varphi_{AKP,\alpha}$ has limiting NRP
bounded by $\alpha$; and the limiting NRP of the new test $\varphi
_{MS-AKP,\alpha}$ is then bounded by $\alpha$ by the assumption that
$\varphi_{Rob,\alpha-\delta}$ has asymptotic size bounded by $\alpha-\delta$.)

From the above, it then follows quite straightforwardly, that the asymptotic
size of $\varphi_{MS-AKP,\alpha}$ is bounded by the nominal size for the
parameter space $\mathcal{F}_{Het}$. Also, the new test is at most as
nonsimilar asymptotically as $\varphi_{Rob,\alpha-\delta}$ which translates
into favorable power properties of the new test.

\begin{theorem}
\label{correct asy size generic}\textbf{ }Suppose Assumptions PS, RT, MS, and
RP\ hold. Then the test $\varphi_{MS-AKP,\delta,c_{n},\alpha}$ defined in
$($\ref{new test}$)$ with $\delta>0$ and $c_{n}$ satisfying the conditions in
$($\ref{sequence cn}$)$ has asymptotic size bounded by the nominal size
$\alpha$ for the parameter space $\mathcal{F}_{Het}$ for $\alpha\in\left\{
1\%,5\%,10\%\right\}  $ and $k-m_{W}\in\left\{  1,...,20\right\}  .$\smallskip
\end{theorem}

\textbf{Comments. 1. }If $\lim\inf_{n\rightarrow\infty}\inf_{(\gamma,\Pi
_{W},\Pi_{Y},F)\in\mathcal{F}_{Het}}E_{(\gamma,\Pi_{W},\Pi_{Y},F)}%
\varphi_{MS-AKP,\delta,c_{n},\alpha}$ is continuous in $\delta$ at $\delta=0$
then as $\delta\rightarrow0$ the new test $\varphi_{MS-AKP,\delta,c_{n}%
,\alpha}$ is asymptotically not more nonsimilar (i.e. less conservative) than
$\varphi_{Rob,\alpha}$, i.e.%
\begin{align}
&  \lim_{\delta\rightarrow0}\lim\inf_{n\rightarrow\infty}\inf_{(\gamma,\Pi
_{W},\Pi_{Y},F)\in\mathcal{F}_{Het}}E_{(\gamma,\Pi_{W},\Pi_{Y},F)}%
\varphi_{MS-AKP,\delta,c_{n},\alpha}\nonumber\\
&  \geq\lim\inf_{n\rightarrow\infty}\inf_{(\gamma,\Pi_{W},\Pi_{Y}%
,F)\in\mathcal{F}_{Het}}E_{(\gamma,\Pi_{W},\Pi_{Y},F)}\varphi_{Rob,\alpha}.
\label{non similarity}%
\end{align}
See the proof of Theorem \ref{correct asy size generic} for a proof. Property
(\ref{non similarity}) should translate into improved power of $\varphi
_{MS-AKP,\delta,c_{n},\alpha}$ relative to $\varphi_{Rob,\alpha}.$

\textbf{2. }The restriction to $\alpha\in\left\{  1\%,5\%,10\%\right\}  $ and
$k-m_{W}\in\left\{  1,...,20\right\}  $ in the formulation of Theorem
\ref{correct asy size generic} is an artifact of Theorem
\ref{correct asy size} where the conditional subvector test $\varphi
_{AKP,\alpha}$ was shown to have correct asymptotic size for these cases only.
The same is true for other theorems formulated below.\smallskip

In the next subsection we specifically use the AR/AR subvector procedure due
to \citet{Andrews2017} as $\varphi_{Rob,\alpha-\delta}.$

\subsection{Model selection methods $\varphi_{MS,c_{n}}\label{2 model sel}$}

In this subsection we discuss two methods that can be used for $\varphi
_{MS,c_{n}}$ as model selection procedures. The first one is akin to the
moment selection method in \citet{AndrewsSoares2010} to check which moment
inequalities are binding in a model defined by moment inequalities. The second
one is based on the test for KP structure introduced in GKM22.

\textbf{Method 1:} Define
\begin{equation}
\widehat{K}_{n}:=n^{1/2}||\widehat{R}_{n}^{-1/2}(\widehat{G}_{n}%
\otimes\widehat{H}_{n}-\widehat{R}_{n})\widehat{R}_{n}^{-1/2}||,
\label{estimate of how far off KPS}%
\end{equation}
with $\widehat{G}_{n}$ and $\widehat{H}_{n}$ defined in (\ref{minprob}), to
evaluate how far the true model is away from KP structure. Define the first
choice for model selection as
\begin{equation}
\varphi_{MS,c_{n}}:=I(\widehat{K}_{n}>c_{n}). \label{AS method}%
\end{equation}
Recall the definition of $\widetilde{\mathcal{F}}_{AKP,a_{n}}$ given in
Assumption PS. Here we take%
\begin{align}
\mathcal{F}_{Het}  &  =\{(\gamma,\Pi_{W},\Pi_{Y},F)\in\widetilde{\mathcal{F}%
}_{AKP,a_{n}},\nonumber\\
&  \text{ }E_{F}((||\overline{Z}_{i}||^{2}||U_{i}||^{2})^{2+\delta_{1}%
})\overset{}{\leq}B,\text{ }\kappa_{\min}(R_{n})\geq\delta_{2}\}.
\label{F het def}%
\end{align}
It is easy to show using the formulae in (\ref{nonsingular transforms}) and
the analogous one $\widehat{R}_{nA}=(I_{p}\otimes T_{A}^{\prime}%
)\widehat{R}_{n}(I_{p}\otimes T_{A})$ for $\widehat{R}_{n}$, orthogonality of
$T_{A}$, and using the fact that the Frobenius norm is invariant to orthogonal
transformations, that $\widehat{K}_{n}$ is invariant to nonsingular
transformations of the instrument vector. Crucial for this result is again
that $f_{i}$ in (\ref{fi}) in the definition of $\widehat{R}_{n}$ (and as a
result in the definition of $\widehat{G}_{n}$ and $\widehat{H}_{n}$ in
(\ref{minprob})) is implemented with the transformed instrument vector $Z_{i}$
(rather than with $\overline{Z}_{i}$).

\textbf{Method 2:} Define%
\begin{equation}
\varphi_{MS,c_{n}}:=I(KPST>c_{n}), \label{GKM2020 method}%
\end{equation}
where $KPST$ is the test statistic introduced in GKM22 to test the null of a
KP structure of $R_{F}.\footnote{The test statistic is defined in (19) and
(22) in GKM20 and not reproduced here for brevity. In their notation our
$f_{i}$ is $\widehat{f}_{i},$ compare the formula below (7) in GKM20 to our
(\ref{fi}).}$ To employ this method, we need to strengthen the moment
restrictions in $\mathcal{F}_{Het}$ to $E_F(||T_i||^2+\delta_1)\leq B,$ for
$T_i\in\{||\overline{Z}_i||^4||U_i||^4,||\overline{Z}_i||^4\},$ see Theorem 3
in GKM22.

We verify Assumption MS in the Appendix, Section \ref{model selection proof},
for these two choices of $\varphi_{MS,c_{n}}$ and for the parameter space
defined in (\ref{F het def}).

\subsection{Choice for $\varphi_{Rob,\alpha}:$ The AR/AR test in
\citet{Andrews2017}\label{robust test AR AR}}

In this subsection we define one particular version of the various weak IVs
and heteroskedasticity robust subvector tests suggested in
\citet{Andrews2017}, namely the so-called AR/AR test and verify that it
satisfies Assumptions RT and RP from the previous subsection. We define it for
nominal size $\alpha.$

To do so, we use the following quantities. For $\theta=\left(  \beta
,\gamma\right)  $ let\footnote{To simplify notation we write $(\beta,\gamma)$
here and in other situations, rather than the more correct $(\beta^{\prime
},\gamma^{\prime})^{\prime}.$}%
\begin{equation}
g_{i}\left(  \theta\right)  :=\overline{Z}_{i}(y_{i}-Y_{i}^{\prime}\beta
-W_{i}^{\prime}\gamma)\text{ and }\widehat{g}_{n}\left(  \theta\right)
:=n^{-1}%
{\textstyle\sum\nolimits_{i=1}^{n}}
g_{i}\left(  \theta\right)  . \label{gdef}%
\end{equation}
Define%
\begin{equation}
\hat{\Sigma}_{n}\left(  \theta\right)  :=n^{-1}%
{\textstyle\sum\nolimits_{i=1}^{n}}
\left(  g_{i}\left(  \theta\right)  -\widehat{g}_{n}\left(  \theta\right)
\right)  \left(  g_{i}\left(  \theta\right)  -\widehat{g}_{n}\left(
\theta\right)  \right)  ^{\prime}. \label{sigmahat def}%
\end{equation}
The heteroskedasticity-robust AR statistic for testing hypotheses involving
the full parameter vector $\theta$, evaluated at $\left(  \beta_{0}%
,\gamma\right)  ,$ is defined as
\begin{equation}
HAR_{n}\left(  \beta_{0},\gamma\right)  :=n\widehat{g}_{n}\left(  \beta
_{0},\gamma\right)  ^{\prime}\hat{\Sigma}_{n}\left(  \beta_{0},\gamma\right)
^{-1}\widehat{g}_{n}\left(  \beta_{0},\gamma\right)  . \label{har def}%
\end{equation}
For $s=1,...,m_{W}$ denote by $W^{s}\in\Re^{n}$ the $s$-th column of $W.$
Next, as in \citet[(7.9)-(7.10)]{Andrews2017} let%
\begin{align}
\tilde{D}_{n}\left(  \theta\right)   &  :=\hat{\Sigma}_{n}\left(
\theta\right)  ^{-1/2}(\widehat{D}_{1n}\left(  \theta\right)  ,...,\widehat{D}%
_{m_{W}n}\left(  \theta\right)  )\in\Re^{k\times m_{W}},\nonumber\\
\widehat{D}_{sn}\left(  \theta\right)   &  :=-n^{-1}\overline{Z}^{\prime}%
W^{s}-\hat{\Gamma}_{sn}\left(  \theta\right)  \hat{\Sigma}_{n}\left(
\theta\right)  ^{-1}\widehat{g}_{n}\left(  \theta\right)  \in\Re
^{k},\nonumber\\
\hat{\Gamma}_{sn}\left(  \theta\right)   &  :=-n^{-1}%
{\textstyle\sum\nolimits_{i=1}^{n}}
\left(  \overline{Z}_{i}W_{i}^{s}-n^{-1}\overline{Z}^{\prime}W^{s}\right)
g_{i}\left(  \theta\right)  ^{\prime}\in\Re^{k\times k},\text{ and}\nonumber\\
HAR_{\beta,n}\left(  \beta_{0},\gamma\right)   &  :=n\widehat{g}_{n}\left(
\beta_{0},\gamma\right)  ^{\prime}\hat{\Sigma}_{n}\left(  \beta_{0}%
,\gamma\right)  ^{-1/2}M_{\tilde{D}_{n}\left(  \beta_{0},\gamma\right)
+an^{-1/2}\zeta_{1}}\hat{\Sigma}_{n}\left(  \beta_{0},\gamma\right)
^{-1/2}\widehat{g}_{n}\left(  \beta_{0},\gamma\right)  , \label{harb def}%
\end{align}
where $HAR_{\beta,n}\left(  \beta_{0},\gamma\right)  $ is a $C\left(
\alpha\right)  $-AR statistic, obtained as a quadratic form in the moment
conditions projected onto the space orthogonal to the orthogonalized Jacobian
with respect to $\gamma$. The random perturbation $an^{-1/2}\zeta_{1}$ (with
$\zeta_{1}\in\Re^{k\times m_{W}}$ a random matrix of independent standard
normal random variables that are independent of all other statistics
considered) in the last line of (\ref{harb def}) is introduced in
\citet[p.23]{Andrews2017}, to guarantee that the space projected on has rank
$m_{W}$ a.s. Here $a\in\Re$ is a tiny positive constant.

Let $\alpha\in(0,1)$. The AR/AR test at nominal size $\alpha$ is defined as follows.

\begin{enumerate}
\item Fix an $\alpha_{1}\in\left(  0,\alpha\right)  .$ As in
\citet[(7.1)]{Andrews2017} define
\begin{equation}
CS_{1n}^{+}:=\{\widetilde{\gamma}\in\Re^{m_{W}}:HAR_{n}\left(  \beta
_{0},\widetilde{\gamma}\right)  <\chi_{k,1-\alpha_{1}}^{2}\}\cup
\widetilde{\Gamma}_{1n}, \label{cs set}%
\end{equation}
where for $\widehat{Q}_{n}\left(  \theta\right)  :=\widehat{g}_{n}\left(
\theta\right)  ^{\prime}(n^{-1}%
{\textstyle\sum\nolimits_{i=1}^{n}}
\overline{Z}_{i}\overline{Z}_{i}^{\prime})^{-1}\widehat{g}_{n}\left(
\theta\right)  ,$
\begin{align}
\widetilde{\Gamma}_{1n}  &  :=\left\{  \gamma\in\Re^{m_{W}}:W^{\prime
}\overline{Z}(%
{\textstyle\sum\nolimits_{i=1}^{n}}
\overline{Z}_{i}\overline{Z}_{i}^{\prime})^{-1}\widehat{g}_{n}\left(
\beta_{0},\gamma\right)  =0^{m_{W}}\text{ \&}\right. \label{est set}\\
&  \text{ }\left.  \widehat{Q}_{n}\left(  \beta_{0},\gamma\right)  \leq
\min_{\widetilde{\gamma}\in\Re^{m_{W}}}\widehat{Q}_{n}\left(  \beta
_{0},\widetilde{\gamma}\right)  +\frac{\ln n}{n}\right\} \nonumber
\end{align}
is the so-called \textquotedblleft estimator set\textquotedblright, see
\citet[p.1 and (7.3)]{Andrews2017}. If $W^{\prime}P_{\overline{Z}}W$ is
invertible (which would happen wpa1 under the assumption (not imposed here)
that $E_{F}\overline{Z}_{i}W_{i}^{\prime}$ is full column rank) then the first
condition in $\widetilde{\Gamma}_{1n}$ has the unique solution $\overline
{\gamma}_{n}:=(W^{\prime}P_{\overline{Z}}W)^{-1}W^{\prime}P_{\overline{Z}%
}(y-Y\beta_{0})$ and therefore $\widetilde{\Gamma}_{1n}=\{\overline{\gamma
}_{n}\}.$ (Note that along certain sequences for which $||\gamma
||\rightarrow\infty$ it follows that $||\widehat{g}_{n}\left(  \beta
_{0},\gamma\right)  ||\rightarrow\infty$ and therefore if the function
$\widehat{Q}_{n}\left(  \beta_{0},\gamma\right)  \geq0$ only has one local
extremum it must be a global minimum.)

\item For $\alpha_{2,n}(\theta)$ defined below (and depending on $\alpha$ and
$\alpha_{1})$, $H_{0}$ in (\ref{eq: hypotheses}) is rejected if \[\inf
_{\widetilde{\gamma}\in CS_{1n}^{+}}\left(HAR_{\beta,n}\left(  \beta_{0}%
,\widetilde{\gamma}\right)  -\chi_{k-m_{W},1-\alpha_{2,n}(\beta_{0}%
,\widetilde{\gamma})}^{2}\right)>0.\]

That is%
\begin{equation}
\varphi_{AR/AR,\alpha,\alpha_{1}}=1_{\left\{  \inf_{\widetilde{\gamma}\in
CS_{1n}^{+}}\left(HAR_{\beta,n}\left(  \beta_{0},\widetilde{\gamma}\right)
-\chi_{k-m_{W},1-\alpha_{2,n}(\beta_{0},\widetilde{\gamma})}^{2}\right)>0\right\}
}, \label{test rejects if}%
\end{equation}
see \citet[(4.2)]{Andrews2017}. We typically write $\varphi_{AR/AR,\alpha}$
instead of $\varphi_{AR/AR,\alpha,\alpha_{1}}.$
\end{enumerate}

The second step size $\alpha_{2,n}(\theta)$ is chosen as
\begin{equation}
\alpha_{2,n}(\theta):=\left\{
\begin{array}
[c]{lc}%
\alpha-\alpha_{1}, & \text{if }ICS_{n}(\theta)\leq K_{L}\\
\alpha, & \text{if }ICS_{n}(\theta)>K_{L},
\end{array}
\right.  \label{def alpha2}%
\end{equation}
for some positive number $K_{L}$, e.g., $K_{L}=0.05$ and $\alpha_{1}=.005,$
see \citet[(7.8) and p.34]{Andrews2017}\footnote{\citet[(7.8)]{Andrews2017}
allows for more involved definitions of $\alpha_{2,n}(\theta).$ We choose the
version that takes $K_{U}=K_{L}$ in the notation of \citet{Andrews2017} that
is also used in the Monte Carlos in \citet{Andrews2017}. Regarding the
definition of $\widehat{\Phi}_{n}(\theta),$ note that it constitutes a slight
modification compared with the definitions in \citet[(7.5)]{Andrews2017}. In
particular, the modification in the definition of $\hat{\sigma}_{sn}^{2}$ is
necessary to make the procedure invariant to nonsingular transformations of
the instrument vector. We thank Donald Andrews for suggesting this updated
version of his test statistic.}, where%
\begin{align}
\widehat{\Phi}_{n}(\theta)  &  :=Diag\{\hat{\sigma}_{1n}^{-1}(\theta
),...,\hat{\sigma}_{m_{W}n}^{-1}(\theta)\}\in\Re^{m_{W}\times m_{W}%
},\nonumber\\
\hat{\sigma}_{sn}^{2}(\theta)  &  :=n^{-1}%
{\textstyle\sum\nolimits_{i=1}^{n}}
\left(  H_{si}(\theta)-\widehat{H}_{sn}(\theta)\right)  ^{2},\text{ for
}s=1,...,m_{W},\nonumber\\
H_{si}(\theta)  &  :=\sqrt{(W_{i}^{s})^{2}\overline{Z}_{i}^{\prime}\hat
{\Sigma}_{n}\left(  \theta\right)  ^{-1}\overline{Z}_{i}},\text{ }%
\widehat{H}_{sn}(\theta):=n^{-1}%
{\textstyle\sum\nolimits_{i=1}^{n}}
H_{si}(\theta),\nonumber\\
ICS_{n}(\theta)  &  :=n^{-1}\kappa_{\min}^{1/2}(\widehat{\Phi}_{n}%
(\theta)W^{\prime}Z\hat{\Sigma}_{n}\left(  \theta\right)  ^{-1}\overline
{Z}^{\prime}W\widehat{\Phi}_{n}(\theta)), \label{more quant for arar test}%
\end{align}
see \citet[(7.4)-(7.5)]{Andrews2017}, where $W_{i}^{s}\in\Re$ denotes the
$s$-th component of $W_{i}$.\medskip

Coming back to the statistic $AR_{AKP,n}(\beta_{0})$ given in
(\ref{eq: eigenprob ARhat}) note that%
\begin{align}
AR_{AKP,n}(\beta_{0})  &  =\inf_{\widetilde{\gamma}\in\mathfrak{R}^{m_{W}}%
}\widetilde{AR}_{AKP,n}(\beta_{0},\widetilde{\gamma}),\text{ where}\nonumber\\
\widetilde{AR}_{AKP,n}(\beta_{0},\widetilde{\gamma})  &  :=\frac{n^{-1}%
\binom{1}{-\widetilde{\gamma}}^{\prime}\left(  \overline{Y}_{0},W\right)
^{\prime}Z\widehat{H}_{n}^{-1}Z^{\prime}\left(  \overline{Y}_{0},W\right)
\binom{1}{-\widetilde{\gamma}}}{\binom{1}{-\widetilde{\gamma}}^{\prime
}\widehat{G}_{n}\binom{1}{-\widetilde{\gamma}}} \label{our test statistic}%
\end{align}
using the fact that the minimal eigenvalue of any symmetric square matrix
$A\in\mathfrak{R}^{p\times p}$ is obtained as $\min_{x\in\mathfrak{R}%
^{p},||x||=1}x^{\prime}Ax.$ Furthermore,%
\begin{align}
\widetilde{AR}_{AKP,n}(\beta_{0},\widetilde{\gamma})  &  =\hspace
{-0.03in}n\widehat{g}_{n}\left(  \beta_{0},\widetilde{\gamma}\right)
^{\prime}\widetilde{\Sigma}_{n}\left(  \beta_{0},\widetilde{\gamma}\right)
^{-1}\widehat{g}_{n}\left(  \beta_{0},\widetilde{\gamma}\right)  ,\text{
where}\nonumber\\
\widetilde{\Sigma}_{n}\left(  \beta_{0},\widetilde{\gamma}\right)   &
:=\hspace{-0.03in}(\left(  1,-\widetilde{\gamma}^{\prime}\right)
\widehat{G}_{n}\left(  1,-\widetilde{\gamma}^{\prime}\right)  ^{\prime
})\otimes(n^{-1}\overline{Z}^{\prime}\overline{Z})^{1/2}\widehat{H}_{n}%
(n^{-1}\overline{Z}^{\prime}\overline{Z})^{1/2}\nonumber\\
&  =\hspace{-0.03in}\left(  \binom{1}{-\widetilde{\gamma}}\otimes
I_{k}\right)  ^{\prime}(\widehat{G}_{n}\otimes(n^{-1}\overline{Z}^{\prime
}\overline{Z})^{1/2}\widehat{H}_{n}(n^{-1}\overline{Z}^{\prime}\overline
{Z})^{1/2})\left(  \binom{1}{-\widetilde{\gamma}}\otimes I_{k}\right)
\label{our test statistic rewritten}%
\end{align}
and $(\widehat{G}_{n},\widehat{H}_{n})$ defined in (\ref{def of KP factors}).

Let $\gamma_{n}^{+}$ be an element in $\arg\min_{\widetilde{\gamma}%
\in\mathfrak{R}^{m_{W}}}\widetilde{AR}_{AKP,n}(\beta_{0},\widetilde{\gamma}).$
We impose a mild technical condition below, namely that
\begin{equation}
\Pi_{Wn}n^{1/2}(\gamma_{n}^{+}-\gamma_{n})=O_{p}(1)
\label{technical condition}%
\end{equation}
and $\gamma_{n}^{+}=O_{p}(1)$ under sequences in $\mathcal{F}_{Het}$ (defined
in (\ref{Hete}) below)\ that are of AKP structure, i.e. under sequences
$\lambda_{n,h}$ for which $h_{9}\in\lbrack0,\infty)$.

Condition (\ref{technical condition}) has been established for several closely
related estimators. E.g. $\gamma_{n}^{+}-\gamma_{n}=O_{p}(1)$ holds under weak
IV sequences $\Pi_{Wn}=C/n^{1/2}$ (for some fixed matrix $C)$ and
homoskedasticity when $\gamma_{n}^{+}$ is the LIML estimator, see
\citet[Theorem 1]{StaigerStock1997}. Results in
\citet[Theorem 1]{HahnKuersteiner2002} imply (\ref{technical condition}) for
the 2SLS\ estimator under a setup where $\Pi_{Wn}=C/n^{\delta}$ for
$\delta>0.$ \citet[Theorem 1(ii)]{StockWright2000} and
\citet[Theorem 2]{GuggenbergerSmith2005} implies (\ref{technical condition})
for the CU estimator under mixed weak/strong IV asymptotics $\Pi
_{Wn}=(C/n^{1/2},D)$ for a fixed full rank matrix $D\in\Re^{k\times
m_{W}^{\prime}}$ with $m_{W}^{\prime}\leq m_{W}$ (using high level
assumptions, such as Assumptions B and D in \citet{StockWright2000}) and
possible CHET.

\citet[Theorem 1(ii)]{StockWright2000} can also be applied in the current
situation to show (\ref{technical condition}) under sequences $\lambda_{n,h}$
for which $h_{9}\in\lbrack0,\infty).$ Given $\varepsilon>0$ we need to show
that for some compact set $K_{\varepsilon},$ $\Pi_{Wn}n^{1/2}(\gamma_{n}%
^{+}-\gamma_{n})\in K_{\varepsilon}$ with probability at least $1-\varepsilon$
for all large enough sample sizes. Assuming $\gamma_{n}^{+}=O_{p}(1),$ then
for all $\varepsilon>0,$ $\gamma_{n}^{+}$ is contained in a compact set
$K_{\varepsilon}$ with probability at least $1-\varepsilon$ for all large
enough sample sizes. Consider the estimator $\gamma_{n}^{K_{\varepsilon}}$
that is defined as a minimizer of $\widetilde{AR}_{AKP,n}(\beta_{0}%
,\widetilde{\gamma})$ in $\widetilde{\gamma}$ over $K_{\varepsilon}.$ Thus
$\gamma_{n}^{K_{\varepsilon}}$ and $\gamma_{n}^{+}$ are numerically identical
for all sample sizes large enough with probability at least $1-\varepsilon.$
Note that $\widetilde{AR}_{AKP,n}(\beta_{0},\widetilde{\gamma})$ has the same
structure as the criterion function $S_{T}(\theta,\theta)$ in (2.2) in
\citet{StockWright2000} with $\widetilde{\Sigma}_{n}\left(  \beta
_{0},\widetilde{\gamma}\right)  ^{-1}$ playing the role of the weighting
matrix $W_{T}(\theta)$ and $n^{1/2}\widehat{g}_{n}\left(  \beta_{0}%
,\widetilde{\gamma}\right)  $ playing the role of $n^{-1/2}%
{\textstyle\sum\nolimits_{s=1}^{T}}
\phi_{s}\left(  \theta\right)  $. Therefore, under drifting sequences of mixed
weak/strong IVs, namely $\Pi_{Wn}=(C/n^{1/2},D),$ the limiting distribution of
$\gamma_{n}^{K_{\varepsilon}}$ is given in
\citet[Theorem 1(ii)]{StockWright2000} if Assumptions B and D in
\citet{StockWright2000} hold for parameter space $K_{\varepsilon}$ for
$\widetilde{\gamma}$ and $\widetilde{AR}_{AKP,n}(\beta_{0},\widetilde{\gamma
})$ has a unique minimum. \citet[Theorem 1(ii)]{StockWright2000} states that
those components of $\gamma_{n}^{K_{\varepsilon}}-\gamma_{n}$ that correspond
to the columns of $C/n^{1/2}$ in $\Pi_{Wn}$ are $O_{p}(1)$ and those that
correspond to the columns of $D$ in $\Pi_{Wn}$ are $O_{p}(n^{-1/2})$ which
establishes $\Pi_{Wn}n^{1/2}(\gamma_{n}^{K_{\varepsilon}}-\gamma_{n}%
)=O_{p}(1)$.

Assumption B in \citet{StockWright2000} holds for $\gamma_{n}^{K_{\varepsilon
}}$ (in fact, Assumption B' in \citet{StockWright2000}, which is sufficient
for Assumption B, holds by linearity of $g_{i}\left(  \theta\right)  ,$ the
moment conditions in $\mathcal{F}_{Het},$ and compactness of $K_{\varepsilon
}).$ To establish Assumption D note that under sequences $\lambda_{n,h},$
$n^{-1}\overline{Z}^{\prime}\overline{Z}\rightarrow_{p}\lim E_{F_{n}}%
\overline{Z}_{i}\overline{Z}_{i}^{\prime},$ $\widehat{G}_{n}\rightarrow
_{p}\lim G_{F_{n}},$ and $\widehat{H}_{n}\rightarrow_{p}\lim H_{F_{n}}$ (note
that the right hand side limits exist by definition of $\lambda_{n,h}).$
Therefore, under sequences $\lambda_{n,h}$
\begin{equation}
\widetilde{\Sigma}_{n}\left(  \beta_{0},\widetilde{\gamma}\right)
^{-1}\rightarrow_{p}(\lim E_{F_{n}}\overline{Z}_{i}\overline{Z}_{i}^{\prime
})^{-1/2}\lim H_{F_{n}}^{-1}(\lim E_{F_{n}}\overline{Z}_{i}\overline{Z}%
_{i}^{\prime})^{-1/2}/(\left(  1,-\widetilde{\gamma}^{\prime}\right)  \lim
G_{F_{n}}\left(  1,-\widetilde{\gamma}^{\prime}\right)  ^{\prime})
\label{limit weight matrix}%
\end{equation}
uniformly over $\widetilde{\gamma}$ (noting that $||\left(
1,-\widetilde{\gamma}^{\prime}\right)  ||\geq1,$ $\lim G_{F_{n}}>0$, and $\lim
H_{F_{n}}>0)$ with the limit matrix being nonrandom, continuous, symmetric,
and pd for all $\widetilde{\gamma}.$ Thus, by
\citet[Theorem 1(ii)]{StockWright2000}, if $\widetilde{AR}_{AKP,n}(\beta
_{0},\widetilde{\gamma})$ has a unique minimum in $K_{\varepsilon}$ and
$\gamma_{n}^{+}=O_{p}(1),$ it follows that $\Pi_{Wn}n^{1/2}(\gamma
_{n}^{K_{\varepsilon}}-\gamma_{n})=O_{p}(1).$ Thus there exists a compact set
$K$ such that $\Pi_{Wn}n^{1/2}(\gamma_{n}^{K_{\varepsilon}}-\gamma_{n})\in K$
at least with probability $1-\varepsilon$ for all $n$ large enough. Because
$\gamma_{n}^{+}$ and $\gamma_{n}^{K_{\varepsilon}}$ coincide at least with
probability $1-\varepsilon$ for all large enough sample sizes, it then follows
that $\Pi_{Wn}n^{1/2}(\gamma_{n}^{+}-\gamma_{n})\in K$ at least with
probability $1-2\varepsilon$ for all $n$ large enough.

Deriving (\ref{technical condition}) under all possible drifting sequences
$\Pi_{Wn}$ is technically tedious and involves e.g. also consideration of
so-called sequences of non-standard weak identification, see {Andrews and
Guggenberger (2019, AG from now on) \nocite{AndrewsGuggenberger2019} }for more
discussion. If (\ref{technical condition}) is not already implied by the
restrictions in the parameter space $\mathcal{F}_{Het}$ below then the
asymptotic size results should simply be interpreted for sequences of
parameter spaces $\mathcal{F}_{Het,n}$ that impose additional restrictions on
$\mathcal{F}_{Het}$ such that (\ref{technical condition}) holds.\smallskip

The null parameter space is restricted by the conditions in $\mathcal{F}%
_{AR/AR}$ of \citet[(8.8)]{Andrews2017} and some weak additional ones, namely,%
\begin{align}
&  \mathcal{F}_{Het}\overset{}{=}\{(\gamma,\Pi_{W},\Pi_{Y},F)\overset{}{\in
}\widetilde{\mathcal{F}}_{AKP,a_{n}}\overset{}{:}\gamma\overset{}{\in}%
\Theta_{\gamma\ast}\overset{}{\subset}\Re^{m_{W}},\nonumber\\
&  E_{F}||U_{ij}\overline{Z}_{il_{1}}\overline{Z}_{il_{2}}\overline{Z}%
_{il_{3}}||^{1+\delta_{1}}\overset{}{\leq}B\text{ for }j\overset{}{=}%
1,...,p,\text{ }l_{1},l_{2},l_{3}\overset{}{=}1,...,k,\nonumber\\
&  E_{F}||\varepsilon_{i}\overline{Z}_{i}||^{2+\delta_{1}}\overset{}{\leq
}B,\text{ }E_{F}||vec(W_{i}^{\prime}\overline{Z}_{i})||^{2+\delta_{1}%
}\overset{}{\leq}B,\text{ }var_{F}||W_{i}^{s}\overline{Z}_{i}%
)||\overset{}{\geq}\delta_{2}\text{ for}\nonumber\\
s  &  =1...,m_{W},\text{ and }\kappa_{\min}(A)\overset{}{\geq}\delta_{2}\text{
for }A\overset{}{\in}\{R_{F},E_{F}\varepsilon_{i}^{2}\overline{Z}_{i}%
\overline{Z}_{i}^{\prime}\}\}, \label{Hete}%
\end{align}
for constants $B<\infty,$ and $\delta_{1},\delta_{2}>0$ and a bounded set
$\Theta_{\gamma\ast}$ such that for some $\epsilon>0$ we have $B(\Theta
_{\gamma\ast},\epsilon)\subset\Theta_{\gamma},$ where $\Theta_{\gamma}$
denotes the null nuisance parameter space for $\gamma$ and $B(\Theta
_{\gamma\ast},\epsilon)$ denotes the union of closed balls in $\Re^{m_{W}}$
with radius $\epsilon$ centered at points in $\Theta_{\gamma\ast}.$\smallskip

\begin{lemma}
\label{ARAR test}Assume that under any sequence of DGPs $(\gamma_{w_{n}}%
,\Pi_{Ww_{n}},\Pi_{Yw_{n}},F_{w_{n}})$ in $\mathcal{F}_{Het}$ defined in
$($\ref{Hete}$)$ for subsequences $w_{n}$ for which $\lambda_{w_{n},h}$
satisfies $h_{9}\in\lbrack0,\infty)$ we have $\gamma_{w_{n}}^{+}=O_{p}(1)$ and
$\Pi_{Ww_{n}}^{1/2}w_{n}(\gamma_{w_{n}}^{+}-\gamma_{w_{n}})=O_{p}(1).$ Then,
for any $\delta>0,$ the AR/AR test $\varphi_{AR/AR,\alpha-\delta,\alpha_{1}}$
in $($\ref{test rejects if}$)$ satisfies Assumptions RT and RP for the
parameter space $\mathcal{F}_{Het}$.
\end{lemma}

\subsection{Main result}

We obtain the following corollary of Lemma \ref{ARAR test}, Theorem
\ref{correct asy size generic}, and the verification of Assumption MS in
subsection \ref{2 model sel} for the two model selection methods
$\varphi_{MS,c_{n}}$ suggested there.

Define the parameter space $\mathcal{F}_{Het}$ as the intersection of the
parameter spaces defined in $($\ref{F het def}$)$ and $($\ref{Hete}$)$ when
the method in (\ref{AS method}) is used as $\varphi_{MS,c_{n}}$ (and a
slightly more restricted parameter space when (\ref{GKM2020 method}) is used,
as explained below (\ref{GKM2020 method}).)\smallskip

\begin{corollary}
\label{New test result}Assume the same condition as in Lemma \ref{ARAR test}.
Then the test $\varphi_{MS-AKP,\alpha}$ defined in $($\ref{new test}$)$ with
$\delta>0$ and $c_{n}$ satisfying the conditions in $($\ref{sequence cn}$)$
implemented with the AR/AR test $\varphi_{AR/AR,\alpha-\delta,\alpha_{1}}$ of
\citet{Andrews2017} playing the role of $\varphi_{Rob,\alpha-\delta}$ and
either of the two model selection methods described above used for
$\varphi_{MS,c_{n}},$ has asymptotic size bounded by the nominal size $\alpha$
for the parameter space $\mathcal{F}_{Het}$ defined in the paragraph above for
$\alpha\in\left\{  1\%,5\%,10\%\right\}  $ and $k-m_{W}\in\left\{
1,...,20\right\}  .$
\end{corollary}

\textbf{Comment. }Note that under the null hypothesis the test does not depend
on the value of the reduced form matrix $\Pi_{Y}.$

\section{Monte Carlo study\label{MC section}}

In this section we investigate the finite sample performance in model
(\ref{struc}) of the suggested new test $\varphi_{MS-AKP,\alpha}$ defined in
(\ref{new test}) and juxtapose it to the performance of alternative methods
suggested in the extant literature, namely the two-step tests AR/AR, AR/LM,
and AR/QLR1 in \citet{Andrews2017}. For the implementation of $\varphi
_{MS-AKP,\alpha}$ we use both methods considered in Section \ref{2 model sel}
and call the resulting tests MS-AKP1 and MS-AKP2 for the remainder of this
section. We also simulate the performance of the test AR$_{AKP,\alpha}$ (which
is of course size distorted in the setups with CHET that are outside of KP structure).

All results below are for nominal size $\alpha=5\%.$ Unless otherwise stated, we take $m_{W}=1.$ We consider the case
$\beta\in\Re$ and $\gamma\in\Re$ and pick $\gamma=0$ and test the null
hypothesis in (\ref{eq: hypotheses}) with $\beta_{0}=0.$ $\smallskip$

\subsubsection*{Choice of tuning parameters}

The implementation of the various tests depends on a large number of user
chosen constants. In particular, to implement the AR/AR, AR/LM,\ and the
AR/QLR1 tests we pick $\alpha_{1}=.005,$ $K_{L}=K_{U}=0.05$ as already
mentioned above after (\ref{def alpha2}). To calculate the estimator set
$\widetilde{\Gamma}_{1n}$ we employ the closed form solution provided below
(\ref{est set}). We choose $a=.001$ and pick the elements of the random matrix
$\zeta_{1}\in\Re^{k\times m_{W}}$ as i.i.d. $N(0,1)$ independent of all other
variables considered, see the last line of (\ref{harb def}).\footnote{Note
that by choosing $a\neq0$ the tests are no longer invariant to nonsingular
transformations of the IV vector. However, for small $a$ the differences after
a transformation are usually very small.} The confidence interval (or region)
for $\gamma$ that appears in (\ref{cs set}) is obtained by grid search over an
interval (or rectangle) of length 20 centered at the true value of $\gamma$
with 100 equally spaced gridpoints.\footnote{When the dimension of $\gamma$
grows then the implementation of that step by grid search will cause an
exponential increase in computation time for each of the two-step methods.} To
implement the AR/QLR1 test, as in \citet{Andrews2017} we pick $K_{L}^{\ast
}=K_{U}^{\ast}=0.005$ and $K_{rk}=1$. We refer to Table II in
\citet{Andrews2017} that provides the results of a comprehensive sensitivity
analysis on most of the user chosen constants above. To calculate the
data-dependent critical values for the AR/QLR1 test we use 10,000 i.i.d
chi-square random variables. There was no noticeable difference between
$\delta=0$ and $\delta=10^{-6}$ for $\delta$ given in (\ref{new test});
therefore, for the sake of computational simplicity, we pick the former in the
simulations. Finally, $c_{n}$ has to be chosen, which we do in the next
subsection.\smallskip

\subsubsection*{Recommended choices for $c_{n}$}

First, we perform a large number of simulations in order to determine
recommendations for the sequence of constants $c_{n}$ satisfying
(\ref{sequence cn}). We make recommendations for $c_{n,k,m_{W}}=c_{n}$ as a
function of the number $k$ of IVs and the subvector dimension $m_{W}$ and consider choices from $k\in\{2,3,4\}$ and $m_{W}\in\{1,2\}$.

For each $k$, sample size $n\in\{250,500\},$ and
$(\Pi_{Y},\Pi_{W})\in\Re^{k\times2}$ with
\begin{equation}
\Pi_{W}=1^{k}\pi_{W}/(nk)^{1/2}\label{piw specification}%
\end{equation}
with $\pi_{W}\in\{2,4,40\},$ corresponding to \textquotedblleft very
weak\textquotedblright, \textquotedblleft weak\textquotedblright,\ and
\textquotedblleft strong\textquotedblright\ identification of $\gamma$ $($and,
relevant for the power results below, $\Pi_{Y}=\widetilde{1}^{k}\pi
_{Y}/(nk)^{1/2}$ with $\pi_{Y}\in\{2,4,40\}$ and $\widetilde{1}^{k}$ equal to
$(1^{k/2\prime},-1^{k/2\prime})^{\prime}$ when $k$ is even and equal to
$(1,-1^{2\prime})^{\prime}$ when $k=3)$ we randomly generate $1,000$ different
DGPs (that is a choice for the covariance matrix) as described below and
simulate the NRPs (using $5,000$ i.i.d samples of each given DGPs) of MS-AKP1
and MS-AKP2 for choices of $c_{n}$ given as
\begin{equation}
c_{n}=c_{n,k,1}=c(k,1)n^{1/2}/\ln\ln n\label{choice of cn}%
\end{equation}
with $c(k,1)$ taken from the set $C:=\{.05,.1,...,3\}.$

In finite sample simulations for the DGPs considered here, the AR/AR test
sometimes slightly overrejects. For example, under CHOM, $n=250,$ $k=3,$
strong IVs, and covariance matrix $\Sigma$ being chosen as below
(\ref{covariance matrix considered}), where $(u_{i},v_{Y,i},v_{W,i})^{\prime
}\sim$ i.i.d. $N(0^{3},\Sigma),$ the AR/AR test has NRP equal to 5.4\%. From
our theory we also know that the test AR$_{AKP,\alpha}$ (at least under AKP
structures) has nonsmaller NRP than the AR/AR test. Define as the
"\emph{simulated size} of a test when there are $k$ IVs" the highest empirical
NRP of the test over all choices of $n$, $\Pi$, and ($1,000$) random DGPs
considered. For each of the two methods MS-AKP1 and MS-AKP2 and for each
$k\in\{2,3,4\},$ our recommendation for $c_{n,k,1}$ then is to take the
largest $c(k,1)$ in $C$ such that the \emph{simulated size} does not exceed
6\% (that is, we allow for a distortion of 1\% in the "\emph{simulated
size}"). It turns out that in our simulations this criterion for $c_{n,k,1}$
always leads to well defined choice of $c(k,1)$ (when a priori it could be
that even for the smallest/largest choice of $c(k,1)$ in $C$ the simulated
size exceeds/is still below 6\%)$.$

To generate random DGPs we consider the following mechanism. Given all tests
considered above, including AR$_{AKP,\alpha},$ have correct asymptotic size
under AKP structure we focus attention on designs with conditional
heteroskedasticity that are not of AKP\ structure. In particular, we choose
\begin{align}
\varepsilon_{i}  &  =(\alpha_{\varepsilon}+||Q_{\varepsilon}\overline{Z}%
_{i}||)u_{i},\nonumber\\
V_{Y,i}  &  =(\alpha_{V}+||Q_{V}\overline{Z}_{i}||)v_{Y,i},\nonumber\\
V_{W,i}  &  =(\alpha_{V}+||Q_{V}\overline{Z}_{i}||)v_{W,i},
\label{design descr}%
\end{align}
with $(u_{i},v_{Y,i},v_{W,i})^{\prime}\sim$ i.i.d.~$N(0^{3},\Sigma)$ and
independent of $\overline{Z}_{i}\sim$ i.i.d.~$N(0^{k},I_{k})$ for $i=1,...,n$.
Each of the 1,000 random DGPs is determined by choosing $\alpha_{\varepsilon
},\alpha_{V}\in\Re,$ $Q_{\varepsilon},Q_{V}\in\Re^{k\times k},$ and $\Sigma
\in\Re^{3\times3},$ where $\Sigma$ has diagonal elements equal to 1. The
scalars $\alpha_{\varepsilon},\alpha_{V}$ and the components of
$Q_{\varepsilon},Q_{V}\in\Re^{k\times k}$ are obtained by i.i.d.~draws from a
$U[0,10],$ and the off-diagonal ones of $\Sigma\in\Re^{3\times3}$ are obtained
by i.i.d.~draws from a $U[0,1]$ (subject to the restriction that the resulting
matrix $\Sigma$ is pd). Note that the setup in (\ref{design descr}) nests KP
structure when e.g. $\alpha_{\varepsilon}=\alpha_{V}=0,$ $Q_{\varepsilon
}=Q_{V}=I_{k}$ and CHOM when e.g. $\alpha_{\varepsilon}=\alpha_{V}=1,$
$Q_{\varepsilon}=Q_{V}=0^{k\times k}.$

For each $k=2,3,4$ the binding constraint on $c(k,1)$ always came from the
combination $n=250$ and \textquotedblleft strong\textquotedblright%
\ identification, while for the \textquotedblleft very
weakly\textquotedblright\ identified scenario even the largest choice of
$c(k,1)\in C$ typically did not yield overrejection for any of the sample
sizes considered. Based on the above setup we recommend the following choices
for $c_{n,k,1}.$ For Method 1 in Section \ref{2 model sel}, MS-AKP1, that is
for $\varphi_{MS-AKP,\alpha}$ based on the distance in Frobenius norm
statistic$,$ we suggest%
\begin{equation}
c(2,1)=.85,\qquad c(3,1)=1.25,\qquad c(4,1)=1.4, \label{recommendation 1 mw=2}%
\end{equation}
while for Method 2, MS-AKP2, that is for $\varphi_{MS-AKP,\alpha}$ based on
the KPST statistic in GKM22, we suggest
\begin{equation}
c(2,1)=.75,\qquad c(3,1)=1.45,\qquad c(4,1)=1.9. \label{recommendation 2 mw=2}%
\end{equation}
Recall that with these choices of $c(k,1)$ and $c_{n}$ chosen as in
(\ref{choice of cn}) the tests and MS-AKP1 and MS-AKP2 have correct asymptotic
size for a parameter space with arbitrary forms of conditional
heteroskedasticity.\smallskip

Next we consider $m_{W}=2.$ We take $\gamma=(0,0)^{\prime}$. As pointed out above already, the computational
effort in the above exercise increases exponentially in the dimension of $m_{W}$ if we use the same number of gridpoints in each dimension in the calculation of the confidence interval for $\gamma$ that appears in (\ref{cs set}). Therefore, we use a grid of a product of two intervals of length 20 centered at the true value of $\gamma$ with only 50 equally spaced gridpoints in each dimension (rather than 100 in the case $m_{W}=1$.) Everything else is the same, mutatis mutandis (e.g. $\Sigma$ is now a 4x4 matrix), as described in the case $m_{W}=1$ except that $\Pi_{W}\in\Re^{k\times2}$ is taken as $(\pi_{W1}e_{1},\pi_{W2}e_{2})/(nk)^{1/2}$ with $\pi_{W1},\pi_{W2} \in\{2,40\}$ and that the search set for $c(k,2)$ is increased to $C:=\{.05,.1,...,7.5\}$.

For Method 1 in Section \ref{2 model sel}, MS-AKP1 we suggest%
\begin{equation}
c(3,2)=1.75,\qquad c(4,2)=3.2,\qquad c(5,2)=3.05, \label{recommendation 1}%
\end{equation}
while for Method 2, MS-AKP2 we suggest
\begin{equation}
c(3,2)=2.9,\qquad c(4,2)=7.2,\qquad c(5,2)=7.5. \label{recommendation 2}%
\end{equation}
Just like in the case $m_{W}=1$ the highest null rejection probabilities occur in the strongly identified case.

\subsubsection*{Size results}

All results below are for the case where $m_{W}=1.$ Under a setup with CHET outside of KP,
the tests MS-AKP1 and MS-AKP2 equal the AR/AR test wpa1. We therefore first
consider the KP setup in \citet{Andrews2017} in Section 9.1 which is obtained
from (\ref{design descr}) with $\alpha_{\varepsilon}=\alpha_{V}=0$ and
$Q_{\varepsilon}=Q_{V}=I_{k}$. We also consider the setup with CHOM obtained
from (\ref{design descr}) with $\alpha_{\varepsilon}=\alpha_{V}=1$ and
$Q_{\varepsilon}=Q_{V}=0^{k\times k}.$ Then, finally, below we also examine
how power is affected as the DGP transitions from CHOM to CHET outside of KP.

In both cases of CHET and CHOM, we take the matrix
\begin{equation}
k\Sigma\in\Re^{3\times3} \label{covariance matrix considered}%
\end{equation}
to have diagonal elements equal to one, and the (1,2) and (1,3) elements equal
to .8 and the (2,3)\ element equal to .3, as in \citet{Andrews2017}. We
consider $\pi_{W}=\pi_{Y}\in\{2,4,40\}$ in (\ref{piw specification}), again,
representing "very weak", "weak", and "strong" IVs, also see
\citet{Andrews2017}. Finally, we take $k\in\{2,3,4\}$ and sample sizes
$n\in\{250,500\}.$ Altogether, that makes for 36 different specifications. In
addition, we also obtain results for certain cases of mixed identification
strength, e.g. when $\pi_{W}\neq\pi_{Y}\in\{2,40\}$ and also some results for
larger sample sizes.

As reported in \citet{Andrews2017}, we also find that in an overall sense the
AR/AR and AR/LM tests are dominated by the AR/QLR1 test. For instance,
regarding the AR/LM test, its power function (even in the strong IV context
under CHOM) is not always U-shaped and suffers from power dips against certain
alternatives. For example, for the KP setup for $n=250,$ $k=4,$ with weak IVs,
the power of the AR/LM and AR/QLR1 tests when $\beta=-2$ are 8.6\% and 75.6\%,
respectively, while in the setup with CHOM when $\beta=-1.43$ the power of the
AR/LM test is 34.9\% while all the other tests have power equal to 100\%. On
the other hand, the AR/AR test fares worse than the AR/QLR1 test in strongly
identified overidentified situations. In what follows, we do not therefore
discuss the AR/LM test in much detail.

We consider rejection probabilities under the null $\beta_{0}=0$ and (for
power) under a grid of seven $\beta$ values on each side of 0 with distances
from the hypothesized value 0 chosen depending on the strength of
identification. For example, in the very weakly, weakly, and strongly
identified cases we take $\beta$ in the interval $[-2,2],$ $[-2,2]$, and
$[-.2,.2]$, respectively, around the true value of 0. Results are obtained
from $10,000$ i.i.d samples from each DGP.

First, we discuss the NRPs. Over the 18 DGPs of the KP setups, the NRPs of
MS-AKP1, MS-AKP2$,$ AR/AR, AR/LM, and AR/QLR1 lie in the intervals (all
numbers in \%): [3.5,5.9], [3.3,6.0], [1.9,5.1], [.6,5.2], and [1.5,4.9]. As
set up above, the tests MS-AKP1 and MS-AKP2 slightly overreject the null for
small sample sizes (especially in the strongly identified case), but the size
distortion disappears as $n$ grows. For example, the NRPs of MS-AKP2 in the KP
setup with $k=3$ and strong identification is 6.0, 5.5, 5.2, and 5.1\%,
respectively, when $n=250,$ $500,$ $1,000,$ and $1,500.$ On the other hand,
the tests AR/AR, AR/LM, and AR/QLR1, while controlling the NRP very well,
underreject the null in weakly identified scenarios. This leads to relatively
poor power properties relative to the tests MS-AKP1 and MS-AKP2 in weakly
identified situations.

Regarding the 18 DGPs with CHOM, the one important difference relative to the
KP\ setup is that the three tests AR/AR, AR/LM, and AR/QLR1 are less
conservative with NRPs over the 18 DGPs in the intervals [4.1,5.4], [3.5,5.4],
and [3.7,5.1], respectively. As a consequence, these tests have relatively
better power properties than in the KP setup.\smallskip

\subsubsection*{Power results}

Next we discuss the power results. We focus again on the case where $m_{W}=1.$
Power for MS-AKP1, MS-AKP2$,$ AR/AR, and AR/QLR1 increases as the IVs become
stronger. On the other hand, by the local-to-zero design considered here (see
(\ref{piw specification}) and below), as $n$ increases, power for these three
tests changes only slightly. We therefore only provide details for the case
where $n=250$. Power of all the tests is much higher in the setting with CHOM
compared to the KP setting and especially so for the AR/QLR1 test (because it
underrejects the null hypothesis less under CHOM than under KP). As one
example, consider the case $n=250,$ $k=2,$ with weak identification. In that
case, when $\beta=-.571$ the tests MS-AKP2, AR/AR, and AR/QLR1 have power
48.7, 46.3, and 45.4\% under KP, but power equal to 95.9, 95.6, and 95.4\%
under CHOM!\bigskip

\begin{figure}[ptb]
\centering
\includegraphics[width=0.75\paperwidth]{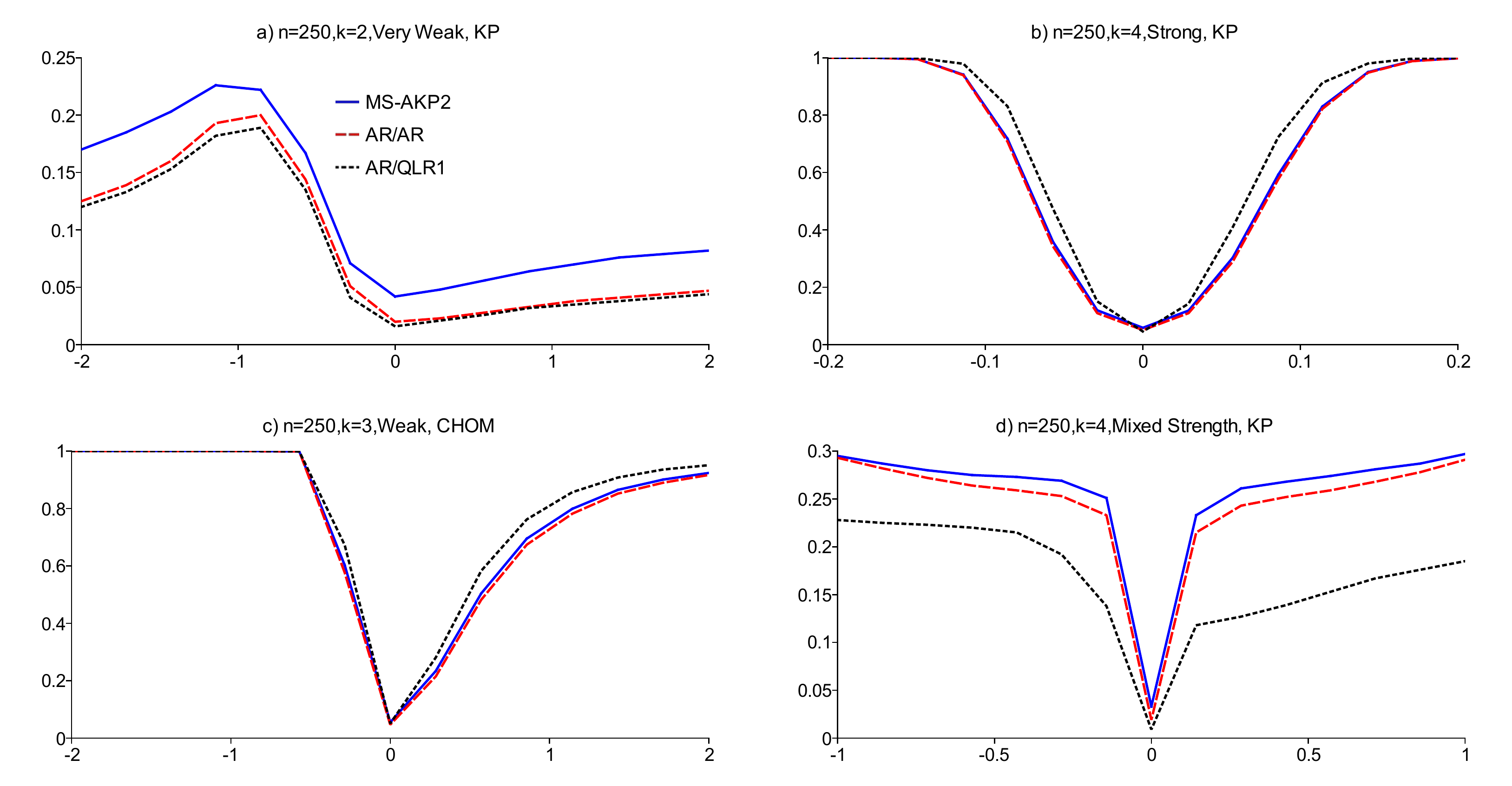}  \caption{Power of
various subvector tests in different cases. Covariance structure: Kronecker
product (KP); CHOM. Identification strength $(\pi_{W},\pi_{Y})$: Very Weak
$(2,2)$; Weak $(4,4)$; Strong $(40,40)$; Mixed strength: $(2,40)$. }%
\label{fig: power}%
\end{figure}

A representative selection of power curves in four different cases is plotted
in Figure \ref{fig: power}. Note that in the figures corresponding to the
different cases, both the scale of the horizontal and the vertical axes vary
by a lot depending on the strength of identification.

The key takeaways from the power study are as follows:

i) Based on the DGPs considered here we cannot make a clear recommendation as
to which one of the two tests MS-AKP1 and MS-AKP2 is preferable. In most
cases, they have virtually identical power. In few cases, one dominates the
other, but only by a small difference. In the Figures below we only report
results for MS-AKP2.

ii) Regarding the comparison between the tests MS-AKP1, MS-AKP2 and AR/AR we
find that the former two virtually uniformly dominate the latter in all the
designs considered. This is not surprising given the construction of the new
tests and given they satisfy Assumption RP above. The relative power advantage
of the tests MS-AKP1, MS-AKP2 over AR/AR partly stems from the underrejection
of the latter test under the null. See e.g. Figure \ref{fig: power}a that
contains power curves for $n=250,$ $k=2,$ very weak identification, and KP
structure for MS-AKP2, AR/AR, and AR/QLR1. (The NRPs of the three tests
reported here are 4.2, 2.0, and 1.6\%, respectively. Note that the x-axis in
Figures I-IV plots the true value $\beta.$)

iii) Regarding the comparison between the tests MS-AKP1, MS-AKP2 and AR/QLR1
in the case of equal identification strength $\pi_{W}=\pi_{Y}$ we find that
the former two are generally more powerful under weak identification and small
$k$ while the reverse is true under strong identification and larger $k,$ see
Figures \ref{fig: power}a and b for the cases \textquotedblleft$k=2$ and very
weak identification\textquotedblright\ and \textquotedblleft$k=4$ and strong
identification,\textquotedblright\ respectively, both for $n=250$ and KP. (In
Figure \ref{fig: power}b, the NRPs of the tests MS-AKP2, AR/AR, and AR/QLR1
are 5.9, 5.1, and 4.6\%, respectively.) These two figures show the best
relative performances for the MS-AKP1, MS-AKP2 and AR/QLR1 tests in the
\textquotedblleft equal identification\textquotedblright\ settings where
$\pi_{W}=\pi_{Y}$. In Figure \ref{fig: power}a the power advantage of MS-AKP2
over AR/QLR1 is as high as 5.2\%, while in Figure \ref{fig: power}b the power
of AR/QLR1 can be up to 13.1\% more powerful than MS-AKP2.

In the \textquotedblleft intermediate\textquotedblright\ case between these
extremes, namely \textquotedblleft$k=3$ and weak
identification\textquotedblright\ (again with $n=250$ and KP; not reported in
Figure \ref{fig: power}), the MS-AKP1 and MS-AKP2 tests have slightly higher
power than AR/QLR1 when $\beta$ is positive while the reverse is true for
negative values of $\beta$. In all cases, the relative performance of the
AR/QLR1 test improves under CHOM; under CHOM, for the \textquotedblleft
intermediate\textquotedblright\ case \textquotedblleft$k=3$ and weak
identification\textquotedblright\ (again with $n=250$) the AR/QLR1 test has
uniformly higher power than the MS-AKP1 and MS-AKP2 tests, see Figure
\ref{fig: power}c. (In Figure \ref{fig: power}c, the NRPs of the tests
MS-AKP2, AR/AR, and AR/QLR1 are 5.5, 4.7, and 5.1\%, respectively.)

In cases of mixed identification strength, $\pi_{W}\neq\pi_{Y}\in\{2,40\},$ we
find that when $\pi_{W}=2$ and $\pi_{Y}=40$ the tests MS-AKP1 and MS-AKP2 have
uniformly higher power than AR/QLR1 for all $k$ considered whereas in the case
$\pi_{W}=40$ and $\pi_{Y}=2$ all tests have comparable power. See Figure
\ref{fig: power}d that contains the case $\pi_{W}=2$ and $\pi_{Y}=40,$
$n=250,$ $k=4,$ with KP structure where the power gap between the new tests
and AR/QLR1 is as high as 13.4\%. (In Figure \ref{fig: power}d, the NRPs of
the tests MS-AKP2, AR/AR, and AR/QLR1 are 3.3, 1.9, and 0.9\%, respectively.)
It seems that in these cases of mixed identification strength the new tests
enjoy their most competitive relative performance.\smallskip

\subsubsection*{Results for non-KP DGPs}

Finally, we examine how rejection probabilities are affected as the DGP
transitions from KP to CHET outside of KP. To do so, we report rejection
probabilities under the null and certain alternatives and probabilities with
which MS-AKP1 and MS-AKP2 equals the AR/AR test in the second stage for a
class of DGPs that under KP coincide with the ones considered in Figure
\ref{fig: power}b ($\pi_{W}=\pi_{Y}=40)$ and Figure
\ref{fig: power}d ($(\pi_{W},\pi_{Y})=(2,40))$.
In particular, we choose $k=4,$ $n=250,$ $\gamma=0,$ and the matrix $\Sigma$
equals the one in (\ref{covariance matrix considered}). In (\ref{design descr}%
) we take%
\begin{equation}
Q_{\varepsilon}=I_{4}+\varrho\left(
\begin{array}
[c]{cccc}%
10 & 8 & 6 & 4\\
3 & 5 & 9 & 3\\
8 & 6 & 9 & 2\\
4 & 3 & 2 & 1
\end{array}
\right)  \label{Qeps}%
\end{equation}
for $\varrho\in\{0,.01,...,.1\},$ $\alpha_{\varepsilon}=\alpha_{V}=0,$ and
$Q_{V}=I_{4}.$ Note that for $\varrho=0,$ the design leads to KP while for
$\varrho>0,$ it leads to CHET outside of KP. In particular, $\arg\min
_{G,H>0}||G\otimes H-\overline{R}_{F}||$ (with $\overline{R}_{F}$ defined in
(\ref{def R_F}) with $U_{i}=(\varepsilon_{i},V_{W,i}^{\prime})^{\prime})$
equals $0,$ $.14,$ $.29,$ $.45,$ $.60,$ $.74,$ $.88,$ $1.01,$ $1.13,$ $1.24$,
and $1.34$ when $\varrho\in\{0,.01,.02,....,.1\},$ respectively. The latter
numbers are found by simulations based on $10^{7}$ simulation repetitions
using Theorem 1 in GKM22.

As before, we report results for $10,000$ simulation repetitions at nominal
size $5\%$.

\textbf{Null rejection probabilities. }Here we report results when $\beta=0,$
that is, we report NRPs$.$

First, in the setup of Figure \ref{fig: power}b the probability with which
MS-AKP1 and MS-AKP2 coincide with AR/AR is strictly increasing in $\varrho$
and e.g. equals 66.2\% and 64.9\%, 82.9\% and 85.1\%, and 98.8\% and 99.5\%,
respectively for $\varrho=0,.03,$ and $.1$, respectively. (We also simulated
these probabilities when $\varrho=0$ for $n=500$ and they equal 20.8\% and
20.2\%, respectively). The highest NRP of both the MS-AKP1 and MS-AKP2 tests
is 5.9\% which occurs when $\varrho=0$ and is caused by a 7.4\% NRP of the
conditional subvector test AR$_{AKP,\alpha}$ (even though by Theorem
\ref{correct asy size} this test has correct asymptotic NRP for this DGP;
interestingly, this test has NRP equal to 5.8\% when $\varrho=.1,$ a case that
is not covered by Theorem \ref{correct asy size}). Given the MS-AKP1 and
MS-AKP2 tests equal the AR/AR test with increasing probability as $\varrho$
increases, their NRPs get closer (but not monotonically so) to 5\% as
$\varrho$ increases. As $\varrho=.1$ both tests have NRP equal to 5.2\%.

Second, in the setup of Figure \ref{fig: power}d the probability with which
MS-AKP1 and MS-AKP2 coincide with AR/AR are identical as just reported for the
setup in Figure \ref{fig: power}b. The highest NRPs of the MS-AKP1 and MS-AKP2
tests are 3.5\% and 3.2\% respectively, which occur when $\varrho=0$. The
AR/AR and AR/QLR1 tests have NRPs in the intervals [1.2\%,1.9\%] and
[.3\%,.8\%], respectively, and therefore, quite substantially underreject the
null hypothesis. As the MS-AKP1 and MS-AKP2 tests equal the AR/AR test with
increasing probability as $\varrho$ increases, their NRPs approach 1.2\% as
$\varrho$ gets closer to $.1.$\smallskip

\textbf{Power results.} Here we examine how power is affected as the DGP
transitions from KP to CHET outside of KP.

First, in the setup of Figure \ref{fig: power}b) we consider the alternative
$\beta=.1.$ The probabilities with which MS-AKP1 and MS-AKP2 coincide with
AR/AR are increasing in $\varrho$ and are very similar to the corresponding
values when $\beta=0;$ e.g. the probabilities equal 68.6\% and 68.9\% when
$\varrho=.01$, respectively, and equal 98.2\% and 99.1\% when $\varrho=.1.$
Power for all tests monotonically decreases as $\varrho$ increases, e.g. for
AR/QLR1, AR/AR, and MS-AKP1 from 83.5\% to 44.4\%, from 71.5\% to 32.4\% and
from 72.5\% to 32.4\%, respectively, when $\varrho$ goes from 0 to .1.

Second, in the setup of Figure \ref{fig: power}d) we consider the alternative
$\beta=-1.$ When $\varrho=.01,$ MS-AKP1 and MS-AKP2 coincide with AR/AR with
probability 78.3\% and 75.5\%, respectively and for $\varrho\geq.04$ both
MS-AKP1 and MS-AKP2 coincide with AR/AR at least 99.6\% of the cases. The
power of the AR/AR and the AR/QLR1 for all values of $\varrho
\in\{0,.01,.02,....,.1\}$ are in the intervals [27.9\%,29.4\%] and
[21.2\%,23.5\%], respectively, with neither test's power being monotonic in
$\varrho.$ While the power of MS-AKP1 and MS-AKP2 slightly exceeds the power
of the AR/AR test for $\varrho<.04$ their power is identical to the one of the
AR/AR test for larger values of $\varrho.\smallskip$

In sum, as one would expect given the construction of MS-AKP1 and MS-AKP2
tests, when moving from KP to CHET outside of KP, their rejection
probabilities get closer and closer to those of the AR/AR test.\smallskip

\section{Conclusion}

We propose the construction of a robust test that improves the power of
another robust test by combining it with a powerful test that is only robust
for a subset of the parameter space. We implement this construction in the
context of the linear IV model applied to the AR$_{AKP,\alpha}$ test that has
correct asymptotic size for a parameter space that imposes AKP structure and
the AR/AR test that is robust even when allowing for arbitrary forms of CHET.
We believe that the particular construction and implementation suggested here,
namely combining a powerful but non fully robust test with a less powerful
fully robust test in order to obtain a fully robust more powerful test, might
be successfully applied in other scenarios and also in the current scenario
based on different choices of testing procedures. For instance, it might be
feasible to combine the LR type subvector test of \citet{Kleibergen2021} with
the AR/QLR1 of \citet{Andrews2017} but it would be technically substantially
more challenging to verify the assumptions given above that are sufficient for
control of the asymptotic size of the resulting test. Other extensions include
improving the power of the AR$_{AKP,\alpha}$ test by making the conditional
critical value depend on more than just the largest eigenvalue.

\appendix

\section{{Appendix}}

{ The Appendix is structured as follows. In Section \ref{s: proof asy size}
the proof of Theorem \ref{correct asy size} is given, prepared for first with
several technical lemmas in Subsection \ref{tech lemmas}. Next in Section
\ref{proof of generic result} the proof of Theorem
\ref{correct asy size generic} is given. We provide verifications of the high
level assumptions for particular implementations of the test including for
both $\varphi_{MS,c_{n}}$ and AR/AR in Sections \ref{model selection proof}
and \ref{AR/AR proof}, respectively. Finally, in Section
\ref{time series section}, we generalize the conditional subvector test to a
time series framework. }

\subsection{{Proof of Theorem \ref{correct asy size}\label{s: proof asy size}%
}}

\subsubsection{{Technical lemmas\label{tech lemmas}}}

{ In what follows below we will require results about solutions to certain
minimization problems involving the Frobenius norm. The next lemma provides a
special case of Corollary 2.2 in \citet{VanLoanPitsianis1993}. Note that
\citet{VanLoanPitsianis1993} point to \citet[p.73]{GolubVanLoan1989} for a
proof of Corollary 2.2. However, the result in \citet[p.73]{GolubVanLoan1989}
is for a minimization problem using the $p$-norm for $p=2$ and not the
Frobenius norm which is used here. }

\begin{lemma}
{ \label{frobenius norm min}Consider the minimization problem
\[
\min_{B\in\Re^{m\times n},\text{ }rk(B)=1}||A-B||^{2}%
\]
for a given nonzero matrix $A\in\Re^{m\times n}$ with singular value
decomposition $A=Udiag(\sigma_{1},...,\sigma_{p})V^{\prime}$ for singular
values $\sigma_{1}\geq\sigma_{2}\geq...\geq\sigma_{p}\geq0$ with
$p=\min\{m,n\}$ and rectangular $diag(\sigma_{1},...,\sigma_{p})\in
\Re^{m\times n}$, orthogonal matrices $U=[u_{1},...,u_{m}]\in\Re^{m\times m},$
and $V=[v_{1},...,v_{n}]\in\Re^{n\times n}.$ Then a minimizing argument is
given by $B=\sigma_{1}u_{1}v_{1}^{\prime}$ and the minimum equals $%
{\textstyle\sum\nolimits_{i=2}^{p}}
\sigma_{i}^{2}.$ If $\sigma_{1}>\sigma_{2}$ then $B=\sigma_{1}u_{1}%
v_{1}^{\prime}$ is the unique minimizer. }
\end{lemma}

{ \textbf{\noindent Proof of Lemma \ref{frobenius norm min}. }Note that
\begin{equation}
\min_{B\in\Re^{m\times n},\text{ }rk(B)=1}||A-B||^{2}=\min_{C\in\Re^{m\times
n},\text{ }rk(C)=1}||diag(\sigma_{1},...,\sigma_{p})-C||^{2} \label{expre0}%
\end{equation}
by viewing $C=U^{\prime}BV$ and because $||D||=||U^{\prime}D||=||DV||$ for any
matrix $D\in\Re^{m\times n}$ and conformable orthogonal matrices $U$ and $V.$
We can write any matrix $C\in\Re^{m\times n}$ with $rk(C)=1$ as
\begin{equation}
C=||c||^{-1}(\alpha_{1}c,...,\alpha_{n}c) \label{expre01}%
\end{equation}
for $c\in\Re^{m}\backslash\{0^{m}\}$ and $\alpha_{k}\in\Re$ for $k=1,...,n.$
Because $||A+B||^{2}=||A||^{2}+||B||^{2}+2<A,B>_{F}$ where $<A,B>_{F}%
:=trace(A^{\prime}B)$ denotes the Frobenius inner product, and $||diag(\sigma
_{1},...,\sigma_{p})||^{2}=%
{\textstyle\sum\nolimits_{i=1}^{p}}
\sigma_{i}^{2},$ $||C||^{2}=%
{\textstyle\sum\nolimits_{i=1}^{n}}
\alpha_{i}^{2},$ $<diag(\sigma_{1},...,\sigma_{p}),C>_{F}=%
{\textstyle\sum\nolimits_{i=1}^{p}}
\sigma_{i}\alpha_{i}c_{i}||c||^{-1}$ for $c=(c_{1},...,c_{m})^{\prime}$ we
have
\begin{equation}
||diag(\sigma_{1},...,\sigma_{p})-C||^{2}=%
{\textstyle\sum\nolimits_{i=1}^{p}}
\sigma_{i}^{2}+%
{\textstyle\sum\nolimits_{i=1}^{n}}
\alpha_{i}^{2}-2%
{\textstyle\sum\nolimits_{i=1}^{p}}
\sigma_{i}\alpha_{i}c_{i}||c||^{-1}. \label{expr1}%
\end{equation}
Viewing (\ref{expr1}) as a function in $\alpha_{k},$ $k=1,...,n,$ and $c,$
taking first order conditions (FOCs) with respect to $\alpha_{k},$ we obtain
$2\alpha_{k}-2\sigma_{k}c_{k}||c||^{-1}=0$ or
\begin{equation}
\alpha_{k}=\sigma_{k}c_{k}||c||^{-1}\text{ for }k=1,...,p\text{ and }%
\alpha_{k}=0\text{ for }k=p+1,...,n. \label{expr2}%
\end{equation}
Taking FOCs with respect to $c_{j},$ $j=1,...,p,$ we obtain $(||c||\sigma
_{j}\alpha_{j}-(%
{\textstyle\sum\nolimits_{i=1}^{p}}
\sigma_{i}\alpha_{i}c_{i})c_{j}||c||^{-1})||c||^{-2}=0$ and thus
\begin{equation}
||c||^{2}\sigma_{j}\alpha_{j}-(%
{\textstyle\sum\nolimits_{i=1}^{p}}
\sigma_{i}\alpha_{i}c_{i})c_{j}=0 \label{expr3}%
\end{equation}
and for $j=p+1,...,m$ we have $(%
{\textstyle\sum\nolimits_{i=1}^{p}}
\sigma_{i}\alpha_{i}c_{i})c_{j}||c||^{-3}=0$ and therefore
\begin{equation}
c_{j}%
{\textstyle\sum\nolimits_{i=1}^{p}}
\sigma_{i}\alpha_{i}c_{i}=0. \label{expr35}%
\end{equation}
The objective is to find $(c_{1},...,c_{p})$ such that the two summands in
(\ref{expr1}) that depend on $C$ are being minimized. Using (\ref{expr2}) we
thus need to find $(c_{1},...,c_{m})$ such that
\begin{equation}%
{\textstyle\sum\nolimits_{i=1}^{p}}
\sigma_{i}^{2}c_{i}^{2}||c||^{-2}-2%
{\textstyle\sum\nolimits_{i=1}^{p}}
\sigma_{i}^{2}c_{i}^{2}||c||^{-2}=-%
{\textstyle\sum\nolimits_{i=1}^{p}}
\sigma_{i}^{2}(\frac{c_{i}}{||c||})^{2} \label{expr5}%
\end{equation}
is minimized. Let $a$ be the largest index for which $\sigma_{1}%
=...=\sigma_{a}.$ Given that $\sigma_{a}>\sigma_{b}$ for $b>a$ it follows that
a vector $c=(c_{1},...,c_{m})^{\prime}$ is a minimizing argument if and only
if $(c_{1},...,c_{a})^{\prime}\neq0^{m-p}$ and $(c_{a+1},...,c_{m})^{\prime
}=0^{m-a}$ and the minimum in (\ref{expr1}) equals%
\begin{equation}%
{\textstyle\sum\nolimits_{i=1}^{p}}
\sigma_{i}^{2}-%
{\textstyle\sum\nolimits_{i=1}^{p}}
\sigma_{i}^{2}(\frac{c_{i}}{||c||})^{2}=%
{\textstyle\sum\nolimits_{i=1}^{p}}
\sigma_{i}^{2}-\sigma_{1}^{2}%
{\textstyle\sum\nolimits_{i=1}^{a}}
(\frac{c_{i}}{||c||})^{2}=%
{\textstyle\sum\nolimits_{i=2}^{p}}
\sigma_{i}^{2}. \label{expre6}%
\end{equation}
For example, one solution is $c=e_{1}:=(1,0,...0)^{\prime}\in\Re^{m}$ for
which the minimizing matrix in (\ref{expre0}) becomes $C=(\sigma_{1}%
e_{1},0^{m},...0^{m}).$ Correspondingly, a minimizing matrix $B$ becomes
$UCV^{\prime}=\sigma_{1}u_{1}v_{1}^{\prime}.$ }

{ If $\sigma_{1}>\sigma_{2}$ then $a=1.$ Therefore, any minimizing $c$ equals
$(c_{1},0,...,0)^{\prime}$ for some $c_{1}\neq0$ and therefore, by
(\ref{expre01}) and (\ref{expr2}), the only minimizing matrix $C$ equals
$||c||^{-1}(\alpha_{1}c,...,\alpha_{n}c)=(\sigma_{1}e_{1},0^{m},...0^{m}).$
And consequently, there can only be a unique minimizer $B=UCV^{\prime}%
=\sigma_{1}u_{1}v_{1}^{\prime}.$ $\square$\bigskip}

{ Let $R\in\Re^{m\times l}$ and $R=U\Sigma V^{\prime}$ be a singular value
decomposition of $R,$ where $\Sigma\in\Re^{m\times l}$ has $\min\{m,l\}$
singular values of $R$ on the diagonal and zeros elsewhere, $U\in\Re^{m\times
m}$ is an orthogonal matrix of eigenvectors of $RR^{\prime},$ and $V\in
\Re^{l\times l}$ is an orthogonal matrix of eigenvectors of $R^{\prime}R.$ In
general, $U,$ $\Sigma,$ and $V$ are not uniquely defined. The matrix $\Sigma$
is uniquely determined by the restriction that the singular values are ordered
nonincreasingly. We assume that this is the case from now on. Let $a$ be the
geometric multiplicity of the largest eigenvalue of $RR^{\prime}.$ Write
$U=[\widetilde{W}:\widetilde{W}^{C}]$ for $\widetilde{W}\in\Re^{m\times a}.$
Thus $\widetilde{W}=(\widetilde{w}_{1},...,\widetilde{w}_{a})$ denotes an
orthogonal basis for the eigenspace associated with the largest eigenvalue of
$RR^{\prime}$. }

\begin{lemma}
{ \label{continuity eigenvector}Let $R$ and $R_{n}$ for $n\geq1$ be
$\Re^{m\times l}$ matrices such that $R_{n}\rightarrow R$ as $n\rightarrow
\infty$. Let $U\Sigma V^{\prime}$ and $U_{n}\Sigma_{n}V_{n}^{\prime}$ be any
singular value decompositions of $R$ and $R_{n},$ respectively, where the
singular values are ordered nonincreasingly. For $j\leq m$, denote by
$\widetilde{w}_{j}$ and $\widetilde{w}_{nj}$ the $j$-th column of $U$ and
$U_{n},$ respectively. Decompose $U=[\widetilde{W}:\widetilde{W}^{C}]\in
\Re^{m\times m},$ where $\widetilde{W}=(\widetilde{w}_{1},...,\widetilde{w}%
_{a})\in\Re^{m\times a}$ is an orthogonal basis for the eigenspace associated
with the largest eigenvalue of $RR^{\prime}.$ Conformingly, let $U_{n}%
=[\widetilde{W}_{n}:\widetilde{W}_{n}^{C}].$\footnote{But note that
$\widetilde{W}_{n}$ does not necessarily correspond to a basis for the
eigenspace of the largest eigenvalue of $R_{n}R_{n}^{\prime}$ but may
represent eigenvectors corresponding to several different eigenvalues because
the multiplicities of eigenvalues of $R_{n}R_{n}^{\prime}$ and $RR^{\prime}$
may not be the same. As a trivial example, consider $RR^{\prime}=I_{2}$ and
$R_{n}R_{n}^{\prime}$ equal to a diagonal matrix with first and second
diagonal elements equal to $1$ and $1-n^{-1}$, respectively.}\ Assume $\Sigma$
does not equal the zero matrix. Then $\widetilde{w}_{nj}^{\prime}%
\widetilde{w}_{l}=o(1)$ for $j>a$ and $l\leq a$.$\smallskip$ }
\end{lemma}

{ \textbf{\noindent Proof of Lemma \ref{continuity eigenvector}.} Wlog we can
assume $m\geq l.$ (If $m<l$ add $l-m$ rows of zeros to the bottom of $R$ and
$R_{n}.$ Then the result for
\[
\left(
\begin{array}
[c]{c}%
R\\
0^{l-m\times l}%
\end{array}
\right)  =\left(
\begin{array}
[c]{cc}%
U & 0^{m\times l-m}\\
0^{l-m\times m} & \widetilde{U}%
\end{array}
\right)  \left(
\begin{array}
[c]{c}%
\Sigma\\
0^{l-m\times l}%
\end{array}
\right)  V^{\prime}%
\]
for any orthogonal matrix $\widetilde{U}$ implies the desired result for
$R=U\Sigma V^{\prime}.$) Denote by $\sigma_{j}$ the $j$-th singular value of
$R$ (i.e. $\sigma_{j}$ equals the $(j,j)$-th element of $\Sigma$) for
$j=1,...,l,$ and likewise $\sigma_{nj}$ denotes the $j$-th singular value of
$R_{n}.$ By definition (and given that the algebraic and geometric
multiplicities coincide for any diagonalizable matrix), $a$ is the largest
index for which $\sigma_{1}=...=\sigma_{a}$. Define
\begin{equation}
\delta_{n}:=\min\{\min_{1\leq j\leq l-a}|\sigma_{a}-\sigma_{n(a+j)}%
|,\sigma_{a}\}. \label{deltan}%
\end{equation}
Then by Wedin's (1972) \nocite{Wedin1972} theorem (see, e.g. \citet{Li1998}
equations (4.4) and (4.8)\footnote{A comprehensive reference for background
reading on Wedin's (1972) \nocite{Wedin1972} theorem is
\citet[p.260, Theorem 4.1]{StewartSun1990}.}), it follows that%
\begin{equation}
||\sin\Theta(\widetilde{W},\widetilde{W}_{n})||=o(1/\delta_{n}),
\label{sin formulation}%
\end{equation}
where $\Theta(\widetilde{W},\widetilde{W}_{n})$ denotes the angle matrix
between $\widetilde{W}$ and $\widetilde{W}_{n}$ (see \citet{Li1998}, equation
(2.3) for a definition). Furthermore, by Lemma 2.1 and equation (2.4) in
\citet{Li1998}, we have
\begin{equation}
||\sin\Theta(\widetilde{W},\widetilde{W}_{n})||=||\widetilde{W}_{n}^{C\prime
}\widetilde{W}||. \label{sin expression}%
\end{equation}
Note that $\delta_{n}$ is bounded away from zero for all large $n$ because (1)
$\sigma_{a}>0$ by the assumption that $\Sigma\neq0,$ (2) if $a<l,$ by
construction $\sigma_{a}>\sigma_{a+1}$ and therefore $\min_{1\leq j\leq
l-a}|\sigma_{a}-\sigma_{n(a+j)}|$ is uniformly bounded away from zero (because
singular values are continuous as functions of the matrix elements and
$R_{n}\rightarrow R)$, and (3) if $a=l$ then $\min_{1\leq j\leq l-a}%
|\sigma_{a}-\sigma_{n(a+j)}|=\infty,$ because we take a minimum of the empty
set. Therefore, by (\ref{sin formulation}) and (\ref{sin expression}) we have%
\begin{equation}
||\widetilde{W}_{n}^{C\prime}\widetilde{W}||=o(1) \label{final result}%
\end{equation}
which implies that $\widetilde{w}_{nj}^{\prime}\widetilde{w}_{l}=o(1)$ for
$j>a$ and $l\leq a.$ $\square$ }

\subsubsection{{Uniformity Reparametrization}}

{ To prove that the new conditional subvector AR$_{AKP}$ test has asymptotic
size bounded by the nominal size $\alpha$ we use a general result in Andrews,
Cheng, and Guggenberger (2020, ACG from now
on)\nocite{AndrewsChengGuggenberger2020}. To describe it, consider a sequence
of arbitrary tests $\{\varphi_{n}:n\geq1\}$ of a certain null hypothesis and
denote by $RP_{n}(\lambda)$ the NRP of $\varphi_{n}$ when the DGP is pinned
down by the parameter vector $\lambda\in\Lambda,$ where $\Lambda$ denotes the
parameter space of $\lambda.$ By definition, the asymptotic size of
$\varphi_{n}$ is defined as
\begin{equation}
AsySz=\lim\sup_{n\rightarrow\infty}\sup_{\lambda\in\Lambda}RP_{n}(\lambda).
\label{asysz}%
\end{equation}
Let $\{h_{n}(\lambda):n\geq1\}$ be a sequence of functions on $\Lambda,$ where
$h_{n}(\lambda)=(h_{n,1}(\lambda),...,h_{n,J}(\lambda))^{\prime}$ with
$h_{n,j}(\lambda)\in\Re$ $\forall j=1,...,J.$ Define
\begin{align}
&  H\overset{}{=}\{h\overset{}{\in}(\Re\cup\{\pm\infty\})^{J}\overset{}{:}%
h_{w_{n}}(\lambda_{w_{n}})\overset{}{\rightarrow}h\text{ for some subsequence
}\{w_{n}\}\nonumber\\
&  \text{of }\{n\}\text{ and some sequence }\{\lambda_{w_{n}}\in\Lambda
:n\geq1\}\}. \label{defn of H}%
\end{align}
}

{ \noindent\textbf{Assumption B in ACG:} For any subsequence $\{w_{n}\}$ of
$\{n\}$ and any sequence $\{\lambda_{w_{n}}\in\Lambda:n\geq1\}$ for which
$h_{w_{n}}(\lambda_{w_{n}})\rightarrow h\in H,$ $RP_{w_{n}}(\lambda_{w_{n}%
})\rightarrow\lbrack RP^{-}(h),RP^{+}(h)]$ for some $RP^{-}(h),RP^{+}%
(h)\in(0,1).$\footnote{By definition, the notation $x_{n}\rightarrow\lbrack
x_{1,\infty},x_{2,\infty}]$ means that $x_{1,\infty}\leq\lim\inf
_{n\rightarrow\infty}x_{n}\leq\lim\sup_{n\rightarrow\infty}x_{n}\leq
x_{2,\infty}.$} }

{ The assumption states, in particular, that along certain drifting sequences
of parameters $\lambda_{w_{n}}$ indexed by a localization parameter $h$ the
NRP of the test cannot asymptotically exceed a certain threshold $RP^{+}(h)$
indexed by $h.$ }

\begin{proposition}
{ \label{Gen Asy Size Prop}$($ACG, Theorem 2.1$($a$)$ and Theorem
2.2$)$\ Suppose Assumption B in ACG holds. Then, $\inf_{h\in H}RP^{-}(h)\leq
AsySz\leq\sup_{h\in H}RP^{+}(h).$ }
\end{proposition}

{ We next verify Assumption B in ACG for the conditional subvector AR$_{AKP}$
test and establish that $\sup_{h\in H}RP^{+}(h)=\alpha$ when the test is
implemented at nominal size $\alpha$. In the setup considered here, the
parameter space $\Lambda$ actually depends on $n$ which does not affect the
conclusion of Theorem 2.1(a) and Theorem 2.2 in ACG. }

{ We use Proposition 16.5 in AG, to derive the joint limiting distribution of
the eigenvalues $\widehat{\kappa}_{in},$ $i=1,...,p$ in
(\ref{eq: eigenprob ARhat}). We reparameterize the null distribution $F$ to a
vector $\lambda.$ The vector $\lambda$ is chosen such that for a subvector of
$\lambda$ convergence of a drifting subsequence of the subvector (after
suitable renormalization) yields convergence of the NRP of the test. For given
$F$ and any $G_{F}\in\Re^{p\times p}$ and $\overline{H}_{F}\in\Re^{k\times k}$
such that $\overline{R}_{F}=G_{F}\otimes\overline{H}_{F}+\Upsilon_{n}$ as in
(\ref{def par spa}) define%
\begin{equation}
U_{F}:=G_{F}^{-1/2}\in\Re^{p\times p}\text{ and }Q_{F}:=H_{F}^{-1/2}%
(E_{F}\overline{Z}_{i}\overline{Z}_{i}^{\prime})^{1/2}\in\Re^{k\times k},
\label{def W and U}%
\end{equation}
where again $H_{F}=(E_{F}\overline{Z}_{i}\overline{Z}_{i}^{\prime}%
)^{-1/2}\overline{H}_{F}(E_{F}\overline{Z}_{i}\overline{Z}_{i}^{\prime
})^{-1/2}$ from (\ref{defH}). Denote by%
\begin{equation}
B_{F}\in\Re^{p\times p}\text{ an orthogonal matrix of eigenvectors of }%
U_{F}^{\prime}(\Pi_{W}\gamma,\Pi_{W})^{\prime}Q_{F}^{\prime}Q_{F}(\Pi
_{W}\gamma,\Pi_{W})U_{F} \label{B_F defn}%
\end{equation}
ordered so that the $p$ corresponding eigenvalues $(\eta_{1F},...,\eta_{pF})$
are nonincreasing. Denote by%
\begin{equation}
C_{F}\in\Re^{k\times k}\text{ an orthogonal matrix of eigenvectors of }%
Q_{F}(\Pi_{W}\gamma,\Pi_{W})U_{F}U_{F}^{\prime}(\Pi_{W}\gamma,\Pi_{W}%
)^{\prime}Q_{F}^{\prime}.\footnote{The matrices $B_{F}$ and $C_{F}$ are not
uniquely defined. We let$\ B_{F}$ denote one choice of the matrix of
eigenvectors of $U_{F}^{\prime}(\Pi_{W}\gamma,\Pi_{W})^{\prime}Q_{F}^{\prime
}Q_{F}(\Pi_{W}\gamma,\Pi_{W})U_{F}$ and analogously for $C_{F}.$
\par
Note that the role of $E_{F}G_{i}$ in AG, Section 16, is played by $(\Pi
_{W}\gamma,\Pi_{W})\in R^{k\times p}$ and the role of $W_{F}$ is played by
$Q_{F}.$} \label{C_F defn}%
\end{equation}
The corresponding $k$ eigenvalues are $(\eta_{1F},...,\eta_{pF},0,...,0).$
Denote by%
\begin{equation}
(\tau_{1F},...,\tau_{pF})\text{ the singular values of }Q_{F}(\Pi_{W}%
\gamma,\Pi_{W})U_{F}\in\Re^{k\times p}, \label{sing val defn}%
\end{equation}
which are nonnegative, ordered so that $\tau_{jF}$ is nonincreasing. (Some of
these singular values may be zero.) As is well-known, the squares of the $p$
singular values of a $k\times p$ matrix $A$ equal the $p$ largest eigenvalues
of $A^{\prime}A$ and $AA^{\prime}.$ In consequence, $\eta_{jF}=\tau_{jF}^{2}$
for $j=1,...,p.$ In addition, $\eta_{jF}=0$ for $j=p+1,...,k.$ }

{ Define the elements of $\lambda$ to be\footnote{For simplicity, as above,
when writing $\lambda=(\lambda_{1,F},...,\lambda_{8,F})$ (and likewise in
similar expressions) we allow the elements to be scalars, vectors, matrices,
and distributions. Note that $\lambda_{5,F}$ is included so that Proposition
16.5 in AG can be applied.}%
\begin{align}
\lambda_{1,F}  &  :=(\tau_{1F},...,\tau_{pF})^{\prime}\in\Re^{p},\nonumber\\
\lambda_{2,F}  &  :=B_{F}\in\Re^{p\times p},\nonumber\\
\lambda_{3,F}  &  :=C_{F}\in\Re^{k\times k},\nonumber\\
\lambda_{4,F}  &  :=E_{F}\overline{Z}_{i}\overline{Z}_{i}^{\prime}\in
\Re^{k\times k},\nonumber\\
\lambda_{5,F}  &  :=(\lambda_{5,1F},...,\lambda_{5,p-1F})^{\prime}:=\left(
\frac{\tau_{2F}}{\tau_{1F}},...,\frac{\tau_{pF}}{\tau_{p-1F}}\right)
^{\prime}\in\lbrack0,1]^{p-1},\text{ where }0/0:=0,\nonumber\\
\lambda_{6,F}  &  :=Q_{F}\in\Re^{k\times k},\nonumber\\
\lambda_{7,F}  &  :=U_{F}\in\Re^{p\times p},\nonumber\\
\lambda_{8,F}  &  :=F,\text{ and }\nonumber\\
\lambda &  :=\lambda_{F}:=(\lambda_{1,F},...,\lambda_{8,F}).
\label{Defn of lambda1}%
\end{align}
Note that by (\ref{def W and U}) we have $G_{F}=U_{F}^{-2}=\lambda_{7,F}^{-2}$
and $H_{F}=(E_{F}\overline{Z}_{i}\overline{Z}_{i}^{\prime})^{1/2}Q_{F}%
^{-1}Q_{F}^{\prime-1}(E_{F}\overline{Z}_{i}\overline{Z}_{i}^{\prime})^{1/2}$
$=\lambda_{4,F}^{1/2}\lambda_{6,F}^{-1}\lambda_{6,F}^{^{\prime}-1}%
\lambda_{4,F}^{1/2}.$ In Section \ref{Het} the additional element
$\lambda_{9,F}$ defined in (\ref{lambda9}) is appended to $\lambda$ with
corresponding changes to several objects below, e.g. $\Lambda_{n}$ and
$h_{n}(\lambda)$ in (\ref{Defn of h_n(lambda)}) and $\lambda_{w_{n},h}$ in
(\ref{Defn of lambda1}) and (\ref{Defn lambda n,h}); e.g. $h_{n}(\lambda)$
becomes $(n^{1/2}\lambda_{1,F},\lambda_{2,F},\lambda_{3,F},...,\lambda
_{7,F},\lambda_{9,F}).$ }

{ The parameter space $\Lambda_{n}$ for $\lambda$ and the function
$h_{n}(\lambda)$ (that appears in Assumption B in ACG) are defined by%
\begin{align}
\Lambda_{n}  &  :=\{\lambda:\lambda=(\lambda_{1,F},...,\lambda_{8,F})\text{
for some }F\text{ st }(\gamma,\Pi_{W},\Pi_{Y},F)\in\mathcal{F}_{AKP,a_{n}%
}\text{ for some }(\gamma,\Pi_{W},\Pi_{Y})\},\nonumber\\
h_{n}(\lambda)  &  :=(n^{1/2}\lambda_{1,F},\lambda_{2,F},\lambda
_{3,F},...,\lambda_{7,F}). \label{Defn of h_n(lambda)}%
\end{align}
}

{ We define $\lambda$ and $h_{n}(\lambda)$ as in (\ref{Defn of lambda1}) and
(\ref{Defn of h_n(lambda)}) because, as shown below, the asymptotic
distributions of the test statistic and conditional critical values under a
sequence $\{F_{n}:n\geq1\}$ for which $h_{n}(\lambda_{F_{n}})\rightarrow h$
depend on $\lim n^{1/2}\lambda_{1,F_{n}}$ and $\lim\lambda_{m,F_{n}}$ for
$m=2,...,7.$ Note that we can view $h\in\left(  \Re\cup\{\pm\infty\}\right)
^{J}$ (for an appropriately chosen finite $J\in N$). }

{ For notational convenience, for any subsequence $\{w_{n}:n\geq1\},$%
\begin{equation}
\{\lambda_{w_{n},h}:n\geq1\}\text{ denotes a sequence }\{\lambda_{w_{n}}%
\in\Lambda_{n}:n\geq1\}\text{ for which }h_{w_{n}}(\lambda_{w_{n}})\rightarrow
h. \label{Defn lambda n,h}%
\end{equation}
It follows that the set $H$ defined in (\ref{defn of H}) is given as the set
of all $h\in(\Re\cup\{\pm\infty\})^{J}$ such that there exists $\{\lambda
_{w_{n},h}:n\geq1\}$ for some subsequence $\{w_{n}:n\geq1\}.$ }

{ We decompose $h$ analogously to the decomposition of the first seven
components of $\lambda$: $h=(h_{1},...,h_{7}),$ where $\lambda_{m,F}$ and
$h_{m}$ have the same dimensions for $m=1,...,7.$ We further decompose the
vector $h_{1}$ as $h_{1}=(h_{1,1},...,h_{1,p})^{\prime},$ where the elements
of $h_{1}$ could equal $\infty.$ Again, by definition, under a sequence
$\{\lambda_{n,h}:n\geq1\},$ we have\
\begin{equation}
n^{1/2}\tau_{jF_{n}}\rightarrow h_{1,j}\geq0\text{ }\forall j=1,...,p,\text{
}\lambda_{m,F_{n}}\rightarrow h_{m}\text{ }\forall m=2,...,7.
\label{h comp defns}%
\end{equation}
Note that $h_{1,p}=\tau_{pF_{n}}=0$ because $\rho(\Pi_{W}\gamma,\Pi_{W})<p,$
where $\rho(A)$ denotes the rank of a matrix $A.$ }

{ By Lyapunov-type WLLNs and CLTs, using the moment restrictions imposed in
(\ref{def par spa}), we have under $\lambda_{n,h}$
\begin{align}
&  \left(
\begin{array}
[c]{c}%
n^{-1/2}\overline{Z}^{\prime}(\varepsilon+V_{W}\gamma_{n})\\
vec\left(  n^{-1/2}\overline{Z}^{\prime}V_{W}\right)
\end{array}
\right)  \overset{}{\underset{d}{\rightarrow}}\left(
\begin{array}
[c]{c}%
\xi_{1,h}\\
\xi_{2,h}%
\end{array}
\right)  \overset{}{\sim}N\left(  0^{kp},\left(  h_{7}^{-2}\otimes(h_{4}%
h_{6}^{-1}h_{6}^{\prime-1}h_{4})\right)  \right)  ,\nonumber\\
&  \lambda_{4,F_{n}}^{-1}(n^{-1}\overline{Z}^{\prime}\overline{Z}%
)\overset{}{\underset{p}{\rightarrow}}I_{k},\text{ }n^{-1}\overline{Z}%
^{\prime}[\varepsilon\overset{}{:}V_{W}]\overset{}{\underset{p}{\rightarrow}%
}0^{k\times p}, \label{CLT and WLLN}%
\end{align}
where the random vector $(\xi_{1,h},\xi_{2,h}^{\prime})^{\prime}$ is defined
here, $F_{n}$ denotes the distribution of $(\varepsilon_{i},\overline{Z}%
_{i}^{\prime},V_{Y,i}^{\prime}V_{W,i}^{\prime})$ under $\lambda_{n,h},$ and,
by definition above, $h_{7}^{-2}$ and $h_{4}h_{6}^{-1}h_{6}^{\prime-1}h_{4}$
denote the limits of $G_{F_{n}}$ and $\overline{H}_{F_{n}}$ under
$\lambda_{n,h}.$ }

{ Let $q=q_{h}\in\{0,...,p-1\}$ be such that
\begin{equation}
h_{1,j}=\infty\text{ for }1\leq j\leq q_{h}\text{ and }h_{1,j}<\infty\text{
for }q_{h}+1\leq j\leq p, \label{q defn}%
\end{equation}
where $h_{1,j}:=\lim n^{1/2}\tau_{jF_{n}}\geq0$ for $j=1,...,p$ by
(\ref{h comp defns}) and the distributions $\{F_{n}:n\geq1\}$ correspond to
$\{\lambda_{n,h}:n\geq1\}$ defined in (\ref{Defn lambda n,h}). This value $q$
exists because $\{h_{1,j}:j\leq p\}$ are nonincreasing in $j$ (since
$\{\tau_{jF}:j\leq p\}$ are nonincreasing in $j,$ as defined in
(\ref{sing val defn})). Note that $q$ is the number of singular values of
$Q_{F_{n}}(\Pi_{Wn}\gamma_{n},\Pi_{Wn})U_{F_{n}}\in\Re^{k\times p}$ that
diverge to infinity when multiplied by $n^{1/2}.$ Note again that $q<p$
because $\rho(\Pi_{Wn}\gamma_{n},\Pi_{Wn})<p.$ }

\subsubsection{{Asymptotic Distributions\label{Asy distn Subsec}}}

{ One might wonder whether the definition of $\widehat{G}_{n}$ in
$($\ref{def of KP factors}$)$ as $vec(\widehat{G}_{n})=\widehat{L}%
(:,1)/\widehat{L}(1,1)$ where $(\widehat{G}_{n},\widehat{H}_{n})$ are
minimizers in (\ref{minprob}) is unique. If for instance the eigenspace
corresponding to the largest eigenvalue was of dimension bigger than one, then
clearly $\widehat{L}(:,1)$ would not be uniquely defined. The following lemma
shows that the definition of $\widehat{G}_{n}$ is unique and derives its
limit. }

{ To simplify notation a bit, we write shorthand $R_{n}$ for $R_{F_{n}}$ and
likewise for other expressions$.$ }

\begin{lemma}
{ \label{CONSISTENCY GHATS}Under sequences $\lambda_{n,h}$ from $\Lambda_{n}$
in $($\ref{Defn of h_n(lambda)}$)$ based on the parameter space $\mathcal{F}%
_{AKP,a_{n}}$, wp1 the definition of $\widehat{G}_{n}\in\Re^{p\times p}$ and
$\widehat{H}_{n}\in\Re^{k\times k}$ in $($\ref{def of KP factors}$)$ is unique
and%
\[
\widehat{G}_{n}\rightarrow\lim_{n\rightarrow\infty}G_{n}\text{ and
}\widehat{H}_{n}\rightarrow\lim_{n\rightarrow\infty}H_{n}\text{ a.s.,}%
\]
where $H_{n}=(E_{F_{n}}\overline{Z}_{i}\overline{Z}_{i}^{\prime}%
)^{-1/2}\overline{H}_{n}(E_{F_{n}}\overline{Z}_{i}\overline{Z}_{i}^{\prime
})^{-1/2}$ is defined in $($\ref{defH}$).$ }
\end{lemma}

{ \textbf{Comment.} Note that under sequences $\lambda_{n,h},$ $\lim
_{n\rightarrow\infty}G_{n}$ and $\lim_{n\rightarrow\infty}H_{n}$ do exist. On
the other hand, the matrices $G_{n}$ and $H_{n}$ may not be uniquely pinned
down by the restrictions in (\ref{def par spa}) in $\mathcal{F}_{AKP,a_{n}}$.
The results $\widehat{G}_{n}\rightarrow\lim_{n\rightarrow\infty}G_{n}$ and
$\widehat{H}_{n}\rightarrow\lim_{n\rightarrow\infty}H_{n}$ a.s. hold for any
possible choice of $G_{n}$ and $H_{n}.\medskip$ }

{ \textbf{\noindent Proof of Lemma \ref{CONSISTENCY GHATS}.} Recall the
definition
\begin{equation}
R_{n}=(I_{p}\otimes(E_{F_{n}}\overline{Z}_{i}\overline{Z}_{i}^{\prime}%
)^{-1/2})E_{F_{n}}(vec(\overline{Z}_{i}U_{i}^{\prime})(vec(\overline{Z}%
_{i}U_{i}^{\prime}))^{\prime})(I_{p}\otimes(E_{F_{n}}\overline{Z}_{i}%
\overline{Z}_{i}^{\prime})^{-1/2}) \label{redef Rn}%
\end{equation}
in (\ref{def RF}). By Theorem 1 in \citet{VanLoanPitsianis1993},
\begin{equation}
||A-B\otimes C||=||\mathcal{R}(A)-vec(B)vec(C)^{\prime}||
\label{Theorem 1 Frob norm}%
\end{equation}
for any conformable matrices $A,B,$ and $C.$ Thus, for
\begin{equation}
\overline{\Upsilon}_{n}:=(I_{p}\otimes(E_{F_{n}}\overline{Z}_{i}\overline
{Z}_{i}^{\prime})^{-1/2})\Upsilon_{n}(I_{p}\otimes(E_{F_{n}}\overline{Z}%
_{i}\overline{Z}_{i}^{\prime})^{-1/2}), \label{bar expression}%
\end{equation}
it follows that $\mathcal{R}(R_{n}-\overline{\Upsilon}_{n})=vec(G_{n}%
)vec(H_{n})^{\prime}$ and because $\kappa_{\min}(E_{F_{n}}\overline{Z}%
_{i}\overline{Z}_{i}^{\prime})^{-1/2}),$ $\kappa_{\min}(G_{n}),$ and
$\kappa_{\min}(\overline{H}_{n})\geq\delta_{2}$ in $\mathcal{F}_{AKP,a_{n}},$
it follows that $\mathcal{R}(R_{n}-\overline{\Upsilon}_{n})$ has rank 1. It
follows also that $\lim_{n\rightarrow\infty}\mathcal{R}(R_{n}-\overline
{\Upsilon}_{n})=\lim_{n\rightarrow\infty}\mathcal{R}(R_{n})$ (which exists
under sequences $\lambda_{n,h}$) has rank 1 (even though the rank of
$\mathcal{R}(R_{n})$ could be larger than 1 for every $n$). By continuity of
the singular values and because the geometric and algebraic multiplicity
coincide for diagonalizable matrices, the dimension of the eigenspace of
$\mathcal{R}(R_{n})\mathcal{R}(R_{n})^{\prime}$ corresponding to the largest
singular value of $\mathcal{R}(R_{n})$ is one for all $n$ large enough. }

{ By the uniform moment restrictions in (\ref{def par spa}) in $\mathcal{F}%
_{AKP,a_{n}},$ namely $E_{F}(||T_{i}||^{2+\delta_{1}})\leq B<\infty,$ for
$T_{i}\in\{vec(\overline{Z}_{i}U_{i}^{\prime}),vec(\overline{Z}_{i}%
\overline{Z}_{i}^{\prime})\}$ and $\kappa_{\min}(E_{F}(\overline{Z}%
_{i}\overline{Z}_{i}^{\prime}))\geq\delta_{2}>0,$ a strong law of large
numbers implies that%
\begin{equation}
\widehat{R}_{n}-R_{n}\rightarrow0^{kp\times kp}\text{ and }\mathcal{R}%
(\widehat{R}_{n})-\mathcal{R}(R_{n})\rightarrow0^{pp\times kk}\text{ a.s.}
\label{CONS RHAT}%
\end{equation}
Therefore, the dimension of the eigenspace of $\mathcal{R}(\widehat{R}%
_{n})\mathcal{R}(\widehat{R}_{n})^{\prime}$ corresponding to the largest
singular value of $\mathcal{R}(\widehat{R}_{n})$ is one for all $n$ large
enough wp1$.$ }

{ By the uniqueness statement of Lemma \ref{frobenius norm min} for the rank 1
case, it follows that the formula for minimizers of the KP approximation
problem in (\ref{minprob}) given in
\citet[Corollary 2 and Theorem 11]{VanLoanPitsianis1993}, namely
\begin{equation}
vec(\widehat{G}_{n})=\widehat{\sigma}_{1}\widehat{L}(:,1)\text{ and
}vec(\widehat{H}_{n})=\widehat{N}(:,1), \label{FORM IN VAN LOAN AND PITSIANIS}%
\end{equation}
yields symmetric pd matrices $\widehat{G}_{n}$ and $\widehat{H}_{n}.$ When
applying Theorem 11, note that $\widehat{R}_{n}>0$ for all large enough $n$
wp1, which holds by (\ref{CONS RHAT}), $\lim_{n\rightarrow\infty}G_{n}\otimes
H_{n}=\lim_{n\rightarrow\infty}R_{n}-\overline{\Upsilon}_{n}=\lim
_{n\rightarrow\infty}R_{n},$ and because $\kappa_{\min}(E_{F_{n}}\overline
{Z}_{i}\overline{Z}_{i}^{\prime})^{-1/2}),$ $\kappa_{\min}(G_{n}),$ and
$\kappa_{\min}(\overline{H}_{n})\geq\delta_{2}$ in $\mathcal{F}_{AKP,a_{n}}$.
Given that $\widehat{G}_{n}>0,$ Sylvester's criterion for positive
definiteness implies that $\widehat{L}(1,1)>0$ for all large enough $n$ wp1$,$
and we can therefore define $\widehat{G}_{n}$ and $\widehat{H}_{n}$ as in
$($\ref{def of KP factors}$)$ with normalization to 1 of the upper left
element of $\widehat{G}_{n}$ for all large enough $n$ wp1. }

{ Next we apply Lemma \ref{continuity eigenvector} with $a=1$ and the roles of
$R_{n}$ and $R$ in Lemma \ref{continuity eigenvector} played by $\mathcal{R}%
(\widehat{R}_{n})$ and $\lim_{n\rightarrow\infty}\mathcal{R}(R_{n})$,
respectively$.$ By (\ref{CONS RHAT}), the lemma implies
\begin{equation}
\widehat{L}(:,j)^{\prime}L_{1}=o(1) \label{key result}%
\end{equation}
wp1. for $j>1$, where $\widehat{L}(:,j)$ denotes the $j$-th column of
$\widehat{L}$ in the singular value decomposition $\widehat{L}^{\prime
}\mathcal{R}(\widehat{R}_{n})\widehat{N}=diag(\widehat{\sigma}_{l})$ of
$\mathcal{R}(\widehat{R}_{n})$ and $L_{1}$ denotes the first column of
$\overline{L}$ in the singular value decomposition $\overline{L}^{\prime
}\mathcal{R}(\lim_{n\rightarrow\infty}\mathcal{R}(R_{n}))\overline
{N}=diag(\overline{\sigma}_{l})$ of $\lim_{n\rightarrow\infty}\mathcal{R}%
(R_{n}).$ For any orthogonal basis $(x_{1},...,x_{p^{2}})$ of $\Re^{p^{2}}$
and $y\in\Re^{p^{2}}$ we have $y=%
{\textstyle\sum\nolimits_{j=1}^{p^{2}}}
(y^{\prime}x_{j})x_{j}.$ In particular, we have $L_{1}=%
{\textstyle\sum\nolimits_{j=1}^{p^{2}}}
(L_{1}^{\prime}\widehat{L}(:,j))\widehat{L}(:,j)=(L_{1}^{\prime}%
\widehat{L}(:,1))\widehat{L}(:,1)+o(1)$ wp1.$,$ where the second equality
holds by (\ref{key result}). Together with the normalization of the upper left
elements of $\widehat{G}_{n}$ and $G_{n}$ to 1, this implies $\widehat{G}%
_{n}-G_{n}\rightarrow0^{p\times p}$ a.s. and $\widehat{H}_{n}-H_{n}%
\rightarrow0^{k\times k}$ a.s. follows analogously. $\square$\bigskip}

{ An analogue to Lemma 16.4 in AG and\ Lemma 1 in GKM19 is given by the
following statement. Define
\begin{equation}
\widehat{D}_{n}:=(\overline{Z}^{\prime}\overline{Z})^{-1}\overline{Z}^{\prime
}\left(  \overline{Y}_{0},W\right)  \text{ and }\widehat{Q}_{n}:=\widehat{H}%
_{n}^{-1/2}(n^{-1}\overline{Z}^{\prime}\overline{Z})^{1/2}%
\text{.\footnote{Note that the quantity defined here differs from
$\widehat{D}_{n}\left(  \theta\right)  $ introduced in (\ref{harb def}).}}
\label{def Dhat}%
\end{equation}
Denote by $vec_{k,m_{W}}^{-1}(\cdot)$ the inverse $vec$ operation that
transforms a $km_{W}$ vector into a $k\times m_{W}$ matrix. }

\begin{lemma}
{ \label{Asy distn Dhat Lem}Under sequences $\{\lambda_{n,h}:n\geq1\}$ with
$\lambda_{n,h}\in\Lambda_{n}$ in $($\ref{Defn of h_n(lambda)}$)$ based on the
parameter space $\mathcal{F}_{AKP,a_{n}},$ $n^{1/2}(\widehat{D}_{n}-(\Pi
_{Wn}\gamma_{n},\Pi_{Wn}))\rightarrow_{d}\overline{D}_{h},$ where%
\[
\overline{D}_{h}\sim h_{4}^{-1}(\xi_{1,h},vec_{k,m_{W}}^{-1}(\xi_{2,h})),
\]
$\xi_{1,h}$ and $\xi_{2,h}$ are defined in $($\ref{CLT and WLLN}$)$, and again
$h_{4}$ is the limit of $\lambda_{4,n}=E_{F_{n}}\overline{Z}_{i}\overline
{Z}_{i}^{\prime}.$ Furthermore, we have $\widehat{Q}_{n}-Q_{n}\rightarrow
_{p}0^{k\times k}$.$\medskip$ }
\end{lemma}

{ \textbf{\noindent Proof of Lemma \ref{Asy distn Dhat Lem}.} We have%
\begin{align}
&  n^{1/2}(\widehat{D}_{n}-(\Pi_{Wn}\gamma_{n},\Pi_{Wn}))\nonumber\\
&  =\hspace{-0.03in}n^{1/2}((\overline{Z}^{\prime}\overline{Z})^{-1}%
\overline{Z}^{\prime}(y-Y\beta_{0},W)-(\Pi_{Wn}\gamma_{n},\Pi_{Wn}%
))\nonumber\\
&  =\hspace{-0.03in}n^{1/2}((\overline{Z}^{\prime}\overline{Z})^{-1}%
\overline{Z}^{\prime}(\overline{Z}\Pi_{Wn}\gamma_{n}+V_{W}\gamma
_{n}+\varepsilon,\overline{Z}\Pi_{Wn}+V_{W})-(\Pi_{Wn}\gamma_{n},\Pi
_{Wn}))\nonumber\\
&  =\hspace{-0.03in}(n^{-1}\overline{Z}^{\prime}\overline{Z})^{-1}%
[n^{-1/2}\overline{Z}^{\prime}(V_{W}\gamma_{n}+\varepsilon,V_{W}%
)]\rightarrow_{d}\overline{D}_{h}, \label{conv in distribution}%
\end{align}
where the first equality uses the definition of $\widehat{D}_{n}$ in
(\ref{def Dhat}), the second equality uses the formulas in (\ref{struc}), and
the convergence results holds by the (triangular array) CLT and WLLN in
(\ref{CLT and WLLN}). The remaining statement holds by the WLLN in
(\ref{CLT and WLLN}) and the consistency of $\widehat{H}_{n}$ for $H_{n}$
proven above$.$ $\square\medskip$ }

{ For notational convenience, write
\begin{equation}
\widehat{U}_{n}:=\widehat{G}_{n}^{-1/2}. \label{uhat}%
\end{equation}
Note that the matrix $n\widehat{U}_{n}\widehat{D}_{n}^{\prime}\widehat{Q}%
_{n}^{\prime}\widehat{Q}_{n}\widehat{D}_{n}\widehat{U}_{n}$ equals
$n^{-1}\widehat{G}_{n}^{-1/2}\left(  \overline{Y}_{0},W\right)  ^{\prime
}Z\widehat{H}_{n}^{-1}Z^{\prime}\left(  \overline{Y}_{0},W\right)
\widehat{G}_{n}^{-1/2}$ which appears in (\ref{eq: eigenprob ARhat}). Thus,
$\widehat{\kappa}_{in}$ for $i=1,...,p$ equals the $i$th eigenvalue of
$n\widehat{U}_{n}^{\prime}\widehat{D}_{n}^{\prime}\widehat{Q}_{n}^{\prime
}\widehat{Q}_{n}\widehat{D}_{n}\widehat{U}_{n},$ ordered nonincreasingly, and
$\widehat{\kappa}_{pn}$ is the subvector AR$_{AKP}$ test statistic. To
describe the limiting distribution of $(\widehat{\kappa}_{1n}%
,...,\widehat{\kappa}_{pn})$ we need additional notation, namely:%
\begin{align}
h_{2}  &  =(h_{2,q},h_{2,p-q}),\text{ }h_{3}=(h_{3,q},h_{3,k-q}),\nonumber\\
h_{1,p-q}^{\diamond}\overset{}{:}  &  =\left[  \hspace{-0.04in}%
\begin{array}
[c]{c}%
0^{q\times(p-q)}\\
Diag\{h_{1,q+1},...,h_{1,p-1},0\}\\
0^{(k-p)\times(p-q)}%
\end{array}
\hspace{-0.04in}\right]  \hspace{-0.04in}\in\Re^{k\times(p-q)},\nonumber\\
\overline{\Delta}_{h}\overset{}{:}  &  =(\overline{\Delta}_{h,q}%
,\overline{\Delta}_{h,p-q})\in\Re^{k\times p},\text{ }\overline{\Delta}%
_{h,q}:=h_{3,q}\in\Re^{k\times q},\text{ }\nonumber\\
\overline{\Delta}_{h,p-q}  &  :=h_{3}h_{1,p-q}^{\diamond}+h_{6}\overline
{D}_{h}h_{7}h_{2,p-q}\in\Re^{k\times(p-q)}, \label{h defns}%
\end{align}
where $h_{2,q}\in\Re^{p\times q},$ $h_{2,p-q}\in\Re^{p\times(p-q)},$
$h_{3,q}\in\Re^{k\times q},$ $h_{3,k-q}\in\Re^{k\times(k-q)},$ $\overline
{\Delta}_{h,q}\in\Re^{k\times q},$ and $\overline{\Delta}_{h,p-q}\in
\Re^{k\times(p-q)}$.\footnote{There is some abuse of notation here. For
example, $h_{2,q}$ and $h_{2,p-q}$ denote different matrices even if $p-q$
equals $q.$} Let $T_{n}:=B_{F_{n}}S_{n}$ and $S_{n}:=Diag\{(n^{1/2}%
\tau_{1F_{n}})^{-1},...,(n^{1/2}\tau_{qF_{n}})^{-1},1,...,1\}\in\Re^{p\times
p}.$ The same proof as the one of Lemma 16.4 in AG shows that $n^{1/2}%
Q_{F_{n}}\widehat{D}_{n}U_{F_{n}}T_{n}\rightarrow_{d}\overline{\Delta}_{h}$
under all sequences $\{\lambda_{n,h}:n\geq1\}$ with $\lambda_{n,h}\in\Lambda$.
The following proposition is an analogue to Proposition 16.5 in AG and to
Proposition 2 in GKM19. }

\begin{proposition}
{ \label{Min Eigenval Prop 1}Under all sequences $\{\lambda_{n,h}:n\geq1\}$
with $\lambda_{n,h}\in\Lambda_{n},$ }

{ $($a$)$ $\widehat{\kappa}_{jn}\rightarrow_{p}\infty$ for all $j\leq q,$ }

{ $($b$)$ the $($ordered$)$ vector of the smallest $p-q$ eigenvalues of
$n\widehat{U}_{n}^{\prime}\widehat{D}_{n}^{\prime}\widehat{Q}_{n}%
\widehat{Q}_{n}\widehat{D}_{n}\widehat{U}_{n},$ i.e., $(\widehat{\kappa
}_{(q+1)n},...,\allowbreak\widehat{\kappa}_{pn})^{\prime},$ converges in
distribution to the $($ordered$)$ $p-q\ $vector of the eigenvalues of
$\overline{\Delta}_{h,p-q}^{\prime}h_{3,k-q}h_{3,k-q}^{\prime}\overline
{\Delta}_{h,p-q}\in\Re^{(p-q)\times(p-q)},$ }

{ $($c$)$ the convergence in parts (a) and\ (b) holds jointly with the
convergence in Lemma \ref{Asy distn Dhat Lem}, and }

{ $($d$)$ under all subsequences $\{w_{n}\}$ and all sequences $\{\lambda
_{w_{n},h}:n\geq1\}$ with $\lambda_{w_{n},h}\in\Lambda_{n},$ the results in
parts $($a$)$-$($c$)$ hold with $n$ replaced with $w_{n}.$ }
\end{proposition}

{ \textbf{Comments.} 1. The proof of the proposition follows from the proof of
Proposition 16.5 in AG. Note that Assumption WU in AG (assumed in their
Proposition 16.5) is fulfilled with the roles of $W_{2F},$ $W_{F},$ $U_{2F},$
and $U_{F}$ in AG played here by $Q_{F},$ $Q_{F},$ $U_{F},$ and $U_{F},$
respectively, while the roles of $W_{1}$ and $U_{1}$ in AG are played by the
identity function. The roles of $\widehat{W}_{2n}$ and $\widehat{W}_{n}$ in AG
are both played by $\widehat{Q}_{n}$ and those of both $\widehat{U}_{2n}$ and
$\widehat{U}_{n}$ by $\widehat{U}_{n}$. Lemma \ref{Asy distn Dhat Lem} then
shows consistency $\widehat{W}_{2n}-W_{2F_{n}}\rightarrow_{p}0^{k\times k}$
and $\widehat{U}_{2n}-U_{2F_{n}}\rightarrow_{p}0^{p\times p}$ under sequences
$\{\lambda_{n,h}:n\geq1\}$ with $\lambda_{n,h}\in\Lambda_{n}$ and trivially
the functions $W_{1}$ and $U_{1}$ are continuous in our case. Note that by the
restrictions in $\mathcal{F}_{AKP,a_{n}}$ in (\ref{def par spa}) the
requirements in the parameter space $F_{WU}$ in AG, namely \textquotedblleft%
$\kappa_{\min}(Q_{F})$ and $\kappa_{\min}(U_{F})$ are uniformly bounded away
from zero and $||Q_{F}||$ and $||U_{F}||$ are uniformly bounded away from
infinity\textquotedblright, are fulfilled. For example, the former follows
because $\kappa_{\min}(Q_{F})=1/\kappa_{\max}(Q_{F}^{-1})=1/\kappa_{\max
}((E_{F}\overline{Z}_{i}\overline{Z}_{i}^{\prime})^{-1/2}H_{F}^{1/2})$ and
$\kappa_{\max}((E_{F}\overline{Z}_{i}\overline{Z}_{i}^{\prime})^{-1/2}%
H_{F}^{1/2})$ is uniformly bounded. }

{ 2. Proposition \ref{Min Eigenval Prop 1} yields the desired joint limiting
distribution of the $p$ eigenvalues in (\ref{eq: eigenprob ARhat}). Using
repeatedly the general formula $(C^{\prime}\otimes A)vec(B)=vec(ABC)$ for
three conformable matrices $A,B,C,$ we have for the expression $h_{6}%
\overline{D}_{h}h_{7}$ that appears in $\overline{\Delta}_{h,p-q}$
\begin{align}
&  vec(h_{6}\overline{D}_{h}h_{7})=vec(h_{6}h_{4}^{-1}(\xi_{1,h},vec_{k,m_{W}%
}^{-1}(\xi_{2,h}))h_{7})=(h_{7}\otimes(h_{4}h_{6}^{-1})^{-1})\left(
\begin{array}
[c]{c}%
\xi_{1,h}\\
\xi_{2,h}%
\end{array}
\right) \nonumber\\
&  \sim\hspace{-0.03in}vec(v_{1},...,v_{p}), \label{calc step}%
\end{align}
where, by definition, $v_{j},$ $j=1,...,p$ are i.i.d.~normal $k$-vectors with
zero mean and covariance matrix $I_{k},$ and the distributional statement
follows by straightforward calculations using (\ref{CLT and WLLN}). Therefore,
by\ Lemma \ref{Asy distn Dhat Lem}, the definition of $\overline{\Delta
}_{h,p-q}$ in (\ref{h defns}), and by noting that
\begin{equation}
h_{3,k-q}^{\prime}h_{3}h_{1,p-q}^{\diamond}=\left(
\begin{array}
[c]{c}%
Diag\{h_{1,q+1},...,h_{1,p-1},0\}\\
0^{(k-p)\times(p-q)}%
\end{array}
\right)  \label{general expression mean}%
\end{equation}
we obtain%
\begin{align}
h_{3,k-q}^{\prime}\overline{\Delta}_{h,p-q}  &  =\hspace{-0.03in}\left(
\begin{array}
[c]{c}%
Diag\{h_{1,q+1},...,h_{1,p-1},0\}\\
0^{(k-p)\times(p-q)}%
\end{array}
\right)  +h_{3,k-q}^{\prime}(v_{1},...,v_{p})h_{2,p-q}\nonumber\\
&  \sim\hspace{-0.03in}\left(
\begin{array}
[c]{c}%
Diag\{h_{1,q+1},...,h_{1,p-1},0\}\\
0^{(k-p)\times(p-q)}%
\end{array}
\right)  +(w_{1},...,w_{p-q}), \label{general case}%
\end{align}
where, by definition, $w_{j},$ $j=1,...,p-q$ are i.i.d.~normal $(k-q)$-vectors
with zero mean and covariance matrix $I_{k-q}.$ The distributional equivalence
in the second line holds because $(v_{1},...,v_{p})h_{2,p-q}\sim
(\widetilde{v}_{1},...,\widetilde{v}_{p-q}),$ where $\widetilde{v}_{j},$
$j=1,...,p-q$ are i.i.d.~$N(0^{k},I_{k})$ as $h_{2,p-q}$ has orthogonal
columns of length 1. Analogously, $h_{3,k-q}^{\prime}(\widetilde{v}%
_{1},...,\widetilde{v}_{p-q})\sim(w_{1},...,w_{p-q})$ because $h_{3,k-q}$ has
orthogonal columns of length 1$.\medskip$ }

{ For example, when $q=p-1=m_{W}$ (which could be called the "strong IV"
case), we obtain from (\ref{general case}) $h_{3,k-q}^{\prime}\overline
{\Delta}_{h,p-q}=w_{1}\in\Re^{k-m_{W}}.$ Therefore $\overline{\Delta}%
_{h,p-q}^{\prime}h_{3,k-q}h_{3,k-q}^{\prime}\overline{\Delta}_{h,p-q}\sim
\chi_{k-m_{W}}^{2}$ and thus by part (b) of Proposition
\ref{Min Eigenval Prop 1} the limiting distribution of the subvector
AR$_{AKP}$ test statistic is $\chi_{k-m_{W}}^{2}$ in that case, while all the
larger roots in (\ref{eq: eigenprob ARhat}) converge in probability to
infinity by part (a).\medskip}

{ \textbf{\noindent Proof of Theorem \ref{correct asy size}.} Given the
discussion in Comment 2 to Proposition \ref{Min Eigenval Prop 1}, the same
proof as for Theorem 5 in GKM19 applies. $\square$ }

\subsection{{Proof of Theorem \ref{correct asy size generic}%
\label{proof of generic result}}}

{ \textbf{\noindent Proof of Theorem \ref{correct asy size generic}.} It is
enough to verify Proposition \ref{Gen Asy Size Prop} above for the parameter
space $\mathcal{F}_{Het}$ and the test $\varphi_{MS-AKP,\alpha}.$ To verify
Assumption B in ACG consider a sequence $\lambda_{w_{n},h}$ defined as in
(\ref{Defn of lambda1}) and (\ref{Defn lambda n,h}) above except that the
component
\begin{equation}
\lambda_{9w_{n}}:=\min||R_{F_{w_{n}}}^{-1/2}(G\otimes H-R_{F_{w_{n}}%
})R_{F_{w_{n}}}^{-1/2}||/c_{w_{n}} \label{new component}%
\end{equation}
is added to $\lambda_{w_{n}}$, where the minimum (here and in similar
expressions below) is taken over $(G,H)$ for $G\in\Re^{p\times p},$ $H\in
\Re^{k\times k}$ being pd, symmetric matrices, normalized such that the upper
left element of $G$ equals 1. In (\ref{Defn of h_n(lambda)}), we replace
$\mathcal{F}_{AKP,a_{w_{n}}}$ by $\mathcal{F}_{Het}$ and define $h_{w_{n}%
}(\lambda_{F}):=(w_{n}^{1/2}\lambda_{1,F},\lambda_{2,F},\lambda_{3,F}%
,\ldots, \allowbreak\lambda_{7,F},w_{n}^{1/2}\lambda_{9,F}).$ To simplify notation, we write
$n$ instead of $w_{n}$ from now on. }

{ Consider first a sequence $\lambda_{n,h}$ with $h_{9}=\infty.$ By Assumption
MS, $\varphi_{MS,c_{n}}=1$ wpa1 and therefore, $\varphi_{MS-AKP,\alpha
}=\varphi_{Rob,\alpha-\delta}$ wpa1. Thus, the new test $\varphi
_{MS-AKP,\alpha}$ has limiting NRP bounded by $\alpha-\delta$ in that case
because $\varphi_{Rob,\alpha-\delta}$ has asymptotic size bounded by its
nominal size by Assumption RT . }

Second, consider a sequence $\lambda_{n,h}$ with $h_{9}\in\lbrack0,\infty).$
In that case, $n^{1/2}/c_{n}\rightarrow\infty$ implies that $\min||R_{F_{n}%
}^{-1/2}(G\otimes H-R_{F_{n}})R_{F_{n}}^{-1/2}||\rightarrow0.$ By
submultiplicativity of the Frobenius norm and $||R_{F_{n}}^{1/2}||$ being
uniformly bounded in $\mathcal{F}_{Het}$ it then follows that $\min||G\otimes
H-R_{F_{n}}||\rightarrow0.$ That is, the covariance matrix $R_{F_{n}}$ has AKP
structure. Therefore, also the covariance matrix $\overline{R}_{F_{n}}$ has
AKP structure. By the proof of Theorem \ref{correct asy size} the test
$\varphi_{AKP,\alpha}$ then has limiting NRP bounded by $\alpha$ under
sequences $\lambda_{n,h}$ with $h_{9}\in\lbrack0,\infty)$. It therefore
follows that
\begin{align}
&  \lim\sup_{n\rightarrow\infty}P_{\lambda_{n,h}}(\varphi_{MS-AKP,\alpha
}\overset{}{=}1)\nonumber\\
&  \overset{}{\leq}\lim\sup_{n\rightarrow\infty}P_{\lambda_{n,h}}%
(\max\{\varphi_{Rob,\alpha-\delta},\varphi_{AKP,\alpha}\}=1)\nonumber\\
&  \overset{}{=}\lim\sup_{n\rightarrow\infty}P_{\lambda_{n,h}}(\varphi
_{AKP,\alpha}=1)\overset{}{\leq}\alpha, \label{tech step}%
\end{align}
where the equality uses Assumption RP, $P_{\lambda_{n,h}}(\varphi
_{Rob,\alpha-\delta}\leq\varphi_{AKP,\alpha})\rightarrow1,$ which implies that
$P_{\lambda_{n,h}}((\max\{\varphi_{Rob,\alpha-\delta},\varphi_{AKP,\alpha
}\}=1)\cap(\varphi_{Rob,\alpha-\delta}>\varphi_{AKP,\alpha}))\rightarrow0$ and
the last inequality follows from the fact that the limiting NRP of the test
$\varphi_{AKP,\alpha}$ is bounded by $\alpha.$

{This establishes Proposition \ref{Gen Asy Size Prop} with $\sup_{h\in
H}RP^{+}(h)\leq\alpha$ and thus} Theorem \ref{correct asy size generic}%
{\textbf{.}}

To prove Comment 1 below Theorem \ref{correct asy size generic}, note that by
the assumed continuity, \begin{align}
\lim_{\delta\rightarrow0}\allowbreak\liminf
_{n\rightarrow\infty}\inf_{(\gamma,\Pi_{W},\Pi_{Y},F)\in\mathcal{F}_{Het}}E_{(\gamma,\Pi_{W},\Pi_{Y},F)}\varphi_{MS-AKP,\delta,c_{n},\alpha} \nonumber \\ =
\liminf_{n\rightarrow\infty}\inf_{(\gamma,\Pi_{W},\Pi_{Y},F)\in
\mathcal{F}_{Het}}E_{(\gamma,\Pi_{W},\Pi_{Y},F)}\varphi_{MS-AKP,0,c_{n}%
,\alpha}.    
\end{align}
But note that
\begin{align}
&  \lim\inf_{n\rightarrow\infty}\inf_{(\gamma,\Pi_{W},\Pi_{Y},F)\in
\mathcal{F}_{Het}}E_{(\gamma,\Pi_{W},\Pi_{Y},F)}\varphi_{MS-AKP,0,c_{n}%
,\alpha}\nonumber\\
&  =\hspace{-0.03in}\lim\inf_{n\rightarrow\infty}E_{(\gamma_{n},\Pi_{Wn}%
,\Pi_{Yn},F_{n})}\varphi_{MS-AKP,0,c_{n},\alpha}\nonumber\\
&  =\hspace{-0.03in}\lim_{n\rightarrow\infty}E_{(\gamma_{w_{n}},\Pi_{Ww_{n}%
},\Pi_{Yw_{n}},F_{w_{n}})}\varphi_{MS-AKP,0,c_{w_{n}},\alpha}\nonumber\\
&  =\hspace{-0.03in}\lim_{n\rightarrow\infty}E_{\lambda_{w_{n},h}}%
\varphi_{MS-AKP,0,c_{w_{n}},\alpha}, \label{eq sequ}%
\end{align}
where in the first equality $(\gamma_{n},\Pi_{Wn},\Pi_{Yn},F_{n}%
)\in\mathcal{F}_{Het}$ is chosen such that $\inf_{(\gamma,\Pi_{W},\Pi
_{Y},F)\in\mathcal{F}_{Het}}$ $E_{(\gamma,\Pi_{W},\Pi_{Y},F)}\varphi
_{MS-AKP,0,c_{n},\alpha}\geq E_{(\gamma_{n},\Pi_{Wn},\Pi_{Yn},F_{n})}%
\varphi_{MS-AKP,0,c_{n},\alpha}-n^{-1},$ in the second equality a subsequence
$\{w_{n}\}$ of $\{n\}$ can be found, and in the third equality $\{w_{n}\}$ may
denote a further subsequence along which $(\gamma_{w_{n}},\Pi_{Ww_{n}}%
,\Pi_{Yw_{n}},F_{w_{n}})$ is of type $\lambda_{w_{n},h}$ for some $h.$ (We are
allowing here for the possibility that $E_{\lambda_{w_{n},h}}\varphi
_{MS-AKP,\delta,c_{w_{n}},\alpha}$ may depend on the particular sequence
$\lambda_{w_{n},h}$ rather than just $h.$) If $h_{9}=\infty$ then
$\varphi_{MS-AKP,0,c_{w_{n}},\alpha}=\varphi_{Rob,\alpha}$ wpa1 by Assumption
MS and
\begin{equation}
\lim_{n\rightarrow\infty}E_{\lambda_{w_{n},h}}\varphi_{Rob,\alpha}\geq\lim
\inf_{n\rightarrow\infty}\inf_{(\gamma,\Pi_{W},\Pi_{Y},F)\in\mathcal{F}_{Het}%
}E_{(\gamma,\Pi_{W},\Pi_{Y},F)}\varphi_{Rob,\alpha}. \label{case infi}%
\end{equation}
On the other hand, if $h_{9}<\infty$ then by Assumption RP, $\varphi
_{Rob,\alpha}\leq\varphi_{AKP,\alpha}$ wpa1 and
\begin{equation}
\lim_{n\rightarrow\infty}E_{\lambda_{w_{n},h}}\varphi_{MS-AKP,0,c_{w_{n}%
},\alpha}\geq\lim_{n\rightarrow\infty}E_{\lambda_{w_{n},h}}\varphi
_{Rob,\alpha} \label{case fini}%
\end{equation}
and the desired conclusion then follows as in (\ref{case infi}). {$\square$ }

\subsection{{Assumption MS for the model selection method $\varphi_{MS,c_{n}}%
$\label{model selection proof}}}

{Here we verify Assumption MS for the two suggested methods for $\varphi
_{MS,c_{n}}.$}

{Method 1, defined as $I(\widehat{K}_{n}>c_{n}):$ To simplify notation we
write again $n$ instead of $w_{n}$ and subscripts $F_{n}$ as $n.$ Consider a
sequence $\lambda_{n,h}$ with $h_{9}=\infty.$ Rewrite
\begin{equation}
\widehat{K}_{n}/c_{n}=n^{1/2}||\widehat{R}_{n}^{-1/2}(\widehat{G}_{n}%
\otimes\widehat{H}_{n}-R_{n}+(R_{n}-\widehat{R}_{n}))\widehat{R}_{n}%
^{-1/2}||/c_{n}.
\end{equation}
In the proof of Lemma \ref{CONSISTENCY GHATS} we use the uniform moment
restrictions in (\ref{def par spa}) in $\mathcal{F}_{AKP,a_{n}}$ to obtain
$\widehat{R}_{n}-R_{n}=o_{p}(1)$; here the stronger uniform moment condition
$E_{F}((||\overline{Z}_{i}||^{2}||U_{i}||^{2})^{2+\delta_{1}})\leq B$ allows
the application of a Lyapunov CLT and to establish that $n^{1/2}%
(\widehat{R}_{n}-R_{n})=O_{p}(1).$ Because by assumption $\kappa_{\min
}(R_{F_{n}})\geq\delta_{2}$ in $\mathcal{F}_{Het},$ we thus have
$n^{1/2}\widehat{R}_{n}^{-1/2}(R_{n}-\widehat{R}_{n})\widehat{R}_{n}%
^{-1/2}/c_{n}=o_{p}(1)$. Furthermore,
\begin{equation}
n^{1/2}||R_{n}^{-1/2}(\widehat{G}_{n}\otimes\widehat{H}_{n}-R_{n})R_{n}%
^{-1/2}||/c_{n}\geq n^{1/2}\lambda_{9n}\rightarrow h_{9}=\infty,
\end{equation}
where the inequality holds by the definition of $\lambda_{9n}$ in
(\ref{lambda9}). Because $\widehat{R}_{n}^{1/2}R_{n}^{-1/2}\rightarrow
_{p}I_{kp}$ and norms are continuous, it thus follows that $\widehat{K}%
_{n}/c_{n}>1$ wpa1. }

{ Method 2: The desired result is obtained using Theorem 3 in GKM22.}

\subsection{{Proofs of Results Involving the AR/AR test\label{AR/AR proof}}}

{ \textbf{Proof of Lemma \ref{ARAR test}.} \textbf{Assumption RT} is satisfied
by the AR/AR test by Theorem 8.1 in \citet{Andrews2017} noting that the
parameter space $\mathcal{F}_{AR/AR}$ in \citet[(8.8)]{Andrews2017} contains
the parameter space $\mathcal{F}_{Het}$ defined in (\ref{Hete}). In
particular, note that }$\xi_{1i}$ defined in (8.2) in \citet{Andrews2017},
equals $0$ in the linear IV model considered here and therefore the condition
in (8.8) $E_{F}\xi_{1i}^{2}$ being bounded holds trivially. Also, Assumption W
in \citet{Andrews2017} holds with the choice $\widehat{W}_{1n}=(n^{-1}%
{\textstyle\sum\nolimits_{i=1}^{n}}
\overline{Z}_{i}\overline{Z}_{i}^{\prime})^{-1}$ considered here.

{ \textbf{Assumption RP} is verified by the following argument that uses Lemma
\ref{Ingredient for ARAR test} below. To simplify notation we write $n$
instead of $w_{n}.$ Let $\widehat{\gamma}_{n}$ be an element in $\arg
\min_{\widetilde{\gamma}\in\mathfrak{R}^{m_{W}}}HAR_{n}\left(  \beta
_{0},\widetilde{\gamma}\right)  .$}

{Consider first the case where $\widehat{\gamma}_{n}\notin CS_{1n}^{+},$
defined in (\ref{cs set})$.$ Then, in particular, it must be that
$HAR_{n}\left(  \beta_{0},\widehat{\gamma}_{n}\right)  >\chi_{k,1-\alpha_{1}%
}^{2}.$ We obtain
\begin{align}
&  AR_{AKP}\left(  \beta_{0}\right)  -c_{1-\alpha}\left(  \hat{\kappa}%
_{1n},k-m_{W}\right) \nonumber\\
&  =HAR_{n}\left(  \beta_{0},\widehat{\gamma}_{n}\right)  -\chi_{k,1-\alpha
_{1}}^{2}+(\chi_{k,1-\alpha_{1}}^{2}-c_{1-\alpha}\left(  \hat{\kappa}%
_{1n},k-m_{W}\right)  )+\widetilde{B}_{n}+o_{p}\left(  1\right)  ,
\label{test stat 2nd case}%
\end{align}
where the equality follows from Lemma \ref{Ingredient for ARAR test}.\ But
$\chi_{k,1-\alpha_{1}}^{2}>\chi_{k-m_{w},1-\alpha}^{2}\geq c_{1-\alpha}\left(
\hat{\kappa}_{1n},k-m_{W}\right)  $ no matter what value $\hat{\kappa}_{1n}$
takes on. Given $m_{W}\geq1$ and }$\alpha_{1}<\alpha$ {we have that
$\chi_{k,1-\alpha_{1}}^{2}-c_{1-\alpha}\left(  \hat{\kappa}_{1n}%
,k-m_{W}\right)  >\epsilon$ wp1 for some $\epsilon>0.$ Because $\widetilde{B}%
_{n}\geq0$ it follows from $HAR_{n}\left(  \beta_{0},\widehat{\gamma}%
_{n}\right)  >\chi_{k,1-\alpha_{1}}^{2}$ that $AR_{AKP}\left(  \beta
_{0}\right)  >c_{1-\alpha}\left(  \hat{\kappa}_{1n},k-m_{W}\right)  $ wpa1. In
other words, the conditional subvector AR$_{AKP}$ test rejects wpa1.}

{Consider second the case where $\widehat{\gamma}_{n}\in CS_{1n}^{+}.$ Recall
the rejection condition of the test $\varphi_{AR/AR,\alpha-\delta,\alpha_{1}}%
$, $\inf_{\widetilde{\gamma}\in CS_{1n}^{+}}(HAR_{\beta,n}\left(  \beta
_{0},\widetilde{\gamma}\right)  -\chi_{k-m_{W},1-\alpha_{2,n}(\beta
_{0},\widetilde{\gamma})}^{2})>0.$ For any $\widetilde{\gamma}\in CS_{1n}%
^{+},$ we have $\alpha_{2,n}(\beta_{0},\widetilde{\gamma})\leq\alpha-\delta$
by (\ref{def alpha2})$.$ Therefore, in particular for $\widehat{\gamma}_{n}\in
CS_{1n}^{+}$
\begin{equation}
\chi_{k-m_{W},1-\alpha_{2,n}(\beta_{0},\widehat{\gamma}_{n})}^{2}%
>\chi_{k-m_{w},1-\alpha}^{2}+\epsilon\geq c_{1-\alpha}\left(  \hat{\kappa
}_{1n},k-m_{W}\right)  +\epsilon\label{ranking cvs}%
\end{equation}
for some $\epsilon>0.$ We thus obtain that%
\begin{align}
&  AR_{AKP,n}(\beta_{0})-c_{1-\alpha}\left(  \hat{\kappa}_{1n},k-m_{W}\right)
\nonumber\\
&  >HAR_{n}\left(  \beta_{0},\widehat{\gamma}_{n}\right)  -\chi_{k-m_{W}%
,1-\alpha_{2,n}(\beta_{0},\widehat{\gamma}_{n})}^{2}+\epsilon+\widetilde{B}%
_{n}+o_{p}\left(  1\right) \nonumber\\
&  \geq HAR_{\beta,n}\left(  \beta_{0},\widehat{\gamma}_{n}\right)
-\chi_{k-m_{W},1-\alpha_{2,n}(\beta_{0},\widehat{\gamma}_{n})}^{2}%
+\epsilon+\widetilde{B}_{n}+o_{p}\left(  1\right) \nonumber\\
&  \geq\min_{\widetilde{\gamma}\in CS_{1n}^{+}}(HAR_{\beta,n}\left(  \beta
_{0},\widetilde{\gamma}\right)  -\chi_{k-m_{W},1-\alpha_{2,n}(\beta
_{0},\widetilde{\gamma})}^{2})+\epsilon+\widetilde{B}_{n}+o_{p}\left(
1\right)  , \label{ranking test statistics}%
\end{align}
where the first inequality follows from Lemma \ref{Ingredient for ARAR test}
and (\ref{ranking cvs}), the second inequality follows from $HAR_{n}\left(
\beta_{0},\widetilde{\gamma}\right)  \geq HAR_{\beta,n}\left(  \beta
_{0},\widetilde{\gamma}\right)  $ for any $\left(  \beta_{0},\widetilde{\gamma
}\right)  $ because $M_{\tilde{D}_{n}\left(  \beta_{0},\widetilde{\gamma
}\right)  +an^{-1/2}\zeta_{1}}$ is a projection matrix, and the last
inequality follows because $\widehat{\gamma}_{n}\in CS_{1n}^{+}.$ Thus, if
$\varphi_{AR/AR,\alpha-\delta,\alpha_{1}}=1$ and $\min_{\widetilde{\gamma}\in
CS_{1n}^{+}}\allowbreak(HAR_{\beta,n}\left(  \beta_{0},\widetilde{\gamma}\right)
-\chi_{k-m_{W},1-\alpha_{2,n}(\beta_{0},\widetilde{\gamma})}^{2})>0,$ it must
also be true that $AR_{AKP,n}(\beta_{0})-\allowbreak c_{1-\alpha}\left(  \hat{\kappa}%
_{1n},k-\allowbreak m_{W}\right) \allowbreak >0$ wpa1.\footnote{Note that it is this derivation that
necessitates using $\varphi_{Rob,\alpha-\delta}$ rather than the more powerful
$\varphi_{Rob,\alpha}$ in the definition of $\varphi_{MS-AKP,\delta
,c_{n},\alpha}.$ The term $\widetilde{B}_{n}$ might go to zero and the
$o_{p}\left(  1\right)  $ term could be negative and dominate and therefore,
without the $\epsilon>0$ term we would not be able to obtain a strict
inequality between the first and second line of (\ref{ranking test statistics}%
) and thus not be able to show that $\varphi_{Rob,\alpha}\leq\varphi
_{AKP,\alpha}$ holds wpa1 under all drifting sequences. Under weak
identification we would still be able to do so; namely, if $q=q_{h}=0,$ see
(\ref{q defn}) above then Proposition \ref{Min Eigenval Prop 1}(b) implies
that $\widehat{\kappa}_{1n}=O_{p}(1)$ and given that the critical values
$c_{1-\alpha}\left(  \hat{\kappa}_{1n},k-m_{W}\right)  $ obtained by linear
interpolation from the tables in the Appendix of GKM19 are strictly increasing
in $\widehat{\kappa}_{1n}$ with $c_{1-\alpha}\left(  \hat{\kappa}_{1n}%
,k-m_{W}\right)  \rightarrow\chi_{k-m_{w},1-\alpha}^{2}$ as $\hat{\kappa}%
_{1n}\rightarrow\infty$ it follows that there is a $\gamma>0$ such that
$\chi_{k-m_{w},1-\alpha}^{2}\geq c_{1-\alpha}\left(  \hat{\kappa}_{1n}%
,k-m_{W}\right)  +\gamma$ wpa1. Then, (\ref{ranking test statistics}) implies
that $\varphi_{Rob,\alpha}\leq\varphi_{AKP,\alpha}$ holds wpa1. But that
argument does not go through when $q=q_{h}\geq1.$} }

{ The inequalities in (\ref{ranking cvs}) and (\ref{ranking test statistics})
immediately imply the desired result%
\begin{align}
&  P_{\lambda_{w_{n},h}}(\varphi_{Rob,\alpha-\delta}\overset{}{\leq}%
\varphi_{AKP,\alpha})\nonumber\\
&  =P_{\lambda_{w_{n},h}}((\varphi_{Rob,\alpha-\delta}\leq\varphi_{AKP,\alpha
})\cap(\widehat{\gamma}_{n}\in CS_{1n}^{+}))+P_{\lambda_{w_{n},h}}%
((\varphi_{Rob,\alpha-\delta}\leq\varphi_{AKP,\alpha})\cap(\widehat{\gamma
}_{n}\notin CS_{1n}^{+}))\nonumber\\
&  \rightarrow1. \label{conclusion}%
\end{align}
$\square$ }

{ Recall that $\widehat{\gamma}_{w_{n}}$ is an element in $\arg\min
_{\widetilde{\gamma}\in\mathfrak{R}^{m_{W}}}HAR_{w_{n}}\left(  \beta
_{0},\widetilde{\gamma}\right)  $ and $\gamma_{w_{n}}^{+}$ is an element in
\newline $\arg\min_{\widetilde{\gamma}\in\mathfrak{R}^{m_{W}}}\widetilde{AR}%
_{AKP,w_{n}}(\beta_{0},\widetilde{\gamma})$. }

\begin{lemma}
{ \label{Ingredient for ARAR test}Consider a sequence $\lambda_{w_{n},h}$
$($of reparameterized elements in $\mathcal{F}_{Het})$ with $h_{9}<\infty$
$($that is, a sequence of AKP structure$).$ If }$\gamma_{w_{n}}^{+}=O_{p}(1)${
and }$\Pi_{Ww_{n}}w_{n}^{1/2}(\gamma_{w_{n}}^{+}-\gamma_{w_{n}})=O_{p}(1)${
then along $\lambda_{w_{n},h}$
\[
AR_{AKP,w_{n}}(\beta_{0})=HAR_{w_{n}}\left(  \beta_{0},\widehat{\gamma}%
_{w_{n}}\right)  +\widetilde{B}_{w_{n}}+o_{p}\left(  1\right)
\]
for some random sequence $\widetilde{B}_{w_{n}}$ that is nonnegative wp1.}
\end{lemma}

{ \textbf{\noindent Proof.} To simplify notation we write $n$ instead of
$w_{n}.$ Recall from (\ref{har def})
\begin{align}
HAR_{n}\left(  \beta_{0},\widetilde{\gamma}\right)   &  =\hspace
{-0.03in}n\widehat{g}_{n}\left(  \beta_{0},\widetilde{\gamma}\right)
^{\prime}\hat{\Sigma}_{n}\left(  \beta_{0},\widetilde{\gamma}\right)
^{-1}\widehat{g}_{n}\left(  \beta_{0},\widetilde{\gamma}\right) \nonumber\\
&  =\hspace{-0.03in}n\binom{1}{-\widetilde{\gamma}}^{\prime}\left(
\overline{Y}_{0},W\right)  ^{\prime}\overline{Z}\hat{\Sigma}_{n}\left(
\beta_{0},\widetilde{\gamma}\right)  ^{-1}\overline{Z}^{\prime}\left(
\overline{Y}_{0},W\right)  \binom{1}{-\widetilde{\gamma}}. \label{HAR stat}%
\end{align}
}

{Defining $b_{n}^{+}:=\left(  1,-\beta_{0}^{\prime},-\gamma_{n}^{+\prime
}\right)  ^{\prime}$ it follows that under the null%
\begin{equation}
\overline{Y}_{0i}-W_{i}^{\prime}\gamma_{n}^{+}=y_{i}-Y_{i}^{\prime}\beta
_{0}-W_{i}^{\prime}\gamma_{n}^{+}=v_{y,i}-V_{Y,i}^{\prime}\beta_{0}%
-V_{W,i}^{\prime}\gamma_{n}^{+}+\overline{Z}_{i}^{\prime}\Pi_{Wn}%
(\gamma-\gamma_{n}^{+})=V_{i}^{\prime}b_{n}^{+}+\overline{Z}_{i}^{\prime}%
\Pi_{Wn}(\gamma-\gamma_{n}^{+}). \label{deriv 1}%
\end{equation}
Define
\begin{equation}
\xi_{in}:=\overline{Z}_{i}\overline{Z}_{i}^{\prime}\Pi_{Wn}(\gamma-\gamma
_{n}^{+})\in\Re^{k}{\ }\text{and }{\overline{\xi}_{n}:=n^{-1}%
{\textstyle\sum\nolimits_{i=1}^{n}}
\xi_{in}.\ } \label{xi}%
\end{equation}
We then have%
\begin{align}
&  n\hat{\Sigma}_{n}\left(  \beta_{0},\gamma_{n}^{+}\right) \nonumber\\
&  =\hspace{-0.03in}%
{\textstyle\sum\nolimits_{i=1}^{n}}
\left[  \overline{Z}_{i}(\overline{Y}_{0i}-W_{i}^{\prime}\gamma_{n}%
^{+})-\overline{Z}^{\prime}\left(  \overline{Y}_{0}-W\gamma_{n}^{+}\right)
/n\right]  \left[  \overline{Z}_{i}(\overline{Y}_{0i}-W_{i}^{\prime}\gamma
_{n}^{+})-\overline{Z}^{\prime}\left(  \overline{Y}_{0}-W\gamma_{n}%
^{+}\right)  /n\right]  ^{\prime}\nonumber\\
&  =\hspace{-0.03in}%
{\textstyle\sum\nolimits_{i=1}^{n}}
(\overline{Y}_{0i}-W_{i}^{\prime}\gamma_{n}^{+})^{2}\overline{Z}_{i}%
\overline{Z}_{i}^{\prime}-\overline{Z}^{\prime}\left(  \overline{Y}%
_{0}-W\gamma_{n}^{+}\right)  \left(  \overline{Y}_{0}-W\gamma_{n}^{+}\right)
^{\prime}\overline{Z}/n\nonumber\\
&  =\hspace{-0.03in}%
{\textstyle\sum\nolimits_{i=1}^{n}}
\left[  \left(  V_{i}^{\prime}b_{n}^{+}\right)  ^{2}+2(V_{i}^{\prime}b_{n}%
^{+}\overline{Z}_{i}^{\prime}\Pi_{Wn}(\gamma-\gamma_{n}^{+}))+\left(
\overline{Z}_{i}^{\prime}\Pi_{Wn}(\gamma-\gamma_{n}^{+})\right)  ^{2}\right]
\overline{Z}_{i}\overline{Z}_{i}^{\prime}\nonumber\\
&  -(\overline{Z}^{\prime}Vb_{n}^{+}b_{n}^{+\prime}V^{\prime}\overline
{Z}+2\overline{Z}^{\prime}Vb_{n}^{+}(\gamma-\gamma_{n}^{+})^{\prime}\Pi
_{Wn}^{\prime}\overline{Z}^{\prime}\overline{Z}+\overline{Z}^{\prime}%
\overline{Z}\Pi_{Wn}(\gamma-\gamma_{n}^{+})(\gamma-\gamma_{n}^{+})^{\prime}%
\Pi_{Wn}^{\prime}\overline{Z}^{\prime}\overline{Z})/n\nonumber\\
&  =\hspace{-0.03in}%
{\textstyle\sum\nolimits_{i=1}^{n}}
\left(  V_{i}^{\prime}b_{n}^{+}\right)  ^{2}\overline{Z}_{i}\overline{Z}%
_{i}^{\prime}+%
{\textstyle\sum\nolimits_{i=1}^{n}}
\left(  \xi_{in}-\overline{\xi}_{n}\right)  \left(  \xi_{in}-\overline{\xi
}_{n}\right)  ^{\prime}\nonumber\\
&  +2%
{\textstyle\sum\nolimits_{i=1}^{n}}
(V_{i}^{\prime}b_{n}^{+}\overline{Z}_{i}^{\prime}\Pi_{Wn}(\gamma-\gamma
_{n}^{+}))\overline{Z}_{i}\overline{Z}_{i}^{\prime}-2\overline{Z}^{\prime
}Vb_{n}^{+}(\gamma-\gamma_{n}^{+})^{\prime}\Pi_{Wn}^{\prime}\overline
{Z}^{\prime}\overline{Z}/n-\overline{Z}^{\prime}Vb_{n}^{+}b_{n}^{+\prime
}V^{\prime}\overline{Z}/n\nonumber\\
&  =\hspace{-0.03in}%
{\textstyle\sum\nolimits_{i=1}^{n}}
\left(  V_{i}^{\prime}b_{n}^{+}\right)  ^{2}\overline{Z}_{i}\overline{Z}%
_{i}^{\prime}+O_{p}(n^{1/2}), \label{variance esti}%
\end{align}
where for the third equality we use (\ref{deriv 1}) and $\overline{Z}^{\prime
}\left(  \overline{Y}_{0}-W\gamma_{n}^{+}\right)  =\overline{Z}^{\prime}%
Vb_{n}^{+}+\overline{Z}^{\prime}\overline{Z}\Pi_{Wn}(\gamma-\gamma_{n}^{+}),$
in the fifth equality we apply a WLLN or a Lyapunov CLT theorem for each of
the last three summands in the second to last line and the second summand in
the third to last line which hold by the moment conditions imposed in the
parameter space $\mathcal{F}_{Het}$ in (\ref{Hete}). In particular, using
}$\gamma_{n}^{+}=O_{p}(1)$ and $\Pi_{Wn}n^{1/2}(\gamma_{n}^{+}-\gamma
_{n})=O_{p}(1),$ {the first summand in the second to last line is
$O_{p}(n^{1/2})$ while the other summands are $O_{p}(1).$}

{The first summand in the last line of (\ref{variance esti}) can be expanded
as follows after normalization by $n^{-1}$.%
\begin{align*}
&  n^{-1}%
{\textstyle\sum\nolimits_{i=1}^{n}}
\left(  V_{i}^{\prime}b_{n}^{+}\right)  ^{2}\overline{Z}_{i}\overline{Z}%
_{i}^{\prime}\\
&  =\left(  b_{n}^{+}\otimes I_{k}\right)  ^{\prime}n^{-1}%
{\textstyle\sum\nolimits_{i=1}^{n}}
\left(  V_{i}\otimes\overline{Z}_{i}\right)  \left(  V_{i}\otimes\overline
{Z}_{i}\right)  ^{\prime}\left(  b_{n}^{+}\otimes I_{k}\right) \\
&  =\left(  \binom{1}{-\gamma_{n}^{+}}\otimes I_{k}\right)  ^{\prime
}\underbrace{n^{-1}%
{\textstyle\sum\nolimits_{i=1}^{n}}
\left(  \binom{v_{yi}-V_{Yi}^{\prime}\beta_{0}}{V_{Wi}}\otimes\overline{Z}%
_{i}\right)  \left(  \binom{v_{yi}-V_{Yi}^{\prime}\beta_{0}}{V_{Wi}}%
\otimes\overline{Z}_{i}\right)  ^{\prime}}_{=:\widehat{\overline{R}}_{F_{n}}%
}\left(  \binom{1}{-\gamma_{n}^{+}}\otimes I_{k}\right)  .
\end{align*}
When $\beta_{0}=\beta$ (which is assumed here) we have%
\begin{equation}
\widehat{\overline{R}}_{F_{n}}=E_{F_{n}}(vec(\overline{Z}_{i}U_{i}^{\prime
})(vec(\overline{Z}_{i}U_{i}^{\prime}))^{\prime})+o_{p}(1)=G_{F_{n}}%
\otimes\overline{H}_{F_{n}}+\Upsilon_{n}+o_{p}(1), \label{Rbarhat}%
\end{equation}
for some $\Upsilon_{n}=o(1),$ where the first equality holds by a WLLN and the
second one holds by the assumption that $n^{1/2}\lambda_{9n}\rightarrow
h_{9}<\infty$ and the argument given in the Proof of Theorem
\ref{correct asy size generic} that establishes that $\overline{R}_{F_{n}}$
has AKP structure.}

{Therefore, by (\ref{our test statistic rewritten})
\begin{align}
&  \hat{\Sigma}_{n}\left(  \beta_{0},\gamma_{n}^{+}\right)  -\widetilde{\Sigma
}\left(  \beta_{0},\gamma_{n}^{+}\right) \nonumber\\
&  =n^{-1}%
{\textstyle\sum\nolimits_{i=1}^{n}}
\left(  V_{i}^{\prime}b_{n}^{+}\right)  ^{2}\overline{Z}_{i}\overline{Z}%
_{i}^{\prime}-(\left(  1,-\gamma_{n}^{+\prime}\right)  \widehat{G}_{n}\left(
1,-\gamma_{n}^{+\prime}\right)  ^{\prime})\otimes(n^{-1}\overline{Z}^{\prime
}\overline{Z})^{1/2}\widehat{H}_{n}(n^{-1}\overline{Z}^{\prime}\overline
{Z})^{1/2}+o_{p}(1)\nonumber\\
&  =o_{p}\left(  1\right)  , \label{var mat diff}%
\end{align}
where the last line follows from }$\gamma_{n}^{+}=O_{p}(1),$ {\ref{Rbarhat}, a
WLLN, and Lemma \ref{CONSISTENCY GHATS}. Therefore,%
\begin{equation}
HAR_{n}\left(  \beta_{0},\gamma_{n}^{+}\right)  =n\widehat{g}\left(  \beta
_{0},\gamma_{n}^{+}\right)  ^{\prime}\left[  \widetilde{\Sigma}\left(
\beta_{0},\gamma_{n}^{+}\right)  +o_{p}\left(  1\right)  \right]
^{-1}\widehat{g}\left(  \beta_{0},\gamma_{n}^{+}\right)  =\widetilde{AR}%
_{AKP,n}(\beta_{0},\gamma_{n}^{+})+o_{p}(1), \label{final outcome}%
\end{equation}
where we use positive definiteness of $\widetilde{\Sigma}\left(  \beta
_{0},\gamma_{n}^{+}\right)  $ in the last equality which holds by the
restrictions on }$E_{F}(\overline{Z}_{i}^{\prime}\overline{Z}_{i}),G_{F},$ and
$\overline{H}_{F}$ {in (\ref{def par spa}).}

{By definition of }$\widehat{\gamma}_{n},$ $HAR_{n}\left(  \beta_{0}%
,\gamma_{n}^{+}\right)  \geq HAR_{n}\left(  \beta_{0},\widehat{\gamma}%
_{n}\right)  $. {By definition of $\gamma_{n}^{+},$ $AR_{AKP,n}(\beta
_{0})=\widetilde{AR}_{AKP,n}(\beta_{0},\gamma_{n}^{+}).$ Thus, by
(\ref{final outcome})
\begin{equation}
{AR_{AKP,n}(\beta_{0})=}HAR_{n}\left(  \beta_{0},\gamma_{n}^{+}\right)
+o_{p}(1)\geq HAR_{n}\left(  \beta_{0},\widehat{\gamma}_{n}\right)  +o_{p}(1),
\label{last step}%
\end{equation}
which is the desired result. $\square$ }

\subsection{{Time series case\label{time series section}}}

{ In this section we drop Assumption B and allow for a stationary time series
setup. In the time series case, $F$ denotes the distribution of the stationary
infinite sequence $\{(\overline{Z}_{i}^{\prime},V_{i}^{\prime})^{\prime
}:i=...,0,1,...\}.$ Recall the definition $U_{i}:=(\varepsilon_{i}%
+V_{W,i}^{\prime}\gamma,V_{W,i}^{\prime})^{\prime}$ and define%
\begin{equation}
\overline{R}_{F,n}:=Var_{F}\left(  n^{-1/2}%
{\textstyle\sum\nolimits_{i=1}^{n}}
vec(\overline{Z}_{i}U_{i}^{\prime})\right)  . \label{Var matr ts}%
\end{equation}
Consider again a sequence $a_{n}=o(1)$ in $\Re_{\geq0}.$ The parameter space
is given by%
\begin{align}
\mathcal{F}_{TS,AKP,a_{n}}\overset{}{:}  &  =\{(\gamma,\Pi_{W},\Pi
_{Y},F):\gamma\overset{}{\in}\Re^{m_{W}},\Pi_{W}\overset{}{\in}\Re^{k\times
m_{W}},\Pi_{Y}\overset{}{\in}\Re^{k\times m_{Y}},\{(\overline{Z}_{i}%
,V_{i}):i=...,0,1,...\}\text{ }\nonumber\\
&  \text{are stationary and strong mixing under }F\text{ with strong mixing
numbers }\nonumber\\
&  \{\alpha_{F}(m)\overset{}{:}m\overset{}{\geq}1\}\text{ that satisfy }%
\alpha_{F}(m)\overset{}{\leq}Cm^{-d},\nonumber\\
&  E_{F}(\overline{Z}_{i}V_{i}^{\prime})\overset{}{=}0^{k\times(m+1)},\text{
}\overline{R}_{F,n}\overset{}{=}G_{F}\otimes\overline{H}_{F}+\Upsilon
_{n},\nonumber\\
&  E_{F}(||T_{i}||^{2+\delta})\overset{}{\leq}B,\text{ for }T_{i}%
\overset{}{\in}\{vec(\overline{Z}_{i}U_{i}^{\prime}),||\overline{Z}_{i}%
||^{2}\}\nonumber\\
&  \kappa_{\min}(A)\overset{}{\geq}\delta\text{ for }A\overset{}{\in}%
\{E_{F}\overline{Z}_{i}\overline{Z}_{i}^{\prime},G_{F},\overline{H}_{F}\}\}
\label{par space ts}%
\end{align}
for some $\delta>0,$ $d>(2+\delta)/\delta$, $B,C<\infty$, for symmetric
matrices $\Upsilon_{n}\in\Re^{kp\times kp}$ such that $||\Upsilon_{n}||\leq
a_{n},$ pd symmetric matrices $G_{F}\in\Re^{p\times p}$ (whose upper left
element is normalized to 1) and $\overline{H}_{F}\in\Re^{k\times k}.$ }

{ In the time series context, the definition of $\widehat{R}_{n}$ in
(\ref{fi}) is replaced by a heteroskedasticity and autocorrelation consistent
(HAC) variance matrix estimator based on $\{f_{i}:i\leq n\}$ for
$R_{F,n}:=(I_{p}\otimes(E_{F}\overline{Z}_{i}\overline{Z}_{i}^{\prime}%
)^{-1/2})\overline{R}_{F,n}(I_{p}\otimes(E_{F}\overline{Z}_{i}\overline{Z}%
_{i}^{\prime})^{-1/2}),$ e.g. see \citet{NeweyWest1987} and
\citet{Andrews1991}\textbf{.} With this modification, the conditional
subvector AR$_{AKP}$ test for the time series case is then defined exactly as
in (\ref{rej condition in asy model}). Theorem \ref{correct asy size} then
holds without Assumption B and with $\mathcal{F}_{AKP,a_{n}}$ replaced by
$\mathcal{F}_{TS,AKP,a_{n}}.\medskip$ }

{ \textbf{Comment. 1.} The proof of the theorem in the time series case
follows the exact same steps as the proof of Theorem \ref{correct asy size} in
the i.i.d.~case in the Appendix with simple modifications. In particular,
define sequences $\{\lambda_{w_{n},h}:n\geq1\}$ as in (\ref{Defn lambda n,h})
but with $\mathcal{F}_{AKP,a_{n}}$ replaced by $\mathcal{F}_{TS,AKP,a_{n}}$ in
(\ref{Defn of h_n(lambda)}). Then, under sequences $\lambda_{n,h}$ (writing
$n$ instead of $w_{n}$ to simplify notation), the HAC estimator $\widehat{R}%
_{n}$ satisfies $\widehat{R}_{n}-R_{F,n}\rightarrow_{p}0^{kp\times kp}$ and
thus $\widehat{R}_{n}\rightarrow_{p}h_{7}^{-2}\otimes h_{4}^{1/2}h_{6}%
^{-1}h_{6}^{^{\prime}-1}h_{4}^{1/2}$ see earlier sections for notation. Also,
the CLT in (\ref{CLT and WLLN}) continues to hold under the mixing conditions
in $\mathcal{F}_{TS,AKP,a_{n}}.$ Then, the exact same proof as for the i.i.d.
case applies. }

{ 2. Again, we obtain the corresponding result for the generalization of the
subvector test in GKMC to the time series KP structure case. This test has
correct asymptotic size for the parameter space $\mathcal{F}_{TS,AKP,a_{n}}$
and the result is obtained fully analytically; its proof does not require any
simulations.\bigskip}

\bibliographystyle{chicago}
\bibliography{AR_AKP}

\end{document}